\documentclass[manuscript]{acmart}
\usepackage{array}
\usepackage{geometry}
\usepackage{longtable}
\usepackage{booktabs}
\usepackage{float}
\usepackage{tabularx}

\renewcommand{\arraystretch}{1.4}
\newcolumntype{L}[1]{>{\raggedright\arraybackslash}p{#1}}
\usepackage{verbatim}

\AtBeginDocument{%
  }

\setcopyright{acmlicensed}
\copyrightyear{2018}
\acmYear{2018}
\acmDOI{XXXXXXX.XXXXXXX}

\acmConference[Conference acronym 'XX]{Make sure to enter the correct
  conference title from your rights confirmation emai}{June 03--05,
  2018}{Woodstock, NY}
\acmISBN{978-1-4503-XXXX-X/18/06}

\begin{document}

\newtoggle{comments}
\togglefalse{comments}
\toggletrue{comments}

\iftoggle{comments} {
\newcommand{\markup}[1]{{#1\normalfont}}
\newcommand{\auste}[1]{{\color{magenta}{AS: #1}\normalfont}}
\newcommand{\viktor}[1]{{\color{cyan}{VK: #1}\normalfont}}
\newcommand{\sean}[1]{{\color{green}{SR: #1}\normalfont}}
}{
  \newcommand {\lev}[1]{}
  \newcommand {\auste}[1]{}
  \newcommand {\viktor}[1]{}
  \newcommand {\sean}[1]{}
 }

 \newcommand{\iquote}[1]{\textit{\textcolor{brown}{#1}}}

\newcommand{\pid}[1]{{\fontfamily{cmss}\selectfont{\footnotesize{{\textcolor{black!50}{#1}}}}}}

\newenvironment{smallquote}
  {\begin{quote}\normalsize}
  {\end{quote}}

\newcommand {\qspace}[1]{\vspace{0.3\baselineskip}}

\title[Generative AI in Higher Education]{The New Calculator? Practices, Norms, and Implications of Generative AI in Higher Education}

\author{Auste Simkute}
\email{a.simkute@sms.ed.ac.uk}
\authornote{Both authors contributed equally to this research.}
\affiliation{%
  \institution{University of Edinburgh}
  \city{Edinburgh}
  \country{United Kingdom}}

\author{Viktor Kewenig}
\email{ucjuvnk@ucl.ac.uk}
\authornotemark[1]
\affiliation{%
  \institution{University College London}
  \city{London}
  \country{United Kingdom}
}

\author{Abigail Sellen}
\email{asellen@microsoft.com}
\affiliation{%
  \institution{Microsoft Research}
  \city{Cambridge}
  \country{United Kingdom}}

\author{Sean Rintel}
\email{serintel@microsoft.com}
\affiliation{%
  \institution{Microsoft Research}
  \city{Cambridge}
  \country{United Kingdom}}

  \author{Lev Tankelevitch}

\email{lev.tankelevitch@microsoft.com}
\affiliation{%
  \institution{Microsoft Research}
  \city{Cambridge}
  \country{United Kingdom}
}

\renewcommand{\shortauthors}{Simkute and Kewenig et al.}

\begin{abstract}
Generative AI (GenAI) has introduced myriad opportunities and challenges for higher education. Anticipating this potential transformation requires understanding students’ contextualised practices and norms around GenAI. We conducted semi-structured interviews with 26 students and 11 educators from diverse departments across two universities. \markup{Grounded in Strong Structuration Theory, we find diversity in students’ uses and motivations for GenAI.} Occurring in the context of unclear university guidelines, institutional fixation on plagiarism, and inconsistent educator communication, students’ practices are informed by unspoken rules around appropriate use, GenAI limitations and reliance strategies, and consideration of agency and skills. Perceived impacts include changes in confidence, and concerns about skill development, relationships with educators, and plagiarism. Both groups envision changes in universities’ attitude to GenAI, responsible use training, assessments, and integration of GenAI into education. We discuss socio-technical implications \markup{in terms of current and anticipated changes in the external and internal structures that contextualise students’ GenAI use.}

\end{abstract}


\keywords{Generative AI, higher education, university, students, educators, teachers, learning, technology adoption, policy, skill development, academic integrity, technology literacy, reliance, trust, impact, norms, etiquette, novelty, future}
\begin{teaserfigure}
\includegraphics[width=\textwidth]{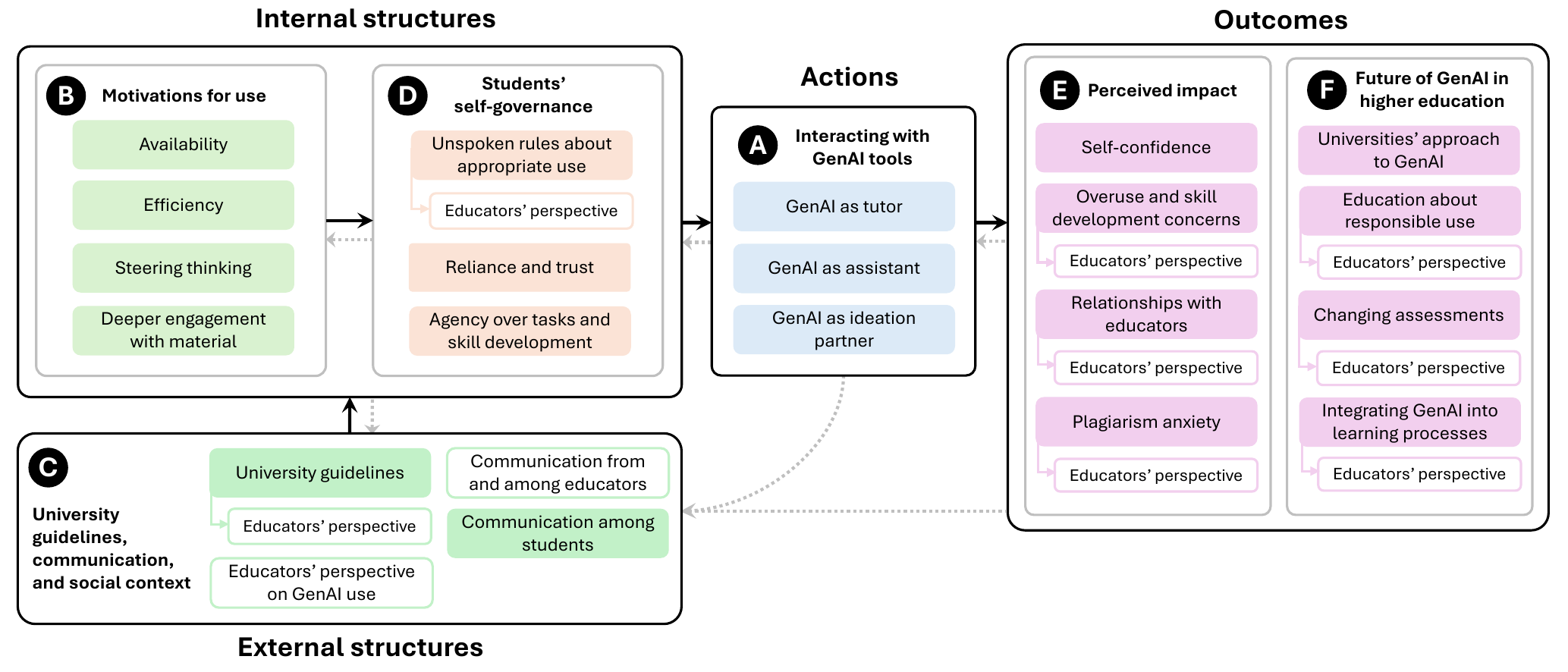}
  \caption{Overview of study findings%
  . Students' \textit{actions} (A) (GenAI tool uses) are driven by a set of \textit{internal structures}, including their motivations (B), and self-governance of GenAI use (D). These internal structures are influenced by a set of \textit{external structures}: %
  including university guidelines, communication from and among educators, and communication among students (C). Finally, students' actions leads to \textit{outcomes}, including perceived impacts on confidence, relationships, etc. %
  (E), as well as students' and educators' visions for the future of GenAI in higher education (F). Although our findings focus on a set of directional influences between these aspects%
  , it is likely that the influence goes both ways. We indicate these currently unexplored influences with dashed grey arrows. Our comparative analysis includes both students' and educators' perspectives where relevant; filled and unfilled boxes indicate these sources, respectively.
  }
  \Description{Overview of study findings, as guided by Strong Structuration Theory.}
  \label{FIG:results-overview}
\end{teaserfigure}

\received{20 February 2007}
\received[revised]{12 March 2009}
\received[accepted]{5 June 2009}

\maketitle

\section{Introduction} \label{SEC:intro}

The rapid diffusion of ChatGPT and other generative artificial intelligence (GenAI) tools %
has raised fundamental questions about teaching and learning models, assessment, and the value of different skills. University groups have responded with stop-gap guiding principles \cite{noauthor_russell_2023}, focusing on %
safe, responsible, fair, and effective use. However, the practical implementation is left to individual universities in the context of student usage, leaving those on the front line grappling with open questions \cite{noauthor_general_2023}. Guiding the potential transformation enabled by GenAI requires understanding more than just its specific effects on learning. Both policymakers and researchers need to understand its practical realities as part of an ecosystem \cite{bennett_how_2023}, balancing its technological potential with recognition of the complexities of teaching and learning, agency of both students and educators, and challenges of its integration. 

\markup{Broadly, %
existing work has provided high-level overviews of higher education students’ perceptions, concerns, and attitudes around GenAI (e.g., \cite{chan_students_2023,smolansky_educator_2023,johnston_student_2024,ghimire_generative_2024,prather_robots_2023}). However, it has under-explored the \textit{real-world practices} that people engage in. As \citet{stone_exploring_2024} concludes, there is a “neglect of ‘what students actually do’”
 with GenAI tools. Other than exploring common use cases for GenAI \cite{hasanein_drivers_2023,chan_students_2023,park_promise_2024}, research has not examined broader practices that likely shape the adoption of GenAI, such as communication about GenAI and the development of AI reliance strategies. Moreover, research has not \textit{situated} practices and attitudes within the educational context---such as university guidelines and communication norms---necessary to understand how these practices and attitudes emerge, and in turn, shape the broader educational context. Finally, HCI research in this space has predominantly focused on computer science education (e.g., \cite{prather_robots_2023, hou_effects_2024,hellas_experiences_2024}), with other disciplines overlooked.\footnote{Other work has explored situated practices and attitudes around GenAI in lower education \cite{han_teachers_2024,tan_more_2024}.}}  

To address this gap, we conducted interviews and surveys with university students (n = 26) and educators (n = 11) from departments across two UK universities, providing a `snapshot in time' of emerging practices, attitudes, norms, and perspectives, guided by Strong Structuration Theory (SST) \cite{stones_structuration_2005,greenhalgh_theorising_2010}. 

We find a diversity in students' uses of GenAI (e.g., as a tutor, assistant) and motivations (e.g., availability, deeper engagement with material). Students' practices occur in the absence of clear university guidelines, institutional fixation on plagiarism, and inconsistent communication from educators. Accordingly, their practices are informed by unspoken rules around appropriate use, GenAI limitations and reliance-mediating strategies, and considerations of task agency and skill development. Corroborated by both students and educators, perceived impacts of students' GenAI use include changes in confidence, concerns about skill development and relationships with educators, and plagiarism anxiety. Both groups expect changes in universities' attitude to GenAI, communication and training around responsible use, formal integration into learning processes, and adaptation of assessments to the realities of GenAI. 

\markup{We discuss our findings from the perspective of SST, outlining issues reflecting the current conjuncture, and anticipated shifts in the external and internal structures within higher education, driven by the evolving adoption of GenAI. In summary, we make the following contributions:
\begin{enumerate}
    \item In-depth findings on higher education students' broader practices around GenAI, including uses, reliance-mediating strategies, and communication norms, covering multiple disciplines and year groups.  
    \item A holistic, comparative understanding of students' and educators' perspectives on (a) the broader structures that shape, and are shaped by, students' practices, including university guidelines, usage norms, and student and educator communication norms; and (b) perceived impacts on students and future expectations for GenAI in higher education.
    \item A guiding model---grounded in Strong Structuration Theory and integrating the micro, meso, and macro perspectives---key for understanding the evolving adoption of GenAI in higher education, and informing both policy and HCI design. 
    
\end{enumerate}}

\section{Related Work} \label{SEC:relatedwork}

\subsection{Integrating AI into higher education}
Before ChatGPT-3.5's release in late 2022 ushered in the current era of GenAI, the integration of AI into higher education was already aiming to democratize education and offer personalized learning at scale. Initial studies showed promise \cite{vanlehn_relative_2011,koedinger_new_2013}, but subsequent research found inconsistent impacts across subjects and limited educational transformation \cite{kulik_effectiveness_2016, reich_best_2015}.
This was attributed to the complexities of learning, the need for diverse pedagogies, and the constraints of algorithmic personalization \cite{winne_psychology_2010,azevedo_lessons_2022,baker_educational_2014}. 

GenAI offers the potential to revolutionize education through its remarkable contextual adaptability and flexible output generation \cite{bubeck_sparks_2023}. 
It enables delivering algorithmic-guided instructions that are more attuned to individual student needs during the learning process \cite{prather_robots_2023}. Studies highlight that GenAI tools should adapt to specific learning objectives, e.g., STEM students' objectives will differ from those in Business \cite{kubullek_understanding_2024}. \markup{Albeit focusing on lower education, \citet{tan_more_2024} identify middle- and high-school teachers' information needs for integrating ChatGPT into their classrooms (e.g., data policies, content restrictions, performance), and develop a framework for designing classroom-based ChatGPT tools, focusing on learning objectives, student needs, technical capabilities, and other considerations. In workshops and interviews \cite{han_teachers_2024}, elementary school teachers were keen to integrate GenAI into teaching and digital literacy development, with students similarly optimistic about GenAI (although parents were more concerned, particularly around safety, privacy, and development of students' critical thinking).}

Research also demonstrates the value of GenAI for co-writing platforms \cite{lee_evidence_2023,gero_social_2023}, image generation for 3D animation courses \cite{kicklighter_empowering_2024}, code generation for computer science education \cite{petrovska_incorporating_2024}, and chatbots for language and creative writing education \cite{han_teachers_2024}. Some work has explored integrating GenAI into assessments \cite{smolansky_educator_2023}, e.g., creating personalized tests \cite{morales-chan_ai-driven_2024}. \markup{A particular focus has been on examining how students learn programming with GenAI tools, often in classroom or lab settings \cite{prather_widening_2024,margulieux_self-regulation_2024,kazemitabaar_how_2023,kazemitabaar_studying_2023}. Grounded in this work, tools are being developed to support the use of GenAI in CS education \cite{liffiton_codehelp_2023,kazemitabaar_improving_2024,denny_prompt_2024,sheese_patterns_2024,denny_explaining_2024}.} In sum, these studies show that GenAI can enhance students' engagement, idea generation, and productivity, but also raise questions around authorship, agency, and critical thinking \cite{han_teachers_2024,prather_robots_2023,ghimire_generative_2024}.

\markup{\subsection{Practices and attitudes around the use of GenAI in higher education}
Akin to the early days of Wikipedia use in education \cite{bayliss_exploring_2013,olsen_i_2012}, research has begun to sketch emerging practices and attitudes about acceptable use of GenAI. Due to the early adoption of GenAI tools in computer science (CS), much work has focused on CS education, with fewer in-depth studies on other disciplines.

\subsubsection{Computer science education}
 \citet{prather_robots_2023} reviewed early research (e.g., \cite{amani_generative_2023,rajabi_exploring_2023, zastudil_generative_2023}) and, in a multi-part study of GenAI in CS education, conducted a mixed-methods survey of students and educators, and interviews with expert educators. Most students used GenAI in coursework (e.g., for debugging or code generation), with non-users citing de-skilling and ethical concerns. A third of educators observed students using GenAI in both legitimate and unethical ways. Students varied in how clear they felt their university policies around GenAI to be, whereas educators consistently reported that policies are unclear---indeed, many university policies focused on delineating the harms and limitations of GenAI use. According to both groups, although GenAI cannot replace human educators, it will play an increasing role in education. Students varied in their feelings about the impact of GenAI on their career prospects. Many educators reported not using GenAI, although expressed an interest in it.  %
Some interviewed educators changed their learning objectives to focus on critical thinking and code comprehension, while others retained a focus on core CS concepts. In assessment, educators report shifting their focus to `process over product'.

Many of the above educator-related findings are echoed in \cite{sheard_instructor_2024}, and \cite{mahon_guidelines_2024}, who propose six descriptive levels at which educators integrate GenAI into their introductory CS courses, ranging from GenAI avoidance to embedding GenAI as a core theme of the course.    

In another study of CS students \cite{hou_effects_2024}, ChatGPT followed internet search and instructors as the preferred source for help-seeking due to its iterative and adaptive nature. This was mediated by perceptions of the trade-off between tools' convenience and output quality, social considerations, and iterative support needs. Even when GenAI was openly available in CS courses, \citet{hellas_experiences_2024} found that most usage came from a few superusers, particularly in software engineering, who used it both for coursework and other purposes. This aligns with early quantitative surveys showing that CS students are predominantly aware of GenAI, find it helpful for learning, but vary in their trust in its accuracy \cite{amoozadeh_trust_2024, shoufan_exploring_2023, singh_exploring_2023}. \citet{park_promise_2024} conducted workshops and participatory design sessions introducing GenAI to an undergraduate AI class. Students recognised the efficiency, availability, and personalisability of GenAI, yet identified challenges such as inaccuracies, and concerns around de-skilling, academic integrity, and social isolation.

\subsubsection{Other educational disciplines}
Outside CS education, \citet{chan_students_2023} found that students across disciplines are aware of, and positive about, GenAI tools due to their personalizability, availability, and ability to support writing, brainstorming, research, and other tasks. However, they also worried about inaccuracies, privacy, ethics, skill development, and career prospects. In \cite{chan_comprehensive_2023}, both students and educators desired clarity on ethical usage of GenAI, AI literacy training, changes in assessments (as per \cite{smolansky_educator_2023}), development of skills like critical thinking, and preparation for future careers. \citet{ghimire_generative_2024} found that over 40\% of surveyed educators across disciplines use GenAI at least periodically, and felt positively about benefits outweighing risks, opportunities for enhancing education, and ease of integration, but had similar concerns as above. 
 \citet{hasanein_drivers_2023}, while outlining similar uses  cases and positive attitudes for ChatGPT among students and educators in business-related disciplines, found that only educators identified potential negative impacts of GenAI, such as over-reliance, compromised academic integrity, and de-skilling. \citet{chan_will_2024} also find that students worried less about the negative impacts of GenAI, yet both groups viewed educators' socio-emotional competencies as irreplaceable. In writing, \citet{johnston_student_2024} found that students' confidence in writing inversely correlates with their using GenAI (aligning with perceived self-confidence changes with GenAI use \cite{hasanein_drivers_2023}), and that students supported using GenAI for writing \textit{assistance}, but not for completing entire assignments, a view echoed by students and educators in \cite{barrett_not_2023}. 

\subsubsection{Summary} Both students and educators in CS and other disciplines are predominantly aware of GenAI. Students use it more often overall, although vary in usage. Both groups tend to recognise GenAI's strengths, weaknesses, and associated concerns, although students in some studies appear to neglect the downsides. Many educators recognise the opportunities of integrating GenAI into education, yet need support in doing so effectively while mitigating risks. 

\section{Motivation and Research Questions}

The reviewed prior studies are undoubtedly valuable, but are limited for three reasons. First, they predominantly reflect students' and educators' perceptions, concerns, and attitudes about GenAI, rather than understanding the \textit{real-world practices} that people engage in. Referring to CS education research, \citet{stone_exploring_2024} concludes that there is a ``neglect of \textit{`what students actually do'}'' with GenAI tools---a conclusion doubly true about other disciplines. The few studies that do describe practices focus on common use cases for GenAI \cite{chan_students_2023, hasanein_drivers_2023, hou_effects_2024} or prompting strategies \cite{ghimire_coding_2024}. Research has not examined broader practices that likely shape the adoption of GenAI, including communication about GenAI tools among students or educators, or the development of AI reliance strategies---our findings address this gap. 

Second, studies thus far have not \textit{situated} practices or attitudes within the relevant educational context \cite{stone_exploring_2024}, including university guidelines and communication norms among and between students and educators. This is because either studies do not cover these issues at all \cite{hasanein_drivers_2023, chan_students_2023, park_promise_2024, chan_will_2024,sheard_instructor_2024}, or cover them using high-level multi-site surveys \cite{prather_robots_2023}, thereby trading off contextual depth for broad patterns. In contrast, our findings examine students' practices and attitudes around GenAI grounded in the context of their university guidelines, unspoken rules, and communication norms among students and educators. %

Third, as evidenced in the above review, most research on GenAI in higher education---particularly in HCI---has focused on CS \cite{prather_robots_2023,park_promise_2024,hou_effects_2024,mahon_guidelines_2024,sheard_instructor_2024,hellas_experiences_2024,ghimire_coding_2024,stone_exploring_2024,amoozadeh_trust_2024, shoufan_exploring_2023, singh_exploring_2023}. Understanding situated practices, attitudes, and contextual norms in broader disciplines is even more limited, as this research has relied on high-level surveys \cite{chan_students_2023,chan_will_2024, chan_comprehensive_2023,ghimire_generative_2024, johnston_student_2024,barrett_not_2023} or, rarely, interviews \cite{hasanein_drivers_2023} documenting de-contextualized beliefs and attitudes. Our interview findings represent experiences in the humanities, social science, and STEM disciplines. 

Thus, at the time of writing, research lacks a holistic, ecological perspective that examines both student and educator \textit{real-world} practices, attitudes, motivations, norms, concerns, and expectations, \textit{situated in the educational context} in which these aspects emerge. This is important because, as \citet{reich_failure_2020} argues, effective educational innovations---such as those from the HCI community---must align with the real-world dynamics of teaching and learning. To address this gap, we scoped our focus to \textit{students'} use of GenAI across disciplines, contextualized by educators' perspectives, and
\textbf{set out to answer the following research questions}:

\begin{enumerate}
    \item How and why do students use GenAI in higher education? 
    \item What are students' norms and etiquette around the use of GenAI in higher education, and how is this shaped by their university context?
    \item What are students' GenAI reliance strategies and what mediates this? 
    \item What are the perceived impacts of GenAI on students? 
    \item What is the future role of GenAI in higher education according to students and educators?
\end{enumerate}

}
\section{Methodology} \label{SEC:methods}

\subsection{Participants} \label{SUBSEC:recruitment}
After ethical approval\footnote{Anonymized for review.}, we used departmental mailing lists and personal contacts to recruit undergraduate students (n = 26; \markup{18 from University A}) and educators (n = 11; \markup{8 from University A}) from two UK universities to complete surveys and semi-structured interviews. %
Participants were from a variety of disciplines and year groups (see \autoref{tab:student-demographics} and \autoref{tab:educator-demographics}%
). 

\markup{Both universities had 25-30k undergraduate students in 2022/23, of which 40-50\% were international. Both included Arts, Humanities, and STEM faculties, and ranked in the top 30 for World University Rankings in 2023. Prior to the start of our data collection (December 2023), University A's guidance (published March 2023) emphasised that assignments should contain students’ original work, highlighted the limitations of GenAI as a source of information, and the need to acknowledge the use of GenAI, noting that generating parts of, or entire, assignments is regarded as academic misconduct. University B's guidance (published August-October 2023) stated that they promote ethical and transparent engagement with GenAI tools, aiming to build critical AI literacy. They provided teaching and learning resources on GenAI, including guidance on appropriate use cases, referencing, plagiarism, using GenAI in assessments, student perspectives on GenAI, guidance on their approved AI tool (Microsoft Copilot), and a non-assessed course on GenAI and academic skills and ethical issues.}

\renewcommand{\arraystretch}{1.4}
\markup{
\begin{table*}[h!]
\footnotesize
  \caption{Student participant demographics}
  \label{tab:student-demographics}
  \centering
  \begin{tabular}{L{0.01\linewidth} L{0.035\linewidth} L{0.04\linewidth} L{0.03\linewidth} L{0.19\linewidth} L{0.1\linewidth} L{0.16\linewidth} L{0.16\linewidth}
  L{0.14\linewidth}}
    \toprule
    \textbf{ID} & \textbf{Age} & \textbf{Gender} & \textbf{Study year} & \textbf{Degree area} & \textbf{Native English speaker} & \textbf{Technology adoption timing} & \textbf{GenAI usage frequency} &
    \textbf{GenAI tools used} \\
    \midrule
    U1 & 18-21 & Woman & 3rd & Law & Yes & Neither early nor late & A few times per month & ChatGPT \\ \hline
    U2 & 22-25 & Man & 1st & Law & Yes & Neither early nor late & Less than once per month & ChatGPT \\ \hline
    U3 & 18-21 & Woman & 1st & Interdisciplinary Futures & No  & Somewhat early adopter & A few times per month & ChatGPT\\ \hline
    U4 & 22-25 & Woman & 4th & History of Art & Yes  & Neither early nor late & A few times per week & ChatGPT \\ \hline
    U5 & 18-21 & Woman & 2nd & Illustration & No & Neither early nor late & A few times per week & ChatGPT, Midjourney\\ \hline
   U6 & 18-21 & Woman & 1st & Cognitive Science & Yes  & Somewhat early adopter & A few times per week & ChatGPT, DALL-E, Midjourney\\ \hline
   U7 & 18-21 & Woman & 1st & Geography & No  & Neither early nor late & Less than once per month & ChatGPT \\ \hline
   U8 & 29-32 & Woman & 1st & Interior Design & Yes & Neither early nor late & Less than once per month & ChatGPT, DALL-E, Midjourney \\ \hline
   U10 & 22-25 & Woman & 4th & Psychology & No & Neither early nor late & Less than once per month & ChatGPT \\ \hline
   U11 & 18-21 & Woman & 1st & Linguistics & Yes & Very early adopter & At least once per day & ChatGPT, Bard \\ \hline
   U12 & 22-25 & Man & 4th & Philosophy and Politics & Yes & Very early adopter & About once per month & ChatGPT \\ \hline
   U13 & 26-29 & Man & 2nd & Philosophy and Psychology & Yes & Somewhat early adopter & A few times per month & ChatGPT, Bard \\ \hline
   U14 & 22-25 & Man & 4th & Philosophy and Economics & Yes  & Very early adopter & At least once per day & ChatGPT, Photoshop \\ \hline
   U15 & 22-25 & Woman & 3rd & Politics, Philosophy and Economics & Yes & Neither early nor late & At least once per day & ChatGPT, DALL-E \\ \hline
   U16 & 18-21 & Man & 2nd & Philosophy and Theology & Yes  & Somewhat early adopter & About once per month & ChatGPT \\ \hline
   U17 & 18-21 & Man & 2nd & Economics & No & Somewhat early adopter & A few times per week & ChatGPT \\ \hline
   U18 & 18-21 & Woman & 2nd & Economics & No  & Neither early nor late & A few times per week & ChatGPT, Copilot \\ \hline
   U19 & 18-21 & Woman & 3rd & Education & No  & Very early adopter & A few times per week & ChatGPT \\ \hline
   U20 & 18-21 & Man & 1st & Japanese & Yes & Very early adopter & A few times per month & ChatGPT \\ \hline
   U21 & 18-21 & Woman & 1st & History & Yes & Neither early nor late & Less than once per month & ChatGPT, DALL-E \\ \hline
   U22 & 18-21 & Man & 1st & History & Yes & Neither early nor late & A few times per month & ChatGPT \\ \hline
   U23 & 22-25 & Man & 3rd & Aeronautical Engineering & No & Neither early nor late & A few times per week & ChatGPT, Bard \\ \hline
   U24 & 18-21 & Woman & 2nd & Medicine & No & Somewhat early adopter & A few times per month & Unstated \\ \hline
   U25 & 18-21 & Woman & 1st & Cognitive Science & Yes  & Neither early nor late & A few times per month & ChatGPT, DALL-E\\ \hline
   U26 & 18-21 & Woman & 1st & History and Politics & No  & Neither early nor late & A few times per week & ChatGPT \\ \hline
   U27 & 26-29 & Man & 3rd & History & Yes & Somewhat early adopter & At least once per day & ChatGPT, Copilot \\ 
    \bottomrule
  \end{tabular}
\end{table*}
}
\renewcommand{\arraystretch}{1.4}

\begin{table*}[h!]
\footnotesize
  \caption{Educator participant demographics}
  \label{tab:educator-demographics}
  \centering
  \begin{tabular}{L{0.01\linewidth} L{0.04\linewidth} L{0.04\linewidth} L{0.09\linewidth} L{0.12\linewidth} L{0.07\linewidth} L{0.19\linewidth} L{0.16\linewidth}
  L{0.14\linewidth}}
    \toprule
    \textbf{ID} & \textbf{Age} & \textbf{Gender} & \textbf{Teaching role} & \textbf{Course Area} & \textbf{Native English speaker} & \textbf{Technology adoption timing} & \textbf{GenAI usage frequency} 
    & \textbf{GenAI tools used} \\
    \midrule
    E1 & 30-44 & Non-binary & Post-doc & Sociology & No  & Somewhat early adopter & At least once per day & ChatGPT \\ \hline
    E2 & 18-29 & Woman & PhD Student & Consultancy & No  & Very early adopter & A few times per week & Snapchat AI \\ \hline
    E3 & 30-44 & Woman & Fellow & Philosophy & No  & Somewhat early adopter & A few times per week & ChatGPT, DALL-E, Midjourney \\ \hline
    E4 & 30-44 & Woman & Post-doc & Education & Yes  & Somewhat early adopter & A few times per week & ChatGPT, Claude AI, Bard, Midjourney, DALL-E \\ \hline
    E5 & 30-44 & Man & PhD Student & Engineering & Yes  & Very early adopter & A few times per week & ChatGPT \\ \hline
    E6 & 18-29 & Man & PhD Student & Engineering & Yes  & Somewhat early adopter & Less than once per month & Snapchat AI \\ \hline
    E7 & 18-29 & Woman & Fellow & Health Economics & No  & Neither early nor late & Less than once per month & ChatGPT \\ \hline
    E8 & 30-44 & Woman & Post-doc & Art History & Yes  & Very early adopter & A few times per week & ChatGPT, DALL-E, Midjourney \\ \hline
    E9 & 18-29 & Woman & PhD Student & Psychology & No  & Somewhat early adopter & At least once per day & ChatGPT \\ \hline
    E10 & 18-29 & Woman & Tutor & Philosophy & No  & Neither early nor late & Never & - \\ \hline
    E11 & 30-44 & Woman & Lecturer & Design Informatics & No & Neither early nor late & A few times per month & ChatGPT \\
    \bottomrule
    
  \end{tabular}
\end{table*}

\renewcommand{\arraystretch}{1}

\subsection{Surveys and Interviews} 
Participants completed an online consent form and pre-interview survey prior to the interview. The survey contained questions about demographics and their use and high-level perspective around GenAI (see \autoref{app:survey-questions}). %
Interviews were conducted from December 2023 to February 2024, and took place via video meeting. The average interview duration was 40 minutes %
and included a 5-minute introduction and a semi-structured interview \markup{covering GenAI use and comparison to traditional methods, experienced limitations, trust in AI, norms around use and communication among students (or educators), and future prospects for GenAI in education (see \autoref{app:interview-protocol}).} Participants were thanked with a gift card.%

\subsection{Data analysis and theoretical framework} \label{SUBSEC:methods_analysis}
The primary data reported is taken from our interviews (descriptive results from the pre-interview survey are reported in \autoref{app:survey-responses}). Interview transcripts were manually reviewed against the video and cleaned.

\subsubsection{Inductive thematic analysis} 
We followed an inductive approach \cite{thomas_general_2006} to perform thematic analysis \cite{braun_using_2006}, seeking patterns of similarity and difference in the perspectives and experiences of educators and students when it came to emerging unspoken rules, etiquettes, the social dynamics of GenAI use in academic settings and expectations of the future role of GenAI tools in education. 
Two researchers independently examined the transcripts for patterns. Then, both researchers used MAXQDA\footnote{https://www.maxqda.com/} to thematically code the interviews separately for educators and students.
After the themes and codes were agreed upon by the two researchers, the main analyst conducted the analysis, using the comparative analysis method \cite{rihoux_configurational_2009} %
using students' data as a basis and comparing educators' data against it. Throughout the findings, to indicate the observed frequency for each code or theme, we include the count in brackets as a proportion of the total number of students or educators (e.g., \pid{(13/26)}), or indicate the participant IDs for counts less than three or specific quotes (e.g., \pid{(U16)} for students; \pid{(E5)} for educators). 

\subsubsection{Theoretical framework: Strong Structuration Theory (SST)} \label{SUBSUBSEC:sst}
After conducting an initial inductive analysis, we sought a framework to cohere the themes. We considered the Technology Acceptance Model (TAM) \cite{davis_perceived_1989, davis_user_1989} and the Diffusion of Innovation Theory \cite{rogers_diffusion_2003} but these were found to be inappropriate to explain granular interactions between student practices, internal motivations and considerations, and external contexts in universities \cite{sackstein_theories_2023}. Instead, we found that \textit{Strong Structuration Theory (SST)} \cite{stones_structuration_2005,greenhalgh_theorising_2010} was a relevant and rich approach \cite{sackstein_theories_2023}. Building on Structuration Theory \cite{giddens_constitution_1984} and developed for empirical studies, SST recognises the mutual interactions between individuals' agency, technology, and their encompassing socio-techno-political structures. Broadly, SST analytically distinguishes between \cite{greenhalgh_theorising_2010}:
\begin{enumerate}
    \item external structures (conditions of action, such as socio-political structures and norms) 
    \item internal structures within the agent (`how and what individuals know')
    \item actions conducted agentically by individuals (drawing on internal structures and constrained by external structures)
    \item resulting outcomes (in which external and internal structures are reproduced or changed)  
\end{enumerate}
\markup{SST has been used to understand technology adoption in healthcare \cite{greenhalgh_theorising_2010, greenhalgh_virtual_2016} and education \cite{sackstein_understanding_2021}. It is valuable here for at least four reasons. First, given our focus on situating students' use of GenAI in its educational context, SST's integration of micro, meso, and macro perspectives is useful for examining how students' individual practices (and motivations and expectations) around GenAI are shaped by wider meso-level structures (e.g., student and educator communication norms) and macro-level structures (e.g., university guidelines). Likewise, SST emphasizes the dynamic interaction between agency and structure, allowing us to examine how students navigate the broader constraints to use or avoid GenAI, and how that in turn shapes these broader structures. Third---and particularly relevant for HCI research---SST recognizes technology as an active element in the broader interaction, whose properties influence students' behavior and attitudes, which in turn influence how the technology is used. This makes it particularly apt for studying GenAI, a uniquely adaptable and multi-purpose technology. Finally, given our `snapshot in time' findings and the rapidly evolving nature of GenAI and related policies,} SST emphasizes the \textit{conjuncture}, a `critical combination of events or circumstances' \cite{greenhalgh_theorising_2010}.

We therefore used SST to refine our analyses and present our findings. \markup{Specifically, our initial inductive analysis produced a set of disjointed themes pertaining to student and educator practices, motivations, norms, university guidelines, and expectations. We mapped these themes onto SST’s analytical framework, categorizing them into: internal and external structures, agency (actions), and outcomes. We were then able to make conceptual links between aspects of our data (e.g., students' unspoken rules on appropriate use being shaped by the lack of clear university guidelines), as well as identify links that are plausible but not evidenced in our data (i.e., dashed grey arrows in \autoref{FIG:results-overview}). Finally, based on this categorization and linking process, we iteratively refined our initial themes and produced \autoref{FIG:results-overview}, summarizing our findings.}

\subsection{Limitations} \label{SUBSEC:limitations}

This study captures a snapshot of experiences that are continuously evolving. Our sample is self-selected, likely reflecting those with an interest in GenAI, and is drawn from only two universities \markup{in the UK (thus also introducing potentially relevant cultural differences in education).} We did not have representative samples of students and educators: Student participants were more from humanities (n = 21) than STEM subjects (n = 5). Educators (n = 11) consisted predominantly of junior teaching staff and only 3 educators were teaching undergraduates since the release of ChatGPT, and one was teaching AI ethics. While we did not have a representative sample of \textit{teaching} educators, %
all educators were marking coursework and tutoring, and interacting with more senior educators, which allowed us to assess emerging social dynamics and expectations. \markup{The sample was also biased towards participants primarily using ChatGPT for text-to-text generation (with a few others using, e.g., Bard or Copilot), and a few also conducting text-to-image generation with, e.g., DALL-E or Midjourney.} Finally, although we found SST to be valuable for understanding our data \textit{post hoc}, the study would have benefited from using SST to guide the study design. 

\markup{\subsection{Researcher Positionality Statement}

This research was conducted by a team comprising graduate academic researchers affiliated with the two UK universities and researchers employed by a large technology company that develops generative AI tools. %
This positionality inevitably shapes our perspectives: the academic researchers are embedded in the cultural and pedagogical contexts of the universities, while the tech-affiliated researchers are directly invested in the evolution and adoption of generative AI technologies. We acknowledge that our roles may influence the framing of our questions and interpretations of the data, and we have sought to mitigate bias through collaborative analysis and reflexivity, examining our assumptions, values, and potential biases.}
\section{Findings} \label{SEC:findings}
Our interview findings explore students' practices and attitudes around GenAI grounded in the surrounding context (§\ref{SUBSEC:tooluse}-\ref{SUBSEC:future}). \markup{We present our interview findings through the lens of Strong Structuration Theory \cite{stones_structuration_2005,greenhalgh_theorising_2010} (see §\ref{SUBSUBSEC:sst}). We first describe \textit{actions}: how students take agency to interact with GenAI tools (\autoref{FIG:results-overview} box A). We then describe their motivations for using GenAI, a key \textit{internal structure} driving their behaviour (\autoref{FIG:results-overview} box B). Students' use of GenAI occurs within a broader set of \textit{external structures}: university guidelines, communication from and among educators, and communication among students (\autoref{FIG:results-overview} box C). Given this context, we then describe how students self-govern their use of GenAI according to various considerations (i.e., another type of \textit{internal structure}; \autoref{FIG:results-overview} box D). Finally, we describe \textit{outcomes}, including perceived changes in confidence, skill development concerns, relationships with educators, and plagiarism anxiety (\autoref{FIG:results-overview} box E); as well as students' and educators' visions for the future of GenAI in higher education (\autoref{FIG:results-overview} box F).} 

Although our findings focus on a set of directional influences between these aspects in the data, it is likely that the influence goes both ways \cite{stones_structuration_2005,greenhalgh_theorising_2010}---e.g., students' actions around GenAI and their perceived impacts influence university guidelines (we indicate these currently unexplored influences with dashed grey arrows in \autoref{FIG:results-overview}).

The pre-interview survey provided additional context on high-level usage and perceptions (see \autoref{app:survey-responses}), but the limited number of participants means that its results should be treated as indicative only. 

\markup{\subsection{\textit{Actions:} Interacting with Generative AI tools} \label{SUBSEC:tooluse}

\markup{Students predominantly conducted text-to-text generation via ChatGPT (or Bard or Copilot for some), with some also conducting text-to-image generation via tools like DALL-E or Midjourney (see \autoref{tab:student-demographics}).} Interview responses revealed three key roles that GenAI tools played in supporting students in education-related tasks: tutor, assistant, or ideation partner \markup{(\autoref{FIG:results-overview} box A)}.

\subsubsection{Generative AI as tutor} \label{SUBSUBSEC:tutor}
Students used GenAI tools for purposes similar to using help from an educator. %
Students \pid{(10/26)} used GenAI tools to \textbf{explain study materials}, such as theories and readings,  %
 for example by simplification \pid{(U26)} or reframing the material \iquote{“in a different way than [they] would normally from [their] lectures“} \pid{(U2)}.
Students \pid{(6/26)} also used AI tools to \textbf{obtain feedback} on their work, %
e.g. by asking for a second opinion \pid{(U4)}, or reviewing their assignment responses for mistakes \pid{(U7)}.
Some \pid{(6/26)} used GenAI tools to \textbf{guide them through tasks}, such as data analysis and coding---%
e.g., by obtaining \iquote{“an outline on how to do it first”} \pid{(U17)}; %
using GenAI tools to \iquote{“highlight where the [code] error is [and] what [they] can do to improve it”} \pid{(U25)}; %
or comparing their code to that generated by ChatGPT \pid{(U23)}.

\subsubsection{Generative AI as assistant} \label{SUBSUBSEC:assistant}

Students also reported using GenAI tools to assist them in completing specific tasks. %
The majority \pid{(16/26)} reported using GenAI tools to \textbf{summarise information}, %
e.g., to \iquote{“get an intuition of [a] topic”} \pid{(U15)} or when \iquote{“crunched for time”} \pid{(U3)}.
Students \pid{(10/26)} also used GenAI to \textbf{search for information}, such as research papers, %
obviating the need to \iquote{“scan through the sea of information”} \pid{(U19)}, and making specific requests:  \iquote{“I've read the stuff that the Uni sent... `what would you suggest reading further?'”} \pid{(U16)}.
Some \pid{(9/26)} also used GenAI to \textbf{adjust their writing}, %
e.g., to correct grammar \pid{(U5)}, paraphrase something \iquote{“so it's not plagiarised”} \pid{(U10)}, or refine their writing style \iquote{“in a way that sounds sophisticated”} \pid{(U23)}. Of the nine students who reported using GenAI in this way, six were non-native English speakers, %
suggesting that this is a valuable use case in that context. 
Most students \pid{(15/26)} also used it to \textbf{structure their work}, %
e.g., by obtaining essay outlines with potential headings and key points to address,  \iquote{“like a surface to work off”} \pid{(U11)}.%

\subsubsection{Generative AI as ideation partner} \label{SUBSUBSEC:ideation}

Students \pid{(8/26)} also used GenAI tools to \textbf{spark ideas} by interacting with them as ideation partners who they could ask questions: %

\begin{smallquote}
    \iquote{“I would sometimes ask ChatGPT or Copilot if they have any thoughts on certain topics because they can list 12345 for certain questions […] kind of like brainstorming”} \pid{(U18)}
\end{smallquote}

Students saw this as a collaborative activity between them and GenAI, which takes on the role of a \iquote{“friend"} \pid{(U4)} or \iquote{“replacement for a partner that you get in, like a seminar or a smaller discussion group”} \pid{(U16)}. Importantly, these students mentioned the value of bringing in their \iquote{“own ideas”} \pid{(U4)}, or preparing their \iquote{“own thoughts on the text”} \pid{(U16)}. In contrast to this more collaborative ideation, students \pid{(4/26)} in some cases used GenAI to \textbf{passively generate ideas} for them, as \pid{U20} used ChatGPT  \iquote{“to sort of brainstorm some ideas... for things to write about”}.  %
Interestingly, only two of the eight students %
who used GenAI as an ideation partner were non-native English speakers. This contrasts with using GenAI to adjust writing (§\ref{SUBSUBSEC:assistant}), suggesting these students found GenAI tools less helpful in an ideation context, or may have prioritised other use cases.}

\subsection{\textit{Internal structures:} Motivations for use} \label{SUBSEC:motivations}
When talking about GenAI tools such as ChatGPT, students’ motivations to use them for education fell into four key areas: \markup{availability}; efficiency; ability to steer thinking; and opportunity to engage deeper with study material (\autoref{FIG:results-overview} box B). \markup{These motivations exemplify how GenAI tools' properties---their flexibility, interactivity, and speed---shape students' educational practices.}

\subsubsection{\markup{Availability}} \label{SUBSUBSEC:accessibility}

Students \pid{(8/26)} referred to the availability of GenAI tools as one of the key benefits of GenAI over other methods, %
\markup{including searching online \pid{(U4)}, asking friends \pid{(U6)}, or reaching out to educators \pid{(U4, U16)}:} 

\begin{smallquote}
    \iquote{“It's in your pockets, in your phone, your home. You don't have your professor always there”} \pid{(U16)} 

\end{smallquote}

GenAI tools were found to be the first port of call or the \iquote{``bottom of the ladder''} \pid{(U6)}, as they enable round-the-clock support, including \iquote{“over the holidays”} \pid{(U25)}, and require no advanced scheduling \pid{(U7, U20)}, or depending on or `bothering others' \pid{(U6,U10)}.

\subsubsection{Efficiency} \label{SUBSUBSEC:efficiency}

Students \pid{(15/26)} found that GenAI tools could make their study processes more efficient. %
They reported that using GenAI saved them time by summarising texts for them \pid{(7/26)}, %
narrowing their search scope \pid{(6/26)}, %
or answering specific queries \pid{(3/26)}. %
For example, students said they saved time that otherwise would have been spent on \iquote{“re-reading [a very dense article or dense book] many times”} \pid{(U14)}, or \iquote{“aimlessly scrolling through Google Scholar”} \pid{(U16)}.

\subsubsection{Steering thinking} \label{SUBSUBSEC:steering}

Beyond availability and efficiency, students \pid{(11/26)}---particularly native English speakers and those in the humanities---noted the helpfulness of GenAI tools for steering their thinking. %
For example, they found that using GenAI tools could help them overcome a block in creativity or problem-solving \pid{(3/26)}: %

\begin{smallquote}
    \iquote{“If you face a block or like, what do I really want to make next or... you can't figure something out, I sometimes use like AI like ChatGPT […] it can sometimes lead you in the right path.”} \pid{(U20)} 
\end{smallquote}

This steering was particularly useful at the beginning of tasks \pid{(5/26)}, %
\iquote{“for just building basic points and to get you sort of thinking”} \pid{(U4)}, or for nudging students in the direction of a solution \pid{(U22, U17)}, for example with maths problems \pid{(U18)} or creative tasks \pid{(U20)}. Even incorrect responses by GenAI tools were able to stimulate students’ thinking about a particular topic \pid{(U22, U11)}, or a creative output \pid{(U8)}:

\begin{smallquote}

    \iquote{``I know it's completely wrong, but then it just gives me like a surface to work off of''} \pid{(U11)}
    \vspace{0.3\baselineskip}

\end{smallquote}
 
\subsubsection{Deeper engagement with material} \label{SUBSUBSEC:deeper}

Students \pid{(12/26)} found that GenAI could support their learning by deepening their engagement with the study material in complementary ways to traditional learning. %
In our sample, this motivation was most commonly expressed by students in the humanities, rather than STEM fields. The ways Students varied in their approach here, suggesting that they discovered these ways individually by matching tool capabilities to their individual learning strategies or study domains. 

Some students used GenAI to \textbf{learn beyond the school-provided material}---\iquote{“you're doing a bit extra and then that supplement your learning to engage with something further”} \pid{(U16)}---or more deeply understand a given topic, which they believed \iquote{“can actually help you in the long run”} \pid{(U20)}. One particular approach was to use GenAI summarisation to identify the key points of a text in order to prioritise further reading \pid{(U4, U26)}.

\markup{Another approach was to use GenAI to \textbf{re-frame the material} from a different perspective \pid{(U2)}. %
In other cases, students \pid{(6/26)} engaged with the material seemingly \textbf{driven by the uncertainty} of the correctness of the outputs or a challenge to find the right answer: }%

\begin{smallquote}
    \iquote{“You can interrogate it to check that it's saying all the right things, and then you can also go back to the original paper and... like cross check it”} \pid{(U13)}
\end{smallquote}

Students considered this additional work worthwhile as it could \iquote{“lead to a more thorough understanding”} \pid{(U13)}. Yet others were driven by sheer \textbf{curiosity}: 

\begin{smallquote}
    \iquote{“[When I've] seen an opinion or statement that I just think is incredulous (sic), [...] it kind of makes me think, right? How is it saying this? Gonna look that up.”} \pid{(U22)}
\end{smallquote}

Some students \pid{(6/26)} found the \textbf{interactivity} of the process particularly valuable, %
including interrogating GenAI answers and asking follow-up questions \pid{(U20)}, or discussing ideas with the GenAI system \pid{(U16)}:

\begin{smallquote}

    \iquote{“If you can use it as almost a partner to bounce ideas off then it might work [...] I think then it could be useful in just helping push your kind of analytical skills a bit further”} \pid{(U16)}
\end{smallquote}

Critically, for the GenAI tools to complement learning, students \pid{(5/26)} saw their own \textbf{engagement with the material} as an important component that should not be automated away: %

\begin{smallquote}
    \iquote{“...anything that removes you from doing something where you're not engaging with the media in any way... that is not useful for learning”} \pid{(U16)}
\end{smallquote}

\subsection{\textit{External structures:} University guidelines, communication, and social context} \label{SUBSEC:norms}
\markup{Regardless of students' individual motivations, their use of GenAI occurred in a wider university context which constrained and influenced their practices (\autoref{FIG:results-overview} box C). We detail this context here, and describe its influence in subsequent sections.} Both students and educators viewed the university guidelines on GenAI as being ineffectively communicated and unclear, predominantly focusing on the risks of plagiarism. Most educators viewed the use of GenAI to complement learning as appropriate within certain boundaries. However, with some exceptions, many did not feel able to talk openly about this among themselves and with students. In contrast, open communication about GenAI among students was normalized.

\subsubsection{University guidelines} \label{SUBSUBSEC:guidelines}

When asked if they were aware of their universities’ written guidelines about the use of GenAI, some students \pid{(9/26)} said they were aware of guidelines being in place but admitted that they had not read them: %

\begin{smallquote}
    \iquote{``I know [they've] got some policies to explain it after ChatGPT was introduced, but honestly, I didn't really check''} \pid{(U18)}
\end{smallquote}

A few students \pid{(4/26)} said that they had never heard of the guidelines, %
and others saw the method by which they were communicated as impractical \pid{(3/26)}: %

\begin{smallquote}
    
    \iquote{``It seems outrageous to me to think that people would read out the actual policy and understand what they are supposed to do and the idea that they’d follow through with it''} \pid{(U23)}
\end{smallquote}

Regardless, some students \pid{(9/26)} expressed low expectations that official guidelines would be useful, expecting them to be focused on the risk of plagiarism: %

\begin{smallquote}
    \iquote{``I never read it thoroughly, but the title is about plagiarism. Yeah. Same, same old.''} \pid{(U17)}
\end{smallquote}

Indeed, students \pid{(9/26)} that were aware of the guidelines said that the key message from their universities was a warning about the potential implications of plagiarism: %

\begin{smallquote}
    \iquote{``We had a lot of like email sent out about how this will be detected as plagiarism and a lot around that''} \pid{(U10)}
\end{smallquote}

Accordingly, when discussing university rules, most students \pid{(17/26)} felt that there was a \textbf{lack of clarity} in communication from their universities on how they should or should not use GenAI tools. %
They reported being confused, in some cases, hesitant about using these tools \pid{(U18)} (see §\ref{SUBSUBSEC:plagiarismanxiety} on `plagiarism anxiety'):

\begin{smallquote}
    \iquote{``We are not quite sure about this Uni's policy about using AI like whether the tendency is gonna be like to check if it's generated by AI...that's why we don't use it to avoid that kind of risk''} \pid{(U18)} 
    \vspace{0.3\baselineskip}

\end{smallquote}

\paragraph{\textbf{Educators’ perspective on university guidelines.}} Most educators \pid{(9/11)} were also unaware of official guidelines on how GenAI tools should be used by them and by students. %
Likewise, they acknowledged that the current rules set by the university are unclear to both them and their students \pid{(7/11)}. %
Accordingly, some were unsure about how to evaluate students’ work \pid{(E2)}, or resorted to the same rules as they would for plagiarism \pid{(E9)}: 

\begin{smallquote}

    \iquote{``Because it's not clearly forbidden... we cannot penalise a student for using it. And we just have to be more conservative and treat it as a traditional case of plagiarism. So I think we would both benefit from having more clarity''} \pid{(E9)}
\end{smallquote}

One educator worried that unclear rules increased the risk of unequal treatment of students: 

\begin{smallquote}
    \iquote{``It's really all over the place [...] one professor like just marks all the essays 0 and the other one just doesn't have a problem with that''} \pid{(E10)}
\end{smallquote}

\subsubsection{Educators' perspective on GenAI use}

Most educators \pid{(7/11)} saw the use of GenAI tools to complement learning as allowed within certain boundaries, %
including students needing to still do the core work themselves:

\begin{smallquote}
    \iquote{``I don't think it would be cheating if you used it in a way that you are only editing the stuff, the work that you have done yourself.''} \pid{(E2)}
\end{smallquote}

Educators also emphasised that students should continue to produce original ideas themselves \pid{(E9, E6)} and use GenAI tools to support their learning rather than replace it \pid{(E9, E8)}. Valid use cases included assistive tasks such as gathering, organising and summarising learning material \pid{(E9, E10)} or using GenAI as a starting point \pid{(E9, E7)}. Many of the educators \pid{(6/11)} saw the use of GenAI for assignments as out of limits. %
They also mentioned the importance of transparency, suggesting that students should reference or disclose the use of GenAI \pid{(E6, E8)}.

\subsubsection{Communication from and among educators} \label{SUBSUBSEC:commswithedu}

Students indicated a lack of open communication from educators about GenAI \pid{(9/26)}. %
If communication did occur, it mainly focused on plagiarism \pid{(4/26)}.

\begin{smallquote}
    \iquote{``Teachers are like going away from it and trying to escape from it, but it's in the reality it's just present.''} \pid{(U11)}
\end{smallquote}

Educators described communicating with students about GenAI as \iquote{``a bit of a dangerous territory''} \pid{(E2)}, with some \pid{(4/11)} reporting feeling cautious about their narratives being misinterpreted. %
For example, they worried about unintentionally nudging students to use these tools inappropriately \pid{(E2)}, or, in their role as educators, were unsure about what information was appropriate to provide to students \pid{(E11)}: 

\begin{smallquote}

    \iquote{``There's a question of when it's my role to talk about these things and filtering kind of what's relevant to my own teaching and what I'm intending students to learn from me''} \pid{(E11)}
\end{smallquote}

\markup{On the other hand, some educators \pid{(5/11)} made different attempts to inform their students about GenAI to encourage responsible use:} %

\begin{itemize}
    \item explaining the importance of studying the course material yourself: \iquote{``I was just trying to explain how these skills that they would learn through philosophy courses be useful in general''} \pid{(E10)}

    \item informing students about the potential limitations of GenAI: \iquote{``I said look... there's this new thing in town. We have to play with it. I know that some of you will be tempted to write essays using it, but I'm telling you I've tried it. These are the problems I have encountered''} \pid{(E1)}
    
    \item explaining how students could use GenAI to support their individual learning needs: \iquote{``He was not an English speaker, so some issues with language, and I was going through all the possibilities within how he could use [AI]''} \pid{(E3)}
    
    \item suggesting potential use cases: \iquote{``I've said, if there are things that you're really stuck on, if there are some generalised concepts you need to just understand, then use that to explain''} \pid{(E8)}.
\end{itemize}

Among educators, although some \pid{(3/11)} did report having open conversations about GenAI, at least within their sphere, %
many \pid{(7/11)} admitted that they do not talk openly about GenAI because of fears of how it would be perceived by colleagues%
---concerns that were salient even during special events on GenAI \pid{(E1)}:   
\begin{smallquote}

    \iquote{``Everyone was interested, but nobody actually touched on the topic. I think everyone was a bit scared.''} \pid{(E1)}
\end{smallquote}

This was despite the fact that some educators \pid{(3/11)} reported using GenAI in their teaching, including for communicating with students, %
such as re-phrasing their student feedback \pid{(E2)}, and preparing lecture slides \pid{(E2, E11)}.\footnote{In other, \textit{non-teaching-related} cases, educators used GenAI in similar ways to students. This included using it to explain concepts and guide them through coding tasks \pid{(3/11)}, %
to spark ideas \pid{(3/11)}, %
to structure presentations or papers \pid{(3/11)}, %
or to adjust their language, e.g. to make it more formal \pid{(E2)}, academic \pid{(E1)}, better tailored to the intended audience \pid{(E4)}, or grammatically correct \pid{(3/11)}.} %
Similarly to students, educators \pid{(6/11)} were motivated by the increased efficiency of GenAI, %
and its ability to steer thinking or inspire ideas \pid{(4/11)}. %

\subsubsection{Communication among students} \label{SUBSUBSEC:studentcomms}

Although a few students \pid{(4/26)} indicated reluctance to discuss the use of GenAI for university work amongst themselves, many \pid{(13/26)} reported having open communication about it. %
GenAI use is \iquote{``very openly talked about''} \pid{(U4)} and indeed \iquote{``normalised''} \pid{(U6)}. Some \pid{(3/26)} perceived that \textit{some} use was ubiquitous among students. %
Conversations could include suggestions to rely on GenAI---\iquote{``I've heard lots of people say `just ChatGPT it'''} \pid{(U6)}--as well as descriptions of its helpfulness \pid{(U2)}. There was evidence of open communication even about uses that may violate academic integrity \pid{(3/26)}, %
although some students \pid{(3/26)} emphasised their personal boundaries around this. %
Indeed, although communication was open, students made their own decisions about when and how to use GenAI---\iquote{``it's still kind of very much individual``} \pid{(U2)}. These decisions were guided by different considerations, which we turn to next.

\subsection{\textit{Internal structures:} Students' self-governance}\label{SUBSEC:selfgov}

Given the perceived lack of clear guidelines from universities and inconsistent communication from educators, students self-governed their use of GenAI. Moreover, some students felt that they should be \textit{allowed} to self-govern how they use these tools \pid{(10/26)}, %
 and that such self-governance will inevitably emerge over time: %

\begin{smallquote}
    \iquote{``I think it's important that students have like a self-regulating idea about this, and that they developed their own feelings about this''} \pid{(U13)}

\end{smallquote}

We observed three ways in which students governed their own use of GenAI: unspoken rules about appropriate and ethical use; reliance strategies for effective use; and, to a lesser extent, considerations of skill development \markup{(\autoref{FIG:results-overview} box D)}.

\subsubsection{Unspoken rules about appropriate use} \label{SUBSUBSEC:unspokenrules}

Students reported adhering to unspoken rules about appropriate use of GenAI. These fell under one broad theme: \textit{It is okay to use GenAI within certain boundaries}, with the boundaries being: (1) \textit{``reference, do not plagiarise''}, (2) \textit{``edit, do not copy and paste''}, and (3) \textit{``use it to assist you, not to do work for you''}. Students felt that using GenAI would be dishonest without referencing it properly \pid{(4/26)}: %

\begin{smallquote}
    \iquote{``If you're using an idea that ChatGPT has given you and you're not properly referencing where ChatGPT got it from or that you got it from ChatGPT, that would count as cheating''} \pid{(U22)}
\end{smallquote}

Likewise, they felt that using GenAI tools is acceptable as long as you have your own input rather than just copying generated output verbatim \pid{(13/26)}: %

\begin{smallquote}
    \iquote{``If someone took the exact wording... and copy-pasted it into their own work and not really read it through or like thought about it...Then yeah, it would be cheating''} \pid{(U20)}
\end{smallquote}

Students also mentioned that GenAI tools could be used to aid--rather than entirely automate---their work \pid{(14/26)}: %

\begin{smallquote}

    \iquote{``If you just sent the question into ChatGPT for it to spit out the whole answer, then that would probably be cheating''} \pid{(U15)}
\end{smallquote}

\subsubsection{Reliance and trust} \label{SUBSUBSEC:reliance}
Students were also predominantly aware of the limitations of GenAI tools, and developed reliance strategies for using them. A particular risk with GenAI, which can generate extensive outputs with often subtle errors or inaccuracies \cite{tankelevitch_metacognitive_2024}, is \textit{overreliance}---the erroneous acceptance of incorrect GenAI output \cite{passi_appropriate_2024,passi_overreliance_2022}. \markup{Students used their awareness of limitations and other mediating considerations %
to guide their reliance on GenAI.}

\paragraph{Perceived limitations and trust}

Students developed awareness about the GenAI limitations based on:

\begin{itemize}
    \item their experience with errors using these tools \pid{(12/26)}: %
    \iquote{``[ChatGPT] gave me a couple of official-sounding theories....I said, hang on, let me just check if there is such a thing...and they don't exist anywhere in any literature''} \pid{(U6)} 

    \item observations of GenAI tool performance \pid{(7/26)}: %
    \iquote{``I find that if the topic is quite a niche AI can't really give you really good answers''} \pid{(U26)}

    \item general knowledge about AI \pid{(5/26)}: %
    \iquote{``The idea that I can just put in enough sort of inputs and data for it to write out like an essay [...] just seemed kind of like just suspicious in the 1st place''} \pid{(U12)} 

    \item others' experiences \pid{(4/26)}:%
    \iquote{``One of my friends, though she said that she asked ChatGPT to recommend some relevant readings or essays, and apparently they don't actually exist''} \pid{(U26)} 
\end{itemize}

\markup{When discussing the limitations of GenAI, the most common type mentioned was inaccurate responses, consisting of: confabulated information (``hallucinations'') \pid{(11/26)}, inaccurate information \pid{(4/26)}, and contextual inaccuracies \pid{(9/26)}.}

GenAI outputs were also considered quite \iquote{``surface level''} \pid{(U1)} or otherwise lacking nuance \pid{(13/26)}. %
Students were also aware of the data-related limitations of GenAI, including potential plagiarism of training data \pid{(3/26)}, the inability to \iquote{``reliably trace the source''} \pid{(U10)} of generated outputs \pid{(4/26)}, %
or privacy and intellectual property concerns \pid{(5/26)}. %

Finally, some students found GenAI’s written outputs unsatisfactory \pid{(7/26)}. %
For example, generated text did not fit students' own writing style: \iquote{``it sort of over-complicates the English''} \pid{(U23)}.

\paragraph{Mitigating strategies}

Students developed strategies to accommodate perceived limitations of GenAI. For example, most of the students said they double-checked GenAI outputs using other sources, such as Google Search or PubMed \pid{(13/26)}. %
They also compared outputs with their own knowledge \pid{(6/26)}: %

\begin{smallquote}
    \iquote{``Try it myself first and then, even for the questions that I don't know, I can see in it when it's like using the same like idea, same strategy and I know it's probably will be right''} \pid{(U17)}
\end{smallquote}

\paragraph{Factors mediating reliance}

Given the perceived limitations of these systems, students considered a range of factors to mediate their reliance decisions. First, they emphasised the importance of having \textbf{domain expertise} in the relevant area \pid{(6/26)}. %
This was implied to be a resource for broadly sense-checking GenAI outputs, without which they would be less likely to trust GenAI outputs \pid{(U1)}, or use the tools at all \pid{(U26)}:

\begin{smallquote}
    \iquote{``If I was to ask it about like computing or mathematics or something I don't know, then, I would be distrustful of that''} \pid{(U1)}
    \vspace{0.3\baselineskip}
    
\end{smallquote}

Students also made reliance-related decisions based on the \textbf{importance of the task}, defined as the potential cost of a mistake \pid{(7/26)}: %

\begin{smallquote}

    \iquote{``If it just notes then I won't really be too bothered... but if it's like a proper essay that I need to submit or for an exam for revision, then I think I'd take it more seriously''} \pid{(U22)}
\end{smallquote}

\textbf{Task-specific GenAI performance expectations} also guided students' decisions to rely on GenAI \pid{(15/26)}. %
Broadly, this corresponded to the perceived simplicity of a given task and how GenAI was expected to handle it: 

\begin{smallquote}
    
    \iquote{``If it's something like really academic or really specific, I wouldn't trust it as much...''} \pid{(U26)} 
\end{smallquote}

Students also made more specific task allocations based on their observed limitations of GenAI in certain domains like \iquote{``complex maths''} \pid{(U25)} or types of questions:  \iquote{``like yes or no questions. I won't trust it that much...''} \pid{(U19)}

\markup{Students also seemed to be \textbf{weighing gained value versus time lost}, with some finding that the time that it takes to verify or adapt the information may not be worth the effort \pid{(6/26)}, whether it was writing \pid{(U8)} or coding \pid{(U14)}: %

\begin{smallquote}
    \iquote{``In terms of writing, I tend to veer away from it... by the time you're going through all the sort of fact-checking, and you may as well just do it yourself''} \pid{(U8)}
\end{smallquote}}

\subsubsection{Agency over tasks and skill development}\label{SUBSUBSEC:agency}

Finally, some students' decisions \textit{not} to use GenAI appeared to be driven by a sense of agency over their tasks and long-term skill development, which was linked to enjoyment of their work \pid{(5/26)}, %
and the desire to make original contributions to it \pid{(U14, U8)}. 

\begin{smallquote}
    
    \vspace{0.3\baselineskip}
    
    \iquote{``I wanna maintain that originality in what I write and to make sure that what I'm writing adds substance to the literature out there.''} \pid{(U14)}
\end{smallquote}

More broadly, students indicated that overusing GenAI undermined the point of education \pid{(4/26)}:%
 \begin{smallquote}
    \iquote{``...they're not really using their own intellect [...] that's the core part of being at uni. And if it's getting done by someone else, then yeah, there's no point.''} \pid{(U23)} 
  \end{smallquote}
  
\subsection{\textit{Outcomes:} impact on confidence, skills, relationships, and plagiarism anxiety} \label{SUBSEC:impacts}

Beyond the direct perceived effects discussed in §\ref{SUBSEC:tooluse} and §\ref{SUBSEC:motivations}, using GenAI tools, as well as the broader university context, led to a range of other perceived impacts on students, including changes in confidence; concerns about overuse leading to skill loss or under-development; worries about negative impacts on relationships with educators; and anxiety about plagiarism \markup{(\autoref{FIG:results-overview} box E)}.

\subsubsection{Self-confidence} \label{SUBSUBSEC:confidence}

Some students felt more confident in their abilities after using GenAI tools. In some cases, they felt that GenAI tools were augmenting their existing abilities \pid{(U22, U27)}---\iquote{``I'm using the resources that are available to me in a way that's kind of academic''} \pid{(U22)}---or supporting them in tasks in which they felt less confident in \pid{(4/26)}:%

\begin{smallquote}

    \iquote{``I'm just good at generating ideas, and I'm not very good at articulating them, so I didn't lose out. I gained a lot from this new technology''} \pid{(U15)}
\end{smallquote}

Paradoxically, some students gained confidence by seeing that they can do tasks \textit{better} than GenAI tools \pid{(4/26)}, %
yet others viewed the impressive performance of GenAI as a dint to their confidence \pid{(3/26)}: %

\begin{smallquote}
    
    \iquote{``I'm so worried. Like it can do most of the work. So it makes me feel useless sometimes.''} \pid{(U17)}
\end{smallquote}

\subsubsection{Overuse and skill development concerns} \label{SUBSUBSEC:skills}

Students---particularly non-native English speakers---were concerned that overusing GenAI tools for certain tasks might lead to them losing or under-developing their existing skills \pid{(9/26)}. %
This included skills like problem solving \pid{(U10)}, writing \pid{(U18)}, and, most commonly, critical thinking \pid{(7/26)}:%

\begin{smallquote}
    
    \iquote{``I started to worry about whether my own ability to generate [written text] would be reduced because of the lack of practice and lack of real applications''} \pid{(U18)}
    \vspace{0.3\baselineskip}
    
\end{smallquote}

 Students also expressed worries about not developing skills that they might need in the future \pid{(6/26)}:%

 \begin{smallquote}
     \iquote{``You cannot rely on AI for your whole life, right? In the future, if you want to be a researcher or want to be a lecturer, you cannot like, let AI do everything for you... it will be harmful to your development''} \pid{(U24)} 
    \vspace{0.3\baselineskip}
    
 \end{smallquote}

\paragraph{\textbf{Educators’ perspective on students' skill development}} Educators \pid{(9/11)} %
also saw the risk of students losing or not developing certain skills if they overused GenAI tools. They noted skills such as development of own writing style \pid{(E11, E8)}, expressing ideas \pid{(E7)}, conducting research \pid{(E9)}, solving problems \pid{(3/11)}, %
and critical thinking \pid{(3/11)}. %
Educators \pid{(6/11)} %
viewed practicing skills as the key factor protecting against the harmful effects of GenAI overuse. Specifically, GenAI overuse was thought to be moderated by whether students engage with the learning material themselves \pid{(E6)}, allocate critical thinking to GenAI \pid{(E4)}, or consistently avoid practising skills such as writing \pid{(E7)}:

 \begin{smallquote}

    \iquote{``If people start outsourcing their ability to critically think and reflect through any topic, I think that's where it's yeah, that's like a very bad place''} \pid{(E4)} 
    \vspace{0.3\baselineskip}
    
 \end{smallquote}
 
\subsubsection{Relationships with educators} \label{SUBSUBSEC:relationships}

Students worried about how relying on GenAI may negatively impact their relationships with educators \pid{(6/26)}:  
 
\begin{smallquote}
    \iquote{``it could potentially harm like the actual relationship between the lecturer and the student [...]  there's no connection been built up between the lecturer and students''} \pid{(U20)}
\end{smallquote}

They saw the benefit of talking to educators over using ChatGPT or similar tools for more in-depth answers \pid{(5/26)}---\iquote{``to better understand the materials during the lectures''} \pid{(U18)}. %
Students also appreciated the \iquote{``interpersonal human interaction''} \pid{(U10)} with educators, and the \iquote{``emotional supportive element''} \pid{(U8)}, which they were skeptical could be replaced by AI \pid{(4/26)}. %
However, they also admitted that they were now more likely to go to search for answers using ChatGPT before approaching an educator \pid{(9/26)}:%

\begin{smallquote}
    \iquote{``It's always ChatGPT first and then if you're not satisfied, then you go for the professors and teachers''} \pid{(U17)}
\end{smallquote}

Notably, of the nine students above that expressed a preference for approaching ChatGPT before educators, five were non-native English speakers. Students suggested that GenAI is an additional resource for those who may be otherwise reluctant to ask questions \pid{(5/26)}---\iquote{``because they're anxious about it or they don't feel comfortable doing it''} \pid{(U16)}%

\paragraph{\textbf{Educators’ perspective on relationships with students.}} Educators anticipated that students might be more likely to approach ChatGPT with their questions before asking them \pid{(5/11)}. %
 Some educators viewed this as a positive, making it easier for students who would otherwise feel reluctant to approach educators \pid{(3/11)}:%

\begin{smallquote}
    \iquote{``...even though they might not understand, they might have questions, but they are shy to ask them. So they can just ask the ChatGPT right in front of them''} \pid{(E5)}
\end{smallquote}

However, like students, some worried about a potential weakening of interpersonal connection \pid{(E11, E8)}.

\subsubsection{Plagiarism anxiety} \label{SUBSUBSEC:plagiarismanxiety}

Potentially due to universities' lack of clear communication around GenAI (§\ref{SUBSUBSEC:guidelines}), students \pid{(7/26)} were feeling anxious about being accused of plagiarism. %
This was the case even for students that were not using GenAI \pid{(U6)}:

\begin{smallquote}
    
    \iquote{``I'm also concerned about universities using AI detection tools [...] I don't want to have worked on a project for a really long time... and then be told, no, this was AI. There's no way of proving, you know, I’ve written this myself''} \pid{(U6)}
\end{smallquote}

There was also a sense that universities’ focus on plagiarism led students to be more wary of having open conversations with educators \pid{(4/26)} %
(see also §\ref{SUBSUBSEC:commswithedu}): 

\begin{smallquote}
    \iquote{``Students are afraid of getting into trouble about plagiarism or just being seen as being lazy and not taking the subject matter seriously''} \pid{(U13)}
\end{smallquote}

\paragraph{\textbf{Educators’ perspective on plagiarism anxiety.}}
Educators \pid{(5/11)} also noticed that students were not talking openly about using GenAI tools with them: %

\begin{smallquote}
    \iquote{``Students are like so scared to tell the teachers that they use ChatGPT''} \pid{(E7)}
\end{smallquote}

They speculated that this may be due to students' \iquote{``hesitation about plagiarism or uncertainty about how it's appropriate to be used''} \pid{(E11)}, and the difficulty of demonstrating original contributions in submitted work \pid{(E2)}:

\begin{smallquote}
    
    \iquote{``If I come to professor and I say listen, I used chat but these thoughts and arguments are originally mine, they wouldn't be able to separate those things...''} \pid{(E2)}
\end{smallquote}

\subsection{\textit{Outcomes:} The future role of generative AI in higher education} \label{SUBSEC:future}

Students pushed back against the university structures outlined in §\ref{SUBSEC:norms}, and wanted universities to change their approach to GenAI, to help students use these tools effectively, ethically, and responsibly rather than prohibiting their use. Both students and educators expected to receive more guidance regarding the use of GenAI tools. 
Both groups recognised the need to change assessments to fit the realities of GenAI, and both also hoped that in the future, GenAI would be more integrated into the learning processes \markup{(\autoref{FIG:results-overview} box F)}.

\subsubsection{Universities' approach to Generative AI} \label{SUBSUBSEC:approach}
When talking about the future role of GenAI in education, many students \pid{(10/26)} shared the same sentiment---universities should \iquote{``catch up''} \pid{(U19)} and embrace these technologies just as other industries are doing, or risk being detached from the \iquote{``real world''} \pid{(U12)}: %

\begin{smallquote}
    \iquote{``It seems mad to me that we're at an institution like this and we're not encouraging intellectual people to be using this and then people outside can use it for all sorts of different things''} \pid{(U13)}
    \vspace{0.3\baselineskip}

\end{smallquote}

Some students expressed worries that universities will continue to lag behind innovation, potentially due to their risk-averse nature \pid{(5/26)}: %
However, some hoped that universities would shift their communication strategy when talking about GenAI tools to a more positive and pragmatic one \pid{(7/26)}:%

\begin{smallquote}

    \iquote{``Less approached as a problem [...] also point towards the positive sides of it and how we can actually make use of it''} \pid{(U25)}
\end{smallquote}

\subsubsection{Education about responsible use of GenAI} \label{SUBSUBSEC:futurecomms}

Students wanted universities to inform them about appropriate and helpful uses of GenAI tools in educational settings \pid{(13/26)}:%

\begin{smallquote}
    \iquote{``Instead of [students] having to figure out themselves and go through trial and error and a minefield almost [...] it could be like...we know AI is really good for this type of learning and we know AI is really good in this topic''} \pid{(U2)} 
\end{smallquote}

Students desired education about the capabilities and potential implications of GenAI tools \pid{(15/26)}. %
This was partly motivated by students' observations of disparity in their relative expertise with GenAI \pid{(3/26)}. %
There was also a hint of urgency in some students' comments \pid{(4/26)}, %
with a sense that GenAI is proceeding \iquote{``quicker than how we are adapting to the use of it''} \pid{(U19)}. Students particularly emphasised the importance of learning how to use GenAI tools responsibly \pid{(15/26)}, 
\markup{including ethical considerations \pid{(U27, U22)}, implications of overuse \pid{(U20, U8)}, potential pitfalls \pid{(U14)}, and the role of critical thinking \pid{(U8)}. They also emphasised the importance of learning to use GenAI effectively to complement traditional forms of learning \pid{(3/26)}.%

Students also wanted practical support about how to effectively harness GenAI’s benefits in their workflows, including training on prompting and verification \pid{(11/26)}:%

\begin{smallquote}
    \iquote{``How can we integrate this so that it makes our lives easier? And how can we integrate this so that people can learn better?''} \pid{(U11)}
\end{smallquote}
}

\paragraph{\textbf{Educators’ perspective on education about responsible GenAI use.}} Like students, educators wished for more clarity about how students are supposed to use these tools, within what boundaries, and how students should understand the capabilities and limitations of these tools \pid{(6/26)}:%

\begin{smallquote}
    
    \iquote{``It's very much a playground... hopefully we'll develop some clearer guidelines and pointers''} \pid{(E11)}
\end{smallquote}

Educators also wished for more open conversations among teaching staff \pid{(5/26)}, including %
e.g., debates about GenAI \pid{(E3)}, or conversations about practical information \pid{(E4)}: 

\begin{smallquote}
    \iquote{``Having more open discussions about it [...] say what are the problems if they are against them, why they are against them and vice versa''} \pid{(E3)}
    
\end{smallquote}

\subsubsection{Changing assessments} \label{SUBSUBSEC:assessments}

Recognising the academic integrity implications of GenAI use, students envisioned different assessments in the future \pid{(5/26)}. %
Some suggested incorporating GenAI tools into assignments, e.g., asking students to challenge GenAI responses \pid{(U13)}, write on topics with and without the support of GenAI \pid{(U8)}, or write reflections on GenAI responses \pid{(U26)}:

\begin{smallquote}
    \iquote{``You can ask it questions, and it could give you like feedback, and maybe it'll intentionally give you false information so that you can apply critical thinking skills''} \pid{(U13)}

\end{smallquote}

\paragraph{\textbf{Educators’ perspective on changing assessments.}} Educators said that they currently do not have sufficient tools to detect the use of GenAI tools in students' work \pid{(6/26)}. %

However, educators also acknowledged that the way students’ work is currently assessed will have to change to ensure it effectively assesses learning \pid{(7/26)}.
They proposed potential changes, such as combining oral and written exams \pid{(E5, E3)}, or in creative domains, using in-depth assessments\markup{that ask students to reflect on their portfolio work \pid{(E9)}, or rationalise their design process \pid{(E4)}. A few educators already implemented GenAI tools in assessments \pid{(3/11)}:

\begin{smallquote}

    \iquote{``I said you will write model cards for your essays. So go ahead, use ChatGPT, but then write me 200 words about your steps. How did you use it? What did you make out of it? What were the problems you encountered?''} \pid{(E1)}

\end{smallquote}
}

\subsubsection{Integrating Generative AI tools into learning processes} \label{SUBSUBSEC:integrating}

Students envisioned a future where GenAI tools would be better integrated into learning processes \pid{(11/26)}.%
This progression was compared by some to prior technological innovations such as calculators or the internet \pid{(4/26)}. %
Some specified that GenAI could enable personalised learning \pid{(U2)} through feedback and exercises \pid{(U13)}, or function like a helpful companion \pid{(U14)}, whereas others noted its promise to reduce workload and enable convenience \pid{(U7, U26)}. 
Notably, students saw a future where GenAI tools would be \textit{complementary} to other learning methods such as teachers \pid{(12/26)}:%

\begin{smallquote}
    \iquote{``The future, I hope, is one where we can use these tools simultaneously along with teaching by the professor and the student''} \pid{(U11)}
    \vspace{0.3\baselineskip}

\end{smallquote}

\paragraph{\textbf{Educators’ perspective on integrating GenAI tools.}} A number of educators \pid{(5/11)} similarly saw GenAI tools as complementary in the future, %
e.g., personalising learning experiences for students \pid{(E3)} or supporting educators during classes \pid{(E5)}:

\begin{smallquote}
    \iquote{``Maybe using it as an additional tutor at home that you can ask questions and it can give you some additional tests, maybe tailored to your own way of learning''} \pid{(E3)}
    
\end{smallquote}

Educators also saw GenAI being more integrated into teaching  \pid{(4/26)}.%
They saw themselves using GenAI tools for marking \pid{(E8)}, preparing lectures \pid{(E11)}, getting ideas for classroom activities \pid{(E2)}, and administrative tasks \pid{(E10)}. However, they also emphasised the human element of teaching that %
GenAI cannot replace \pid{(E7, E8)}.

\section{Discussion} \label{SEC:discussion}

This study provides an ecological overview of early-stage practices and perceptions of GenAI tools in higher education. \markup{Our findings demonstrate the value of SST for understanding the complex interaction---at the macro, meso, and micro levels---between the external and internal structures (university guidelines, communication norms) that shape, and are shaped by, students' agency in using and communicating about GenAI, educators' agency in avoiding communication about it, and the properties of the technology.}

Since the landscape is evolving, we structure our discussion into \markup{current issues (this `conjuncture' \cite{greenhalgh_theorising_2010}), and anticipated changes in external and internal structures (i.e., reflecting the dashed grey arrows in \autoref{FIG:results-overview}).} A condensed version of future research questions is also provided in \autoref{app:futurequestions}.

\subsection{\markup{Current \textit{conjuncture}: GenAI, university context, and student agency}}
\markup{GenAI tools' novelty means that---as students rapidly and independently adopt this technology---universities are still developing guidelines for their use, and educators are grappling with key questions. We discuss key aspects of this particular `conjuncture' below.}

\subsubsection{Unclear guidelines and a fixation on plagiarism} \label{SUBSUBSEC:unclearguidelines}
\markup{We found a lack of clarity around university guidelines on GenAI, their inconsistent communication (echoing \cite{johnston_student_2024,rajabi_exploring_2023}), and a fixation on plagiarism. Students were confused about appropriate uses of GenAI and evaluation of assignments (echoing \cite{ditta_what_2023}). Educators were also unsure about the rules and their communication to students. %
This created an atmosphere of fear and disconnect between students and educators. Accordingly, students expressed `plagiarism anxiety' while taking agency to self-govern their use of GenAI (see §\ref{SUBSUBSEC:self-regulation}). Educators reported concerns about `playing detective' and expressed reluctance to discuss the topic, even amongst themselves (§\ref{SUBSUBSEC:eduexp}). }

Universities' fixation on plagiarism could impede creating an open dialogue and informed regulatory environment \cite{smolansky_educator_2023, huallpa_exploring_2023, hadi_mogavi_chatgpt_2024}. University policies should be communicated more systematically and proactively to avoid sending mixed messages \cite{barrett_not_2023}, e.g., through dedicated town-hall sessions. Guidelines should also be relevant to students' and educators' realities of GenAI use, \textbf{covering issues beyond plagiarism and proposing balanced and pragmatic approaches} \cite{prather_robots_2023}. One approach is to involve both groups in the development of guidelines using participatory design methods \cite{dindler_computational_2020,demirbas_re-designing_2020}, or via student and educator working groups. %

\subsubsection{Educators' lack of communication} \label{SUBSUBSEC:eduexp}

Despite desiring open discussion, educators found it difficult to communicate about GenAI with both students and colleagues. This stems partly from uncertainty about colleagues' views of GenAI, with early career staff potentially worried about judgment from senior colleagues \cite{chan_ai_2023}. \markup{This under-communication may also negatively influence communication with students, perpetuating the taboo around GenAI, reducing the sharing of experiences on how to effectively integrate GenAI, and leading to inconsistent teaching practices \cite{barrett_not_2023,tlili_what_2023,ghimire_coding_2024}. }

Communication among educators around GenAI needs to be central to how universities approach GenAI literacy (§\ref{SUBSUBSEC:genailiteracy}), innovating assessments (§\ref{SUBSUBSEC:innovassessments}), and students' skill development (§\ref{SUBSUBSEC:skillsdev}). However, while universities are still developing such policy and processes, \textbf{educators should be empowered to harness the power of their professional communities to bootstrap relevant knowledge} about the latest advancements and best practices to integrate these tools into their teaching \cite{smolansky_educator_2023,rajabi_exploring_2023,barrett_not_2023}. This can be followed by continuing professional development when universities are ready.

\subsubsection{Students' self-governance of GenAI use} \label{SUBSUBSEC:self-regulation}
In the absence of clear guidelines and communication from educators, and motivated by GenAI's capabilities, \markup{students exercised agency in self-governing their usage in three distinct ways.} Students followed unspoken rules defining appropriate use, confirming earlier work on students' perceptions of GenAI assistance vs. automation, appropriate attribution, and academic integrity \cite{johnston_student_2024,barrett_not_2023,chan_ai_2023, rajabi_exploring_2023}. Students also developed concrete reliance strategies for effective use, echoing \cite{chan_students_2023, shoufan_exploring_2023,rajabi_exploring_2023}. Finally, they considered their agency over tasks and long-term skill development \cite{prather_robots_2023} (see §\ref{SUBSUBSEC:unclearguidelines} on skill development). 

However, our findings are based on students' self-reports, so the \textit{objective} extent of over- or under- reliance in real-world education contexts remains to be studied, and over-reliance remains a real risk \cite{tankelevitch_metacognitive_2024,hasanein_drivers_2023, prather_robots_2023} (indeed as observed in lab studies \cite{prather_its_2023, prather_widening_2024,drosos_its_2024}). Moreover, our self-selected sample of students likely reflects those with an interest in GenAI. As the technology spreads and norms evolve, how a wider population of students will understand and regulate their use of GenAI remains to be seen. Students' \textit{subjective} experience may or may not be sharply honed. This has implications for how students consider ethical norms (§\ref{SUBSUBSEC:unclearguidelines}) and skills development (§\ref{SUBSUBSEC:skillsdev}). In line with \cite{amoozadeh_trust_2024}, we suggest that a balanced approach is needed, starting with \textbf{allowing students to explore small-scale test cases of using GenAI to support their learning, within certain boundaries}, which might later form part of fostering broader GenAI literacy (§\ref{SUBSUBSEC:genailiteracy}).\markup{This channels students' agency and knowledge of the technology to positively shape \textit{external structures} such as university guidelines and training.}

\markup{\subsection{Anticipated changes in \textit{external} structures}
As students continue to exercise agency with GenAI tools, we, and our participants, anticipate broader changes to external structures including university guidelines, training, assessment approaches, and relationship norms among students and educators. }

\subsubsection{GenAI literacy} \label{SUBSUBSEC:genailiteracy}

Students wanted universities to embrace GenAI (echoing \cite{chan_students_2023, johnston_student_2024}). 
Broadly, they sought a common ground where students adhered to academic integrity rules but were not forbidden to use GenAI tools or left to discover them independently. Students also wanted  support in harnessing GenAI in their workflows. Indeed, 58\% of surveyed students do not feel that they have sufficient AI knowledge and skills, per the Digital Education Council\cite{digital_education_council_what_2024}. Training could reduce inappropriate use among students \cite{lan_analyzing_2023}, alleviate disparities in students' GenAI literacy, and support better learning outcomes \cite{holland_qualitative_2024}. \markup{Concrete examples from HCI include interactive and pedagogical approaches to improving prompt crafting, such as Prompt Problems \cite{denny_prompt_2024} and others \cite{denny_explaining_2024}.}
Educators echoed this need. Our findings therefore indicate \textbf{a need for multi-faceted GenAI literacy training in universities} \cite{tlili_what_2023, hadi_mogavi_chatgpt_2024}.

Given likely ongoing disparities in GenAI literacy, it may be helpful to formalise collaborative GenAI learning environments (e.g., peer-assisted learning groups), where students could share their experiences among themselves. Such approaches have been shown to be more engaging for students \cite{carvalho_developing_2022}.
\markup{To prevent over-reliance, training that includes exposure to and recognition of errors in GenAI output is a promising strategy long deployed in, e.g., aviation \cite{skitka_automation_2000, prather_widening_2024}. %
\textbf{Recognising over-reliance issues will need to built into educational practice as a foundational skill}.}

\subsubsection{Academic integrity and innovating assessments} \label{SUBSUBSEC:innovassessments}
Maintaining academic integrity in the face of advanced AI tools will remain a challenge \cite{hasanein_drivers_2023, hadi_mogavi_chatgpt_2024, firat_how_2023, rajabi_exploring_2023,singh_students_2023, smolansky_educator_2023,ghimire_generative_2024,shoufan_exploring_2023}, particularly for coding tasks and written assessments \cite{smolansky_educator_2023}. The effectiveness of current assessments is subject to the quality and accuracy of plagiarism or GenAI detection tools \cite{pudasaini_survey_2024}. An important caveat here is the potential for false positives, leading to false accusations of plagiarism and increasing assessment-related stress \cite{rajabi_exploring_2023}. Moreover, even with accurate detection tools, \textbf{the extent of allowed GenAI use needs precise definition}. 

Institutions must develop strategies that assess relevant aspects of students' learning while upholding standards of academic integrity \cite{johnston_student_2024,lan_analyzing_2023, singh_students_2023, prather_robots_2023,ghimire_generative_2024}. Both students and educators suggested incorporating GenAI into assignments, such as writing on topics with and without GenAI, or explaining the steps taken to use GenAI for a task. This approach has the added benefit of increasing students’ experience with using GenAI tools effectively and responsibly (§\ref{SUBSUBSEC:genailiteracy}). 
Assessments could also focus on creativity and critical thinking, hands-on and (simulated) real-world tasks, oral assessments, and collaborative tasks \cite{kooli_chatbots_2023, smolansky_educator_2023, eke_chatgpt_2023, tlili_what_2023, rajabi_exploring_2023}. \markup{Achieving this requires understanding educators' and students' needs and the relevant capabilities of GenAI---interactive frameworks like those in \citet{tan_more_2024} provide a way of doing this systematically. \citet{mahon_guidelines_2024} provide six descriptive levels at which educators may consider integrating GenAI into introductory programming courses---future work should extend this to other disciplines.}
Beyond the challenge of assessing traditional skills, students must be prepared for the evolving future of GenAI-augmented work. This means \textbf{assessing the breadth of acquired skills and knowledge relevant for using GenAI}.

\subsubsection{Interpersonal educational relationships} 

The rise of GenAI raises questions about the future of the relationship between students and teachers, as well as among students \cite{prather_robots_2023, chan_will_2024}. \markup{\citet{park_promise_2024} find that students value the socio-emotional and physical aspects of human tutors and peers and don't want GenAI tools to replace this.}
It will be important to \textbf{intentionally (re)design the role of human relationships in education}. \markup{One approach may be to design GenAI-enabled personalised learning to explicitly include roles for educators and student-led collaborative learning \cite{kharrufa_potential_2024}, or projects or assessments that revolve around interpersonal interaction (§\ref{SUBSUBSEC:innovassessments}). Tutor CoPilot is an illustrative example: as a collaborative human-AI system, it uses LLM-generated guidance to augment and scale human tutors' expertise and feedback for students in real time, improving student learning, while retaining human tutors' importance \cite{wang_tutor_2024}.}    

\subsection{Anticipated changes in \textit{internal} structures}
\markup{Changes in external structures and the potential normalization of GenAI may lead to concomitant changes and concerns in internal structures, including skill development and the integration of personalized learning.}

\subsubsection{Skill development} \label{SUBSUBSEC:skillsdev}
\markup{As implied by SST \cite{greenhalgh_theorising_2010}, students are \textit{exercising their agency} with GenAI tools in the absence of helpful guidelines and educator support. Indeed, we observed a wide range in students' motivations, use cases, and interaction approaches for GenAI. Accordingly,} GenAI's long-term impact on students' skill development may be multi-faceted. On one hand, some used GenAI as a tutor and ideation partner, potentially strengthening their skills in general problem solving and their respective study domains. Further, some students' GenAI reliance strategies involved reflecting on GenAI accuracy and verification. It is plausible that these experiences may promote long-term development of domain-specific skills and potentially critical thinking \cite{fahim_critical_2012}. 

On the other hand, students' GenAI use was also motivated by the availability and efficiency of GenAI. We also found that some students used GenAI to support them in tasks in which they were less confident, such as writing \cite{johnston_student_2024}. Students expressed concerns that overusing GenAI for certain tasks might lead to underdevelopment or loss of existing skills such as problem solving, critical thinking, and writing, and worried about the impact on their careers (echoing \cite{chan_students_2023,prather_robots_2023}). Such concerns were mirrored by educators, as also found in \cite{hasanein_drivers_2023,ghimire_generative_2024}.  

\markup{Understanding the differences in students' motivations for, and use of, GenAI, and mediating factors is necessary for anticipating impact on skills. For example, we observed anecdotally that non-native English speakers particularly valued GenAI as an assistant (and indeed, particularly benefit from this \cite{usdan_generative_2024}), using it less for ideation. Relatedly, a key strength yet limitation of our study is its skew towards humanities and social science disciplines which may engender different motivations and experiences compared to STEM fields. Future work should explore such potential differences systematically with larger samples.}

Safeguards may be necessary to protect skills that are key for students’ education goals. Safeguards might include setting maximum requirements on the extent to which students can use GenAI, or enabling students to exercise those skills most at risk of being affected by over-reliance, such as critical thinking. \markup{AI-enabled questioning exercises can be used to improve critical thinking in individual \cite{danry_dont_2023} and group settings \cite{lee_conversational_2024}. Recent work proposes to use LLMs as `provocateurs' \cite{sarkar_ai_2024} or `antagonists' \cite{cai_antagonistic_2024}, providing novel design opportunities for such interventions.} 
When designing AI systems for learning, cognitive forcing functions may be used to slow users down and elicit analytical thinking \cite{bucinca_trust_2021, kazemitabaar_exploring_2024}. %
\markup{In programming, CodeAid\cite{kazemitabaar_codeaid_2024} and CodeHelp \cite{liffiton_codehelp_2023} both use LLMs to provide assistance to students while encouraging their own thinking and without revealing solutions. Recent work explores offers design space for such approaches \cite{kazemitabaar_exploring_2024}.} 
A broader question is \textbf{\textit{which} skills should be preserved and developed}. Students did not express concerns about losing or under-developing  skills in summarization or search, indicating the potential de-valuation of these skills in the long term. The rapid pace of GenAI development and its integration into the workplace requires continuous re-assessment of the socio-economic value of skills.

\subsubsection{Integrating personalised learning}

Both students and educators saw opportunities for GenAI to support personalised learning that complements traditional methods \cite{jurenka_towards_nodate,prather_robots_2023}. Students used GenAI to engage deeper with their studies, including reading beyond the assigned material, seeking explanations, or using GenAI for ideation. This deeper learning focuses on comprehension and making connections rather than retaining facts \cite{draper_catalytic_2009, czerkawski_designing_2014}. \markup{Tools designed to promote reasoning, analysis, creativity, and nonlinear thinking could lead to better learning experiences and promote problem-solving \cite{rahman_21st_2019,czerkawski_designing_2014, shoufan_exploring_2023, abdelshiheed_power_2023}.}

Although students discovered various methods for deeper learning, they will need guidance to engage with the GenAI outputs and study material. Research in multimedia learning shows that while students' motivation benefits from having control over the sequence and pace of content \cite{chung_instructional_1992}), increased agency does not guarantee \textit{effective} learning. \markup{Indeed, students who attempted math problems themselves before consulting LLMs show improved learning \cite{kumar_math_2023}, and guiding students to do this encourages more clarification questions \cite{kumar_guiding_2024}.} \textbf{Student-centred interface design should therefore balance self-regulation by the learner and external support} \cite{lowyck_design_2001}. For example, GenAI tools could provide prompts to promote self-evaluation \cite{tankelevitch_metacognitive_2024}\markup{---e.g., on learning in classroom settings \cite{kumar_supporting_2024}---}and time-on-task \cite{lowyck_design_2001}.

\section{Conclusion} \label{conclusion}

Our study captures emerging practices, attitudes, and norms around GenAI in higher education approximately two years after GenAI came to public attention. \markup{By framing our understanding in Strong Structuration Theory, we were able to begin teasing apart the complex interactions between the technology, individuals, and the broader encompassing structures, thereby integrating the micro, meso, and macro perspectives. Critically, SST emphasizes the \textit{conjuncture}---``at this time and in this place, what
does this agent, with this technology, do and why – and what
happens as a result?'' \cite{greenhalgh_theorising_2010}. Accordingly, we hope that our work establishes a starting point for future research to build on and adapt as practices, attitudes, and norms undoubtedly evolve.}

\bibliographystyle{ACM-Reference-Format}
\bibliography{references}


\begin{thebibliography}{108}


\ifx \showCODEN    \undefined \def \showCODEN     #1{\unskip}     \fi
\ifx \showDOI      \undefined \def \showDOI       #1{#1}\fi
\ifx \showISBNx    \undefined \def \showISBNx     #1{\unskip}     \fi
\ifx \showISBNxiii \undefined \def \showISBNxiii  #1{\unskip}     \fi
\ifx \showISSN     \undefined \def \showISSN      #1{\unskip}     \fi
\ifx \showLCCN     \undefined \def \showLCCN      #1{\unskip}     \fi
\ifx \shownote     \undefined \def \shownote      #1{#1}          \fi
\ifx \showarticletitle \undefined \def \showarticletitle #1{#1}   \fi
\ifx \showURL      \undefined \def \showURL       {\relax}        \fi
\providecommand\bibfield[2]{#2}
\providecommand\bibinfo[2]{#2}
\providecommand\natexlab[1]{#1}
\providecommand\showeprint[2][]{arXiv:#2}

\bibitem[noa(2023a)]%
        {noauthor_general_2023}
 \bibinfo{year}{2023}\natexlab{a}.
\newblock \showarticletitle{General {Principles} for use of {Generative} {AI}, {Department} of {Education}, {University} of {Cambridge}}.
\newblock  (\bibinfo{year}{2023}).
\newblock
\urldef\tempurl%
\url{https://blendedlearning.cam.ac.uk/guidance-support/ai-and-education/using-generative-ai}
\showURL{%
\tempurl}


\bibitem[noa(2023b)]%
        {noauthor_russell_2023}
 \bibinfo{year}{2023}\natexlab{b}.
\newblock \showarticletitle{Russell {Group} principles on the use of generative {AI} tools in education}.
\newblock \bibinfo{journal}{\emph{2023}} (\bibinfo{year}{2023}).
\newblock
\urldef\tempurl%
\url{https://russellgroup.ac.uk/media/6137/rg_ai_principles-final.pdf}
\showURL{%
\tempurl}


\bibitem[Abdelshiheed et~al\mbox{.}(2023)]%
        {abdelshiheed_power_2023}
\bibfield{author}{\bibinfo{person}{Mark Abdelshiheed}, \bibinfo{person}{John~Wesley Hostetter}, \bibinfo{person}{Preya Shabrina}, \bibinfo{person}{Tiffany Barnes}, {and} \bibinfo{person}{Min Chi}.} \bibinfo{year}{2023}\natexlab{}.
\newblock \bibinfo{title}{The {Power} of {Nudging}: {Exploring} {Three} {Interventions} for {Metacognitive} {Skills} {Instruction} across {Intelligent} {Tutoring} {Systems}}.
\newblock
\newblock
\urldef\tempurl%
\url{http://arxiv.org/abs/2303.11965}
\showURL{%
\tempurl}
\newblock
\shownote{arXiv:2303.11965 [cs]}.


\bibitem[Amani et~al\mbox{.}(2023)]%
        {amani_generative_2023}
\bibfield{author}{\bibinfo{person}{Sara Amani}, \bibinfo{person}{Lance White}, \bibinfo{person}{Trini Balart}, \bibinfo{person}{Laksha Arora}, \bibinfo{person}{Kristi~J. Shryock}, \bibinfo{person}{Kelly Brumbelow}, {and} \bibinfo{person}{Karan~L. Watson}.} \bibinfo{year}{2023}\natexlab{}.
\newblock \bibinfo{title}{Generative {AI} {Perceptions}: {A} {Survey} to {Measure} the {Perceptions} of {Faculty}, {Staff}, and {Students} on {Generative} {AI} {Tools} in {Academia}}.
\newblock
\newblock
\urldef\tempurl%
\url{https://doi.org/10.48550/arXiv.2304.14415}
\showDOI{\tempurl}
\newblock
\shownote{arXiv:2304.14415}.


\bibitem[Amoozadeh et~al\mbox{.}(2024)]%
        {amoozadeh_trust_2024}
\bibfield{author}{\bibinfo{person}{Matin Amoozadeh}, \bibinfo{person}{David Daniels}, \bibinfo{person}{Daye Nam}, \bibinfo{person}{Aayush Kumar}, \bibinfo{person}{Stella Chen}, \bibinfo{person}{Michael Hilton}, \bibinfo{person}{Sruti Srinivasa~Ragavan}, {and} \bibinfo{person}{Mohammad~Amin Alipour}.} \bibinfo{year}{2024}\natexlab{}.
\newblock \showarticletitle{Trust in {Generative} {AI} among {Students}: {An} exploratory study}. In \bibinfo{booktitle}{\emph{Proceedings of the 55th {ACM} {Technical} {Symposium} on {Computer} {Science} {Education} {V}. 1}} \emph{(\bibinfo{series}{{SIGCSE} 2024})}. \bibinfo{publisher}{Association for Computing Machinery}, \bibinfo{address}{New York, NY, USA}, \bibinfo{pages}{67--73}.
\newblock
\showISBNx{9798400704239}
\urldef\tempurl%
\url{https://doi.org/10.1145/3626252.3630842}
\showDOI{\tempurl}


\bibitem[Azevedo et~al\mbox{.}(2022)]%
        {azevedo_lessons_2022}
\bibfield{author}{\bibinfo{person}{Roger Azevedo}, \bibinfo{person}{François Bouchet}, \bibinfo{person}{Melissa Duffy}, \bibinfo{person}{Jason Harley}, \bibinfo{person}{Michelle Taub}, \bibinfo{person}{Gregory Trevors}, \bibinfo{person}{Elizabeth Cloude}, \bibinfo{person}{Daryn Dever}, \bibinfo{person}{Megan Wiedbusch}, \bibinfo{person}{Franz Wortha}, {and} \bibinfo{person}{Rebeca Cerezo}.} \bibinfo{year}{2022}\natexlab{}.
\newblock \showarticletitle{Lessons {Learned} and {Future} {Directions} of {MetaTutor}: {Leveraging} {Multichannel} {Data} to {Scaffold} {Self}-{Regulated} {Learning} {With} an {Intelligent} {Tutoring} {System}}.
\newblock \bibinfo{journal}{\emph{Frontiers in Psychology}}  \bibinfo{volume}{13} (\bibinfo{year}{2022}).
\newblock
\showISSN{1664-1078}
\urldef\tempurl%
\url{https://www.frontiersin.org/articles/10.3389/fpsyg.2022.813632}
\showURL{%
\tempurl}


\bibitem[Baker and Inventado(2014)]%
        {baker_educational_2014}
\bibfield{author}{\bibinfo{person}{Ryan~Shaun Baker} {and} \bibinfo{person}{Paul~Salvador Inventado}.} \bibinfo{year}{2014}\natexlab{}.
\newblock \showarticletitle{Educational {Data} {Mining} and {Learning} {Analytics}}.
\newblock In \bibinfo{booktitle}{\emph{Learning {Analytics}: {From} {Research} to {Practice}}}, \bibfield{editor}{\bibinfo{person}{Johann~Ari Larusson} {and} \bibinfo{person}{Brandon White}} (Eds.). \bibinfo{publisher}{Springer}, \bibinfo{address}{New York, NY}, \bibinfo{pages}{61--75}.
\newblock
\showISBNx{978-1-4614-3305-7}
\urldef\tempurl%
\url{https://doi.org/10.1007/978-1-4614-3305-7_4}
\showDOI{\tempurl}


\bibitem[Barrett and Pack(2023)]%
        {barrett_not_2023}
\bibfield{author}{\bibinfo{person}{Alex Barrett} {and} \bibinfo{person}{Austin Pack}.} \bibinfo{year}{2023}\natexlab{}.
\newblock \showarticletitle{Not quite eye to {A}.{I}.: student and teacher perspectives on the use of generative artificial intelligence in the writing process}.
\newblock \bibinfo{journal}{\emph{International Journal of Educational Technology in Higher Education}} \bibinfo{volume}{20}, \bibinfo{number}{1} (\bibinfo{date}{Nov.} \bibinfo{year}{2023}), \bibinfo{pages}{59}.
\newblock
\showISSN{2365-9440}
\urldef\tempurl%
\url{https://doi.org/10.1186/s41239-023-00427-0}
\showDOI{\tempurl}


\bibitem[Bayliss(2013)]%
        {bayliss_exploring_2013}
\bibfield{author}{\bibinfo{person}{Gemma Bayliss}.} \bibinfo{year}{2013}\natexlab{}.
\newblock \showarticletitle{Exploring the {Cautionary} {Attitude} {Toward} {Wikipedia} in {Higher} {Education}: {Implications} for {Higher} {Education} {Institutions}}.
\newblock \bibinfo{journal}{\emph{New Review of Academic Librarianship}} (\bibinfo{date}{Jan.} \bibinfo{year}{2013}).
\newblock
\showISSN{1361-4533}
\urldef\tempurl%
\url{https://www.tandfonline.com/doi/abs/10.1080/13614533.2012.740439}
\showURL{%
\tempurl}
\newblock
\shownote{Publisher: Taylor \& Francis Group}.


\bibitem[Bennett et~al\mbox{.}(2023)]%
        {bennett_how_2023}
\bibfield{author}{\bibinfo{person}{Dan Bennett}, \bibinfo{person}{Oussama Metatla}, \bibinfo{person}{Anne Roudaut}, {and} \bibinfo{person}{Elisa Mekler}.} \bibinfo{year}{2023}\natexlab{}.
\newblock \bibinfo{title}{How does {HCI} {Understand} {Human} {Autonomy} and {Agency}?}
\newblock
\newblock
\urldef\tempurl%
\url{https://doi.org/10.1145/3544548.3580651}
\showDOI{\tempurl}
\newblock
\shownote{arXiv:2301.12490 [cs]}.


\bibitem[Braun and Clarke(2006)]%
        {braun_using_2006}
\bibfield{author}{\bibinfo{person}{Virginia Braun} {and} \bibinfo{person}{Victoria Clarke}.} \bibinfo{year}{2006}\natexlab{}.
\newblock \showarticletitle{Using thematic analysis in psychology}.
\newblock \bibinfo{journal}{\emph{Qualitative Research in Psychology}} \bibinfo{volume}{3}, \bibinfo{number}{2} (\bibinfo{date}{Jan.} \bibinfo{year}{2006}), \bibinfo{pages}{77--101}.
\newblock
\showISSN{1478-0887, 1478-0895}
\urldef\tempurl%
\url{https://doi.org/10.1191/1478088706qp063oa}
\showDOI{\tempurl}


\bibitem[Bubeck et~al\mbox{.}(2023)]%
        {bubeck_sparks_2023}
\bibfield{author}{\bibinfo{person}{Sébastien Bubeck}, \bibinfo{person}{Varun Chandrasekaran}, \bibinfo{person}{Ronen Eldan}, \bibinfo{person}{Johannes Gehrke}, \bibinfo{person}{Eric Horvitz}, \bibinfo{person}{Ece Kamar}, \bibinfo{person}{Peter Lee}, \bibinfo{person}{Yin~Tat Lee}, \bibinfo{person}{Yuanzhi Li}, \bibinfo{person}{Scott Lundberg}, \bibinfo{person}{Harsha Nori}, \bibinfo{person}{Hamid Palangi}, \bibinfo{person}{Marco~Tulio Ribeiro}, {and} \bibinfo{person}{Yi Zhang}.} \bibinfo{year}{2023}\natexlab{}.
\newblock \bibinfo{title}{Sparks of {Artificial} {General} {Intelligence}: {Early} experiments with {GPT}-4}.
\newblock
\newblock
\urldef\tempurl%
\url{https://doi.org/10.48550/arXiv.2303.12712}
\showDOI{\tempurl}
\newblock
\shownote{arXiv:2303.12712 [cs]}.


\bibitem[Buçinca et~al\mbox{.}(2021)]%
        {bucinca_trust_2021}
\bibfield{author}{\bibinfo{person}{Zana Buçinca}, \bibinfo{person}{Maja~Barbara Malaya}, {and} \bibinfo{person}{Krzysztof~Z. Gajos}.} \bibinfo{year}{2021}\natexlab{}.
\newblock \showarticletitle{To {Trust} or to {Think}: {Cognitive} {Forcing} {Functions} {Can} {Reduce} {Overreliance} on {AI} in {AI}-assisted {Decision}-making}.
\newblock \bibinfo{journal}{\emph{Proceedings of the ACM on Human-Computer Interaction}} \bibinfo{volume}{5}, \bibinfo{number}{CSCW1} (\bibinfo{date}{April} \bibinfo{year}{2021}), \bibinfo{pages}{1--21}.
\newblock
\showISSN{2573-0142}
\urldef\tempurl%
\url{https://doi.org/10.1145/3449287}
\showDOI{\tempurl}


\bibitem[Cai et~al\mbox{.}(2024)]%
        {cai_antagonistic_2024}
\bibfield{author}{\bibinfo{person}{Alice Cai}, \bibinfo{person}{Ian Arawjo}, {and} \bibinfo{person}{Elena~L. Glassman}.} \bibinfo{year}{2024}\natexlab{}.
\newblock \bibinfo{title}{Antagonistic {AI}}.
\newblock
\newblock
\urldef\tempurl%
\url{https://doi.org/10.48550/arXiv.2402.07350}
\showDOI{\tempurl}
\newblock
\shownote{arXiv:2402.07350 [cs] version: 1}.


\bibitem[Carvalho and Santos(2022)]%
        {carvalho_developing_2022}
\bibfield{author}{\bibinfo{person}{Ana~Raquel Carvalho} {and} \bibinfo{person}{Carlos Santos}.} \bibinfo{year}{2022}\natexlab{}.
\newblock \showarticletitle{Developing peer mentors’ collaborative and metacognitive skills with a technology-enhanced peer learning program}.
\newblock \bibinfo{journal}{\emph{Computers and Education Open}}  \bibinfo{volume}{3} (\bibinfo{date}{Dec.} \bibinfo{year}{2022}), \bibinfo{pages}{100070}.
\newblock
\showISSN{2666-5573}
\urldef\tempurl%
\url{https://doi.org/10.1016/j.caeo.2021.100070}
\showDOI{\tempurl}


\bibitem[Chan(2023)]%
        {chan_comprehensive_2023}
\bibfield{author}{\bibinfo{person}{Cecilia Ka~Yuk Chan}.} \bibinfo{year}{2023}\natexlab{}.
\newblock \showarticletitle{A comprehensive {AI} policy education framework for university teaching and learning}.
\newblock \bibinfo{journal}{\emph{International Journal of Educational Technology in Higher Education}} \bibinfo{volume}{20}, \bibinfo{number}{1} (\bibinfo{date}{July} \bibinfo{year}{2023}), \bibinfo{pages}{38}.
\newblock
\showISSN{2365-9440}
\urldef\tempurl%
\url{https://doi.org/10.1186/s41239-023-00408-3}
\showDOI{\tempurl}


\bibitem[Chan and Hu(2023)]%
        {chan_students_2023}
\bibfield{author}{\bibinfo{person}{Cecilia Ka~Yuk Chan} {and} \bibinfo{person}{Wenjie Hu}.} \bibinfo{year}{2023}\natexlab{}.
\newblock \bibinfo{title}{Students' {Voices} on {Generative} {AI}: {Perceptions}, {Benefits}, and {Challenges} in {Higher} {Education}}.
\newblock
\newblock
\urldef\tempurl%
\url{https://doi.org/10.48550/arXiv.2305.00290}
\showDOI{\tempurl}
\newblock
\shownote{arXiv:2305.00290 [cs]}.


\bibitem[Chan and Lee(2023)]%
        {chan_ai_2023}
\bibfield{author}{\bibinfo{person}{Cecilia Ka~Yuk Chan} {and} \bibinfo{person}{Katherine K.~W. Lee}.} \bibinfo{year}{2023}\natexlab{}.
\newblock \showarticletitle{The {AI} generation gap: {Are} {Gen} {Z} students more interested in adopting generative {AI} such as {ChatGPT} in teaching and learning than their {Gen} {X} and millennial generation teachers?}
\newblock \bibinfo{journal}{\emph{Smart Learning Environments}} \bibinfo{volume}{10}, \bibinfo{number}{1} (\bibinfo{date}{Nov.} \bibinfo{year}{2023}), \bibinfo{pages}{60}.
\newblock
\showISSN{2196-7091}
\urldef\tempurl%
\url{https://doi.org/10.1186/s40561-023-00269-3}
\showDOI{\tempurl}


\bibitem[Chan and Tsi(2024)]%
        {chan_will_2024}
\bibfield{author}{\bibinfo{person}{Cecilia Ka~Yuk Chan} {and} \bibinfo{person}{Louisa H.~Y. Tsi}.} \bibinfo{year}{2024}\natexlab{}.
\newblock \showarticletitle{Will generative {AI} replace teachers in higher education? {A} study of teacher and student perceptions}.
\newblock \bibinfo{journal}{\emph{Studies in Educational Evaluation}}  \bibinfo{volume}{83} (\bibinfo{date}{Dec.} \bibinfo{year}{2024}), \bibinfo{pages}{101395}.
\newblock
\showISSN{0191-491X}
\urldef\tempurl%
\url{https://doi.org/10.1016/j.stueduc.2024.101395}
\showDOI{\tempurl}


\bibitem[Chung and Reigeluth(1992)]%
        {chung_instructional_1992}
\bibfield{author}{\bibinfo{person}{Jaesam Chung} {and} \bibinfo{person}{Charles~M. Reigeluth}.} \bibinfo{year}{1992}\natexlab{}.
\newblock \showarticletitle{Instructional {Prescriptions} for {Learner} {Control}}.
\newblock \bibinfo{journal}{\emph{Educational Technology}} \bibinfo{volume}{32}, \bibinfo{number}{10} (\bibinfo{year}{1992}), \bibinfo{pages}{14--20}.
\newblock
\showISSN{0013-1962}
\urldef\tempurl%
\url{https://www.jstor.org/stable/44427631}
\showURL{%
\tempurl}
\newblock
\shownote{Publisher: Educational Technology Publications, Inc.}.


\bibitem[Czerkawski(2014)]%
        {czerkawski_designing_2014}
\bibfield{author}{\bibinfo{person}{Betul~C. Czerkawski}.} \bibinfo{year}{2014}\natexlab{}.
\newblock \showarticletitle{Designing deeper learning experiences for online instruction}.
\newblock \bibinfo{journal}{\emph{Journal of Interactive Online Learning}} \bibinfo{volume}{13}, \bibinfo{number}{2} (\bibinfo{year}{2014}), \bibinfo{pages}{29--40}.
\newblock
\showISSN{1541-4914}
\urldef\tempurl%
\url{http://www.scopus.com/inward/record.url?scp=84914156270&partnerID=8YFLogxK}
\showURL{%
\tempurl}


\bibitem[Danry et~al\mbox{.}(2023)]%
        {danry_dont_2023}
\bibfield{author}{\bibinfo{person}{Valdemar Danry}, \bibinfo{person}{Pat Pataranutaporn}, \bibinfo{person}{Yaoli Mao}, {and} \bibinfo{person}{Pattie Maes}.} \bibinfo{year}{2023}\natexlab{}.
\newblock \showarticletitle{Don’t {Just} {Tell} {Me}, {Ask} {Me}: {AI} {Systems} that {Intelligently} {Frame} {Explanations} as {Questions} {Improve} {Human} {Logical} {Discernment} {Accuracy} over {Causal} {AI} explanations}. In \bibinfo{booktitle}{\emph{Proceedings of the 2023 {CHI} {Conference} on {Human} {Factors} in {Computing} {Systems}}} \emph{(\bibinfo{series}{{CHI} '23})}. \bibinfo{publisher}{Association for Computing Machinery}, \bibinfo{address}{New York, NY, USA}, \bibinfo{pages}{1--13}.
\newblock
\showISBNx{978-1-4503-9421-5}
\urldef\tempurl%
\url{https://doi.org/10.1145/3544548.3580672}
\showDOI{\tempurl}


\bibitem[Davis(1989)]%
        {davis_perceived_1989}
\bibfield{author}{\bibinfo{person}{Fred~D. Davis}.} \bibinfo{year}{1989}\natexlab{}.
\newblock \showarticletitle{Perceived {Usefulness}, {Perceived} {Ease} of {Use}, and {User} {Acceptance} of {Information} {Technology}}.
\newblock \bibinfo{journal}{\emph{MIS Quarterly}} \bibinfo{volume}{13}, \bibinfo{number}{3} (\bibinfo{year}{1989}), \bibinfo{pages}{319--340}.
\newblock
\showISSN{0276-7783}
\urldef\tempurl%
\url{https://doi.org/10.2307/249008}
\showDOI{\tempurl}
\newblock
\shownote{Publisher: Management Information Systems Research Center, University of Minnesota}.


\bibitem[Davis et~al\mbox{.}(1989)]%
        {davis_user_1989}
\bibfield{author}{\bibinfo{person}{Fred~D. Davis}, \bibinfo{person}{Richard~P. Bagozzi}, {and} \bibinfo{person}{Paul~R. Warshaw}.} \bibinfo{year}{1989}\natexlab{}.
\newblock \showarticletitle{User {Acceptance} of {Computer} {Technology}: {A} {Comparison} of {Two} {Theoretical} {Models}}.
\newblock \bibinfo{journal}{\emph{Management Science}} \bibinfo{volume}{35}, \bibinfo{number}{8} (\bibinfo{date}{Aug.} \bibinfo{year}{1989}), \bibinfo{pages}{982--1003}.
\newblock
\showISSN{0025-1909}
\urldef\tempurl%
\url{https://doi.org/10.1287/mnsc.35.8.982}
\showDOI{\tempurl}
\newblock
\shownote{Publisher: INFORMS}.


\bibitem[Demirbas and Timur~Ogut(2020)]%
        {demirbas_re-designing_2020}
\bibfield{author}{\bibinfo{person}{Duysal Demirbas} {and} \bibinfo{person}{Sebnem Timur~Ogut}.} \bibinfo{year}{2020}\natexlab{}.
\newblock \showarticletitle{Re-{Designing} the {Design} {Brief} as a {Digital} {Learning} {Tool} with {Participatory} {Design} {Approach}}.
\newblock \bibinfo{journal}{\emph{Turkish Online Journal of Distance Education}} \bibinfo{volume}{21}, \bibinfo{number}{1} (\bibinfo{date}{Jan.} \bibinfo{year}{2020}), \bibinfo{pages}{83--100}.
\newblock
\showISSN{1302-6488}
\urldef\tempurl%
\url{https://doi.org/10.17718/tojde.690356}
\showDOI{\tempurl}


\bibitem[Denny et~al\mbox{.}(2024a)]%
        {denny_prompt_2024}
\bibfield{author}{\bibinfo{person}{Paul Denny}, \bibinfo{person}{Juho Leinonen}, \bibinfo{person}{James Prather}, \bibinfo{person}{Andrew Luxton-Reilly}, \bibinfo{person}{Thezyrie Amarouche}, \bibinfo{person}{Brett~A. Becker}, {and} \bibinfo{person}{Brent~N. Reeves}.} \bibinfo{year}{2024}\natexlab{a}.
\newblock \showarticletitle{Prompt {Problems}: {A} {New} {Programming} {Exercise} for the {Generative} {AI} {Era}}. In \bibinfo{booktitle}{\emph{Proceedings of the 55th {ACM} {Technical} {Symposium} on {Computer} {Science} {Education} {V}. 1}}. \bibinfo{publisher}{ACM}, \bibinfo{address}{Portland OR USA}, \bibinfo{pages}{296--302}.
\newblock
\showISBNx{9798400704239}
\urldef\tempurl%
\url{https://doi.org/10.1145/3626252.3630909}
\showDOI{\tempurl}


\bibitem[Denny et~al\mbox{.}(2024b)]%
        {denny_explaining_2024}
\bibfield{author}{\bibinfo{person}{Paul Denny}, \bibinfo{person}{David~H. Smith}, \bibinfo{person}{Max Fowler}, \bibinfo{person}{James Prather}, \bibinfo{person}{Brett~A. Becker}, {and} \bibinfo{person}{Juho Leinonen}.} \bibinfo{year}{2024}\natexlab{b}.
\newblock \showarticletitle{Explaining {Code} with a {Purpose}: {An} {Integrated} {Approach} for {Developing} {Code} {Comprehension} and {Prompting} {Skills}}. In \bibinfo{booktitle}{\emph{Proceedings of the 2024 on {Innovation} and {Technology} in {Computer} {Science} {Education} {V}. 1}}. \bibinfo{publisher}{ACM}, \bibinfo{address}{Milan Italy}, \bibinfo{pages}{283--289}.
\newblock
\showISBNx{9798400706004}
\urldef\tempurl%
\url{https://doi.org/10.1145/3649217.3653587}
\showDOI{\tempurl}


\bibitem[{Digital Education Council}(2024)]%
        {digital_education_council_what_2024}
\bibfield{author}{\bibinfo{person}{{Digital Education Council}}.} \bibinfo{year}{2024}\natexlab{}.
\newblock \bibinfo{title}{What {Students} {Want}: {Key} {Results} from {DEC} {Global} {AI} {Student} {Survey} 2024}.
\newblock
\newblock
\urldef\tempurl%
\url{https://www.digitaleducationcouncil.com/post/what-students-want-key-results-from-dec-global-ai-student-survey-2024}
\showURL{%
\tempurl}


\bibitem[Dindler et~al\mbox{.}(2020)]%
        {dindler_computational_2020}
\bibfield{author}{\bibinfo{person}{Christian Dindler}, \bibinfo{person}{Rachel Smith}, {and} \bibinfo{person}{Ole~Sejer Iversen}.} \bibinfo{year}{2020}\natexlab{}.
\newblock \showarticletitle{Computational empowerment: participatory design in education}.
\newblock \bibinfo{journal}{\emph{CoDesign}} \bibinfo{volume}{16}, \bibinfo{number}{1} (\bibinfo{date}{Jan.} \bibinfo{year}{2020}), \bibinfo{pages}{66--80}.
\newblock
\showISSN{1571-0882}
\urldef\tempurl%
\url{https://doi.org/10.1080/15710882.2020.1722173}
\showDOI{\tempurl}
\newblock
\shownote{Publisher: Taylor \& Francis \_eprint: https://doi.org/10.1080/15710882.2020.1722173}.


\bibitem[Ditta et~al\mbox{.}(2023)]%
        {ditta_what_2023}
\bibfield{author}{\bibinfo{person}{Annie~S. Ditta}, \bibinfo{person}{Julia~S. Soares}, {and} \bibinfo{person}{Benjamin~C. Storm}.} \bibinfo{year}{2023}\natexlab{}.
\newblock \showarticletitle{What happens to memory for lecture content when students take photos of the lecture slides?}
\newblock \bibinfo{journal}{\emph{Journal of Applied Research in Memory and Cognition}} \bibinfo{volume}{12}, \bibinfo{number}{3} (\bibinfo{date}{Sept.} \bibinfo{year}{2023}), \bibinfo{pages}{421--430}.
\newblock
\showISSN{2211-369X, 2211-3681}
\urldef\tempurl%
\url{https://doi.org/10.1037/mac0000069}
\showDOI{\tempurl}


\bibitem[Draper(2009)]%
        {draper_catalytic_2009}
\bibfield{author}{\bibinfo{person}{Stephen~W. Draper}.} \bibinfo{year}{2009}\natexlab{}.
\newblock \showarticletitle{Catalytic assessment: understanding how {MCQs} and {EVS} can foster deep learning}.
\newblock \bibinfo{journal}{\emph{British Journal of Educational Technology}} \bibinfo{volume}{40}, \bibinfo{number}{2} (\bibinfo{year}{2009}), \bibinfo{pages}{285--293}.
\newblock
\showISSN{1467-8535}
\urldef\tempurl%
\url{https://doi.org/10.1111/j.1467-8535.2008.00920.x}
\showDOI{\tempurl}
\newblock
\shownote{\_eprint: https://onlinelibrary.wiley.com/doi/pdf/10.1111/j.1467-8535.2008.00920.x}.


\bibitem[Drosos et~al\mbox{.}(2024)]%
        {drosos_its_2024}
\bibfield{author}{\bibinfo{person}{Ian Drosos}, \bibinfo{person}{Advait Sarkar}, \bibinfo{person}{Xiaotong Xu}, \bibinfo{person}{Carina Negreanu}, \bibinfo{person}{Sean Rintel}, {and} \bibinfo{person}{Lev Tankelevitch}.} \bibinfo{year}{2024}\natexlab{}.
\newblock \showarticletitle{"{It}'s like a rubber duck that talks back": {Understanding} {Generative} {AI}-{Assisted} {Data} {Analysis} {Workflows} through a {Participatory} {Prompting} {Study}}. In \bibinfo{booktitle}{\emph{Proceedings of the 3rd {Annual} {Meeting} of the {Symposium} on {Human}-{Computer} {Interaction} for {Work}}}. \bibinfo{publisher}{ACM}, \bibinfo{address}{Newcastle upon Tyne United Kingdom}, \bibinfo{pages}{1--21}.
\newblock
\showISBNx{9798400710179}
\urldef\tempurl%
\url{https://doi.org/10.1145/3663384.3663389}
\showDOI{\tempurl}


\bibitem[Eke(2023)]%
        {eke_chatgpt_2023}
\bibfield{author}{\bibinfo{person}{Damian~Okaibedi Eke}.} \bibinfo{year}{2023}\natexlab{}.
\newblock \showarticletitle{{ChatGPT} and the rise of generative {AI}: {Threat} to academic integrity?}
\newblock \bibinfo{journal}{\emph{Journal of Responsible Technology}}  \bibinfo{volume}{13} (\bibinfo{date}{April} \bibinfo{year}{2023}), \bibinfo{pages}{100060}.
\newblock
\showISSN{2666-6596}
\urldef\tempurl%
\url{https://doi.org/10.1016/j.jrt.2023.100060}
\showDOI{\tempurl}


\bibitem[Fahim and Masouleh(2012)]%
        {fahim_critical_2012}
\bibfield{author}{\bibinfo{person}{Mansoor Fahim} {and} \bibinfo{person}{Nima~Shakouri Masouleh}.} \bibinfo{year}{2012}\natexlab{}.
\newblock \showarticletitle{Critical thinking in higher education: {A} pedagogical look.}
\newblock \bibinfo{journal}{\emph{Theory \& Practice in Language Studies (TPLS)}} \bibinfo{volume}{2}, \bibinfo{number}{7} (\bibinfo{year}{2012}).
\newblock
\urldef\tempurl%
\url{http://academypublication.com/issues/past/tpls/vol02/07/tpls0207.pdf#page=56}
\showURL{%
\tempurl}


\bibitem[Firat(2023)]%
        {firat_how_2023}
\bibfield{author}{\bibinfo{person}{Mehmet Firat}.} \bibinfo{year}{2023}\natexlab{}.
\newblock \bibinfo{booktitle}{\emph{How {Chat} {GPT} {Can} {Transform} {Autodidactic} {Experiences} and {Open} {Education}?}}
\newblock
\urldef\tempurl%
\url{https://doi.org/10.31219/osf.io/9ge8m}
\showDOI{\tempurl}


\bibitem[Gero et~al\mbox{.}(2023)]%
        {gero_social_2023}
\bibfield{author}{\bibinfo{person}{Katy~Ilonka Gero}, \bibinfo{person}{Tao Long}, {and} \bibinfo{person}{Lydia~B Chilton}.} \bibinfo{year}{2023}\natexlab{}.
\newblock \showarticletitle{Social {Dynamics} of {AI} {Support} in {Creative} {Writing}}. In \bibinfo{booktitle}{\emph{Proceedings of the 2023 {CHI} {Conference} on {Human} {Factors} in {Computing} {Systems}}} \emph{(\bibinfo{series}{{CHI} '23})}. \bibinfo{publisher}{Association for Computing Machinery}, \bibinfo{address}{New York, NY, USA}, \bibinfo{pages}{1--15}.
\newblock
\showISBNx{978-1-4503-9421-5}
\urldef\tempurl%
\url{https://doi.org/10.1145/3544548.3580782}
\showDOI{\tempurl}


\bibitem[Ghimire and Edwards(2024)]%
        {ghimire_coding_2024}
\bibfield{author}{\bibinfo{person}{Aashish Ghimire} {and} \bibinfo{person}{John Edwards}.} \bibinfo{year}{2024}\natexlab{}.
\newblock \showarticletitle{Coding with {AI}: {How} {Are} {Tools} {Like} {ChatGPT} {Being} {Used} by {Students} in {Foundational} {Programming} {Courses}}. In \bibinfo{booktitle}{\emph{Artificial {Intelligence} in {Education}}}, \bibfield{editor}{\bibinfo{person}{Andrew~M. Olney}, \bibinfo{person}{Irene-Angelica Chounta}, \bibinfo{person}{Zitao Liu}, \bibinfo{person}{Olga~C. Santos}, {and} \bibinfo{person}{Ig~Ibert Bittencourt}} (Eds.). \bibinfo{publisher}{Springer Nature Switzerland}, \bibinfo{address}{Cham}, \bibinfo{pages}{259--267}.
\newblock
\showISBNx{978-3-031-64299-9}
\urldef\tempurl%
\url{https://doi.org/10.1007/978-3-031-64299-9_20}
\showDOI{\tempurl}


\bibitem[Ghimire et~al\mbox{.}(2024)]%
        {ghimire_generative_2024}
\bibfield{author}{\bibinfo{person}{Aashish Ghimire}, \bibinfo{person}{James Prather}, {and} \bibinfo{person}{John Edwards}.} \bibinfo{year}{2024}\natexlab{}.
\newblock \bibinfo{title}{Generative {AI} in {Education}: {A} {Study} of {Educators}' {Awareness}, {Sentiments}, and {Influencing} {Factors}}.
\newblock
\newblock
\urldef\tempurl%
\url{http://arxiv.org/abs/2403.15586}
\showURL{%
\tempurl}
\newblock
\shownote{arXiv:2403.15586 [cs]}.


\bibitem[Giddens(1984)]%
        {giddens_constitution_1984}
\bibfield{author}{\bibinfo{person}{Anthony Giddens}.} \bibinfo{year}{1984}\natexlab{}.
\newblock \bibinfo{booktitle}{\emph{The {Constitution} of {Society}: {Outline} of the {Theory} of {Structuration}}}.
\newblock \bibinfo{publisher}{University of California Press}.
\newblock
\showISBNx{978-0-520-05292-5}
\newblock
\shownote{Google-Books-ID: x2bf4g9Z6ZwC}.


\bibitem[Greenhalgh and Stones(2010)]%
        {greenhalgh_theorising_2010}
\bibfield{author}{\bibinfo{person}{Trisha Greenhalgh} {and} \bibinfo{person}{Rob Stones}.} \bibinfo{year}{2010}\natexlab{}.
\newblock \showarticletitle{Theorising {Big} {IT} {Programmes} in {Healthcare}: {Strong} {Structuration} {Theory} {Meets} {Actor}-{Network} {Theory}}.
\newblock \bibinfo{journal}{\emph{Social science \& medicine (1982)}}  \bibinfo{volume}{70} (\bibinfo{date}{Feb.} \bibinfo{year}{2010}), \bibinfo{pages}{1285--94}.
\newblock
\urldef\tempurl%
\url{https://doi.org/10.1016/j.socscimed.2009.12.034}
\showDOI{\tempurl}


\bibitem[Greenhalgh et~al\mbox{.}(2016)]%
        {greenhalgh_virtual_2016}
\bibfield{author}{\bibinfo{person}{Trisha Greenhalgh}, \bibinfo{person}{Shanti Vijayaraghavan}, \bibinfo{person}{Joe Wherton}, \bibinfo{person}{Sara Shaw}, \bibinfo{person}{Emma Byrne}, \bibinfo{person}{Desirée Campbell-Richards}, \bibinfo{person}{Satya Bhattacharya}, \bibinfo{person}{Philippa Hanson}, \bibinfo{person}{Seendy Ramoutar}, \bibinfo{person}{Charles Gutteridge}, \bibinfo{person}{Isabel Hodkinson}, \bibinfo{person}{Anna Collard}, {and} \bibinfo{person}{Joanne Morris}.} \bibinfo{year}{2016}\natexlab{}.
\newblock \showarticletitle{Virtual online consultations: advantages and limitations ({VOCAL}) study}.
\newblock \bibinfo{journal}{\emph{BMJ Open}} \bibinfo{volume}{6}, \bibinfo{number}{1} (\bibinfo{date}{Jan.} \bibinfo{year}{2016}), \bibinfo{pages}{e009388}.
\newblock
\showISSN{2044-6055, 2044-6055}
\urldef\tempurl%
\url{https://doi.org/10.1136/bmjopen-2015-009388}
\showDOI{\tempurl}
\newblock
\shownote{Publisher: British Medical Journal Publishing Group Section: Health informatics}.


\bibitem[Hadi~Mogavi et~al\mbox{.}(2024)]%
        {hadi_mogavi_chatgpt_2024}
\bibfield{author}{\bibinfo{person}{Reza Hadi~Mogavi}, \bibinfo{person}{Chao Deng}, \bibinfo{person}{Justin Juho~Kim}, \bibinfo{person}{Pengyuan Zhou}, \bibinfo{person}{Young D.~Kwon}, \bibinfo{person}{Ahmed Hosny Saleh~Metwally}, \bibinfo{person}{Ahmed Tlili}, \bibinfo{person}{Simone Bassanelli}, \bibinfo{person}{Antonio Bucchiarone}, \bibinfo{person}{Sujit Gujar}, \bibinfo{person}{Lennart~E. Nacke}, {and} \bibinfo{person}{Pan Hui}.} \bibinfo{year}{2024}\natexlab{}.
\newblock \showarticletitle{{ChatGPT} in education: {A} blessing or a curse? {A} qualitative study exploring early adopters’ utilization and perceptions}.
\newblock \bibinfo{journal}{\emph{Computers in Human Behavior: Artificial Humans}} \bibinfo{volume}{2}, \bibinfo{number}{1} (\bibinfo{date}{Jan.} \bibinfo{year}{2024}), \bibinfo{pages}{100027}.
\newblock
\showISSN{2949-8821}
\urldef\tempurl%
\url{https://doi.org/10.1016/j.chbah.2023.100027}
\showDOI{\tempurl}


\bibitem[Han et~al\mbox{.}(2024)]%
        {han_teachers_2024}
\bibfield{author}{\bibinfo{person}{Ariel Han}, \bibinfo{person}{Xiaofei Zhou}, \bibinfo{person}{Zhenyao Cai}, \bibinfo{person}{Shenshen Han}, \bibinfo{person}{Richard Ko}, \bibinfo{person}{Seth Corrigan}, {and} \bibinfo{person}{Kylie~A Peppler}.} \bibinfo{year}{2024}\natexlab{}.
\newblock \showarticletitle{Teachers, {Parents}, and {Students}' perspectives on {Integrating} {Generative} {AI} into {Elementary} {Literacy} {Education}}. In \bibinfo{booktitle}{\emph{Proceedings of the {CHI} {Conference} on {Human} {Factors} in {Computing} {Systems}}} \emph{(\bibinfo{series}{{CHI} '24})}. \bibinfo{publisher}{Association for Computing Machinery}, \bibinfo{address}{New York, NY, USA}, \bibinfo{pages}{1--17}.
\newblock
\showISBNx{9798400703300}
\urldef\tempurl%
\url{https://doi.org/10.1145/3613904.3642438}
\showDOI{\tempurl}


\bibitem[Hasanein and Sobaih(2023)]%
        {hasanein_drivers_2023}
\bibfield{author}{\bibinfo{person}{Ahmed~M. Hasanein} {and} \bibinfo{person}{Abu Elnasr~E. Sobaih}.} \bibinfo{year}{2023}\natexlab{}.
\newblock \showarticletitle{Drivers and {Consequences} of {ChatGPT} {Use} in {Higher} {Education}: {Key} {Stakeholder} {Perspectives}}.
\newblock \bibinfo{journal}{\emph{European Journal of Investigation in Health, Psychology and Education}} \bibinfo{volume}{13}, \bibinfo{number}{11} (\bibinfo{date}{Nov.} \bibinfo{year}{2023}), \bibinfo{pages}{2599--2614}.
\newblock
\showISSN{2254-9625}
\urldef\tempurl%
\url{https://doi.org/10.3390/ejihpe13110181}
\showDOI{\tempurl}
\newblock
\shownote{Number: 11 Publisher: Multidisciplinary Digital Publishing Institute}.


\bibitem[Hellas et~al\mbox{.}(2024)]%
        {hellas_experiences_2024}
\bibfield{author}{\bibinfo{person}{Arto Hellas}, \bibinfo{person}{Juho Leinonen}, {and} \bibinfo{person}{Leo Leppänen}.} \bibinfo{year}{2024}\natexlab{}.
\newblock \bibinfo{title}{Experiences from {Integrating} {Large} {Language} {Model} {Chatbots} into the {Classroom}}.
\newblock
\newblock
\urldef\tempurl%
\url{https://doi.org/10.48550/arXiv.2406.04817}
\showDOI{\tempurl}
\newblock
\shownote{arXiv:2406.04817}.


\bibitem[Holland and Ciachir(2024)]%
        {holland_qualitative_2024}
\bibfield{author}{\bibinfo{person}{Anna Holland} {and} \bibinfo{person}{Constantin Ciachir}.} \bibinfo{year}{2024}\natexlab{}.
\newblock \showarticletitle{A qualitative study of students’ lived experience and perceptions of using {ChatGPT}: immediacy, equity and integrity}.
\newblock \bibinfo{journal}{\emph{Interactive Learning Environments}} (\bibinfo{date}{May} \bibinfo{year}{2024}).
\newblock
\showISSN{1049-4820}
\urldef\tempurl%
\url{https://www.tandfonline.com/doi/abs/10.1080/10494820.2024.2350655}
\showURL{%
\tempurl}
\newblock
\shownote{Publisher: Routledge}.


\bibitem[Hou et~al\mbox{.}(2024)]%
        {hou_effects_2024}
\bibfield{author}{\bibinfo{person}{Irene Hou}, \bibinfo{person}{Sophia Mettille}, \bibinfo{person}{Owen Man}, \bibinfo{person}{Zhuo Li}, \bibinfo{person}{Cynthia Zastudil}, {and} \bibinfo{person}{Stephen MacNeil}.} \bibinfo{year}{2024}\natexlab{}.
\newblock \showarticletitle{The {Effects} of {Generative} {AI} on {Computing} {Students}’ {Help}-{Seeking} {Preferences}}. In \bibinfo{booktitle}{\emph{Proceedings of the 26th {Australasian} {Computing} {Education} {Conference}}}. \bibinfo{publisher}{ACM}, \bibinfo{address}{Sydney NSW Australia}, \bibinfo{pages}{39--48}.
\newblock
\showISBNx{9798400716195}
\urldef\tempurl%
\url{https://doi.org/10.1145/3636243.3636248}
\showDOI{\tempurl}


\bibitem[Huallpa and Al(2023)]%
        {huallpa_exploring_2023}
\bibfield{author}{\bibinfo{person}{Jorge~Jinchuña Huallpa} {and} \bibinfo{person}{Et Al}.} \bibinfo{year}{2023}\natexlab{}.
\newblock \showarticletitle{Exploring the ethical considerations of using {Chat} {GPT} in university education}.
\newblock \bibinfo{journal}{\emph{Periodicals of Engineering and Natural Sciences}} \bibinfo{volume}{11}, \bibinfo{number}{4} (\bibinfo{date}{Aug.} \bibinfo{year}{2023}), \bibinfo{pages}{105--115}.
\newblock
\showISSN{2303-4521}
\urldef\tempurl%
\url{https://doi.org/10.21533/pen.v11i4.3770}
\showDOI{\tempurl}
\newblock
\shownote{Number: 4}.


\bibitem[Johnston et~al\mbox{.}(2024)]%
        {johnston_student_2024}
\bibfield{author}{\bibinfo{person}{Heather Johnston}, \bibinfo{person}{Rebecca~F. Wells}, \bibinfo{person}{Elizabeth~M. Shanks}, \bibinfo{person}{Timothy Boey}, {and} \bibinfo{person}{Bryony~N. Parsons}.} \bibinfo{year}{2024}\natexlab{}.
\newblock \showarticletitle{Student perspectives on the use of generative artificial intelligence technologies in higher education}.
\newblock \bibinfo{journal}{\emph{International Journal for Educational Integrity}} \bibinfo{volume}{20}, \bibinfo{number}{1} (\bibinfo{date}{Feb.} \bibinfo{year}{2024}), \bibinfo{pages}{2}.
\newblock
\showISSN{1833-2595}
\urldef\tempurl%
\url{https://doi.org/10.1007/s40979-024-00149-4}
\showDOI{\tempurl}


\bibitem[Jurenka et~al\mbox{.}({[n.\,d.]})]%
        {jurenka_towards_nodate}
\bibfield{author}{\bibinfo{person}{Irina Jurenka}, \bibinfo{person}{Markus Kunesch}, \bibinfo{person}{Kevin McKee}, \bibinfo{person}{Daniel Gillick}, \bibinfo{person}{Shaojian Zhu}, \bibinfo{person}{Sara Wiltberger}, \bibinfo{person}{Shubham~Milind Phal}, \bibinfo{person}{Katherine Hermann}, \bibinfo{person}{Daniel Kasenberg}, \bibinfo{person}{Avishkar Bhoopchand}, \bibinfo{person}{Ankit Anand}, \bibinfo{person}{Miruna Pîslar}, \bibinfo{person}{Stephanie Chan}, \bibinfo{person}{Lisa Wang}, \bibinfo{person}{Jennifer She}, \bibinfo{person}{Parsa Mahmoudieh}, \bibinfo{person}{Aliya Rysbek}, \bibinfo{person}{Andrea Huber}, \bibinfo{person}{Brett Wiltshire}, \bibinfo{person}{Gal Elidan}, \bibinfo{person}{Roni Rabin}, \bibinfo{person}{Jasmin Rubinovitz}, \bibinfo{person}{Amit Pitaru}, \bibinfo{person}{Julia Wilkowski}, \bibinfo{person}{David Choi}, \bibinfo{person}{Roee Engelberg}, \bibinfo{person}{Lidan Hackmon}, \bibinfo{person}{Adva Levin}, \bibinfo{person}{Rachel Griffin}, \bibinfo{person}{Michael Sears},
  \bibinfo{person}{Filip Bar}, \bibinfo{person}{Mia Mesar}, \bibinfo{person}{Mana Jabbour}, \bibinfo{person}{Arslan Chaudhry}, \bibinfo{person}{James Cohan}, \bibinfo{person}{Nir Levine}, \bibinfo{person}{Ben Brown}, \bibinfo{person}{Dilan Gorur}, \bibinfo{person}{Svetlana Grant}, \bibinfo{person}{Rachel Hashimoshoni}, \bibinfo{person}{Jieru Hu}, \bibinfo{person}{Dawn Chen}, \bibinfo{person}{Kuba Dolecki}, \bibinfo{person}{Canfer Akbulut}, \bibinfo{person}{Maxwell Bileschi}, \bibinfo{person}{Laura Culp}, \bibinfo{person}{Wen-Xin Dong}, \bibinfo{person}{Nahema Marchal}, \bibinfo{person}{Kelsie~Van Deman}, \bibinfo{person}{Hema~Bajaj Misra}, \bibinfo{person}{Michael Duah}, \bibinfo{person}{Moran Ambar}, \bibinfo{person}{Avi Caciularu}, \bibinfo{person}{Sandra Lefdal}, \bibinfo{person}{Chris Summerfield}, \bibinfo{person}{James An}, \bibinfo{person}{Pierre-Alexandre Kamienny}, \bibinfo{person}{Abhinit Mohdi}, \bibinfo{person}{Theofilos Strinopoulous}, \bibinfo{person}{Annie Hale}, \bibinfo{person}{Wayne
  Anderson}, \bibinfo{person}{Luis~C Cobo}, \bibinfo{person}{Niv Efron}, \bibinfo{person}{Muktha Ananda}, \bibinfo{person}{Shakir Mohamed}, \bibinfo{person}{Maureen Heymans}, \bibinfo{person}{Zoubin Ghahramani}, \bibinfo{person}{Yossi Matias}, \bibinfo{person}{Ben Gomes}, {and} \bibinfo{person}{Lila Ibrahim}.} \bibinfo{year}{[n.\,d.]}\natexlab{}.
\newblock \showarticletitle{Towards {Responsible} {Development} of {Generative} {AI} for {Education}: {An} {Evaluation}-{Driven} {Approach}}.
\newblock  (\bibinfo{year}{[n.\,d.]}).
\newblock


\bibitem[Kazemitabaar et~al\mbox{.}(2023a)]%
        {kazemitabaar_studying_2023}
\bibfield{author}{\bibinfo{person}{Majeed Kazemitabaar}, \bibinfo{person}{Justin Chow}, \bibinfo{person}{Carl Ka~To Ma}, \bibinfo{person}{Barbara~J. Ericson}, \bibinfo{person}{David Weintrop}, {and} \bibinfo{person}{Tovi Grossman}.} \bibinfo{year}{2023}\natexlab{a}.
\newblock \showarticletitle{Studying the effect of {AI} {Code} {Generators} on {Supporting} {Novice} {Learners} in {Introductory} {Programming}}. In \bibinfo{booktitle}{\emph{Proceedings of the 2023 {CHI} {Conference} on {Human} {Factors} in {Computing} {Systems}}} \emph{(\bibinfo{series}{{CHI} '23})}. \bibinfo{publisher}{Association for Computing Machinery}, \bibinfo{address}{New York, NY, USA}, \bibinfo{pages}{1--23}.
\newblock
\showISBNx{978-1-4503-9421-5}
\urldef\tempurl%
\url{https://doi.org/10.1145/3544548.3580919}
\showDOI{\tempurl}


\bibitem[Kazemitabaar et~al\mbox{.}(2023b)]%
        {kazemitabaar_how_2023}
\bibfield{author}{\bibinfo{person}{Majeed Kazemitabaar}, \bibinfo{person}{Xinying Hou}, \bibinfo{person}{Austin Henley}, \bibinfo{person}{Barbara~J. Ericson}, \bibinfo{person}{David Weintrop}, {and} \bibinfo{person}{Tovi Grossman}.} \bibinfo{year}{2023}\natexlab{b}.
\newblock \bibinfo{title}{How {Novices} {Use} {LLM}-{Based} {Code} {Generators} to {Solve} {CS1} {Coding} {Tasks} in a {Self}-{Paced} {Learning} {Environment}}.
\newblock
\newblock
\urldef\tempurl%
\url{https://doi.org/10.48550/arXiv.2309.14049}
\showDOI{\tempurl}
\newblock
\shownote{arXiv:2309.14049 [cs]}.


\bibitem[Kazemitabaar et~al\mbox{.}(2024a)]%
        {kazemitabaar_exploring_2024}
\bibfield{author}{\bibinfo{person}{Majeed Kazemitabaar}, \bibinfo{person}{Oliver Huang}, \bibinfo{person}{Sangho Suh}, \bibinfo{person}{Austin~Z. Henley}, {and} \bibinfo{person}{Tovi Grossman}.} \bibinfo{year}{2024}\natexlab{a}.
\newblock \bibinfo{title}{Exploring the {Design} {Space} of {Cognitive} {Engagement} {Techniques} with {AI}-{Generated} {Code} for {Enhanced} {Learning}}.
\newblock
\newblock
\urldef\tempurl%
\url{https://doi.org/10.48550/arXiv.2410.08922}
\showDOI{\tempurl}
\newblock
\shownote{arXiv:2410.08922}.


\bibitem[Kazemitabaar et~al\mbox{.}(2024b)]%
        {kazemitabaar_improving_2024}
\bibfield{author}{\bibinfo{person}{Majeed Kazemitabaar}, \bibinfo{person}{Jack Williams}, \bibinfo{person}{Ian Drosos}, \bibinfo{person}{Tovi Grossman}, \bibinfo{person}{Austin Henley}, \bibinfo{person}{Carina Negreanu}, {and} \bibinfo{person}{Advait Sarkar}.} \bibinfo{year}{2024}\natexlab{b}.
\newblock \showarticletitle{Improving {Steering} and {Verification} in {AI}-{Assisted} {Data} {Analysis} with {Interactive} {Task} {Decomposition}}.
\newblock
\urldef\tempurl%
\url{https://www.microsoft.com/en-us/research/publication/improving-steering-and-verification-in-ai-assisted-data-analysis-with-interactive-task-decomposition/}
\showURL{%
\tempurl}


\bibitem[Kazemitabaar et~al\mbox{.}(2024c)]%
        {kazemitabaar_codeaid_2024}
\bibfield{author}{\bibinfo{person}{Majeed Kazemitabaar}, \bibinfo{person}{Runlong Ye}, \bibinfo{person}{Xiaoning Wang}, \bibinfo{person}{Austin~Zachary Henley}, \bibinfo{person}{Paul Denny}, \bibinfo{person}{Michelle Craig}, {and} \bibinfo{person}{Tovi Grossman}.} \bibinfo{year}{2024}\natexlab{c}.
\newblock \showarticletitle{{CodeAid}: {Evaluating} a {Classroom} {Deployment} of an {LLM}-based {Programming} {Assistant} that {Balances} {Student} and {Educator} {Needs}}. In \bibinfo{booktitle}{\emph{Proceedings of the 2024 {CHI} {Conference} on {Human} {Factors} in {Computing} {Systems}}} \emph{(\bibinfo{series}{{CHI} '24})}. \bibinfo{publisher}{Association for Computing Machinery}, \bibinfo{address}{New York, NY, USA}, \bibinfo{pages}{1--20}.
\newblock
\showISBNx{9798400703300}
\urldef\tempurl%
\url{https://doi.org/10.1145/3613904.3642773}
\showDOI{\tempurl}


\bibitem[Kharrufa and Johnson(2024)]%
        {kharrufa_potential_2024}
\bibfield{author}{\bibinfo{person}{Ahmed Kharrufa} {and} \bibinfo{person}{Ian Johnson}.} \bibinfo{year}{2024}\natexlab{}.
\newblock \showarticletitle{The {Potential} and {Implications} of {Generative} {AI} on {HCI} {Education}}. In \bibinfo{booktitle}{\emph{Proceedings of the 6th {Annual} {Symposium} on {HCI} {Education}}} \emph{(\bibinfo{series}{{EduCHI} '24})}. \bibinfo{publisher}{Association for Computing Machinery}, \bibinfo{address}{New York, NY, USA}, \bibinfo{pages}{1--8}.
\newblock
\showISBNx{9798400716591}
\urldef\tempurl%
\url{https://doi.org/10.1145/3658619.3658627}
\showDOI{\tempurl}


\bibitem[Kicklighter et~al\mbox{.}(2024)]%
        {kicklighter_empowering_2024}
\bibfield{author}{\bibinfo{person}{Caleb Kicklighter}, \bibinfo{person}{Jinsil~Hwaryoung Seo}, \bibinfo{person}{Mayet Andreassen}, {and} \bibinfo{person}{Emily Bujnoch}.} \bibinfo{year}{2024}\natexlab{}.
\newblock \showarticletitle{Empowering {Creativity} with {Generative} {AI} in {Digital} {Art} {Education}}. In \bibinfo{booktitle}{\emph{{ACM} {SIGGRAPH} 2024 {Educator}'s {Forum}}} \emph{(\bibinfo{series}{{SIGGRAPH} '24})}. \bibinfo{publisher}{Association for Computing Machinery}, \bibinfo{address}{New York, NY, USA}, \bibinfo{pages}{1--2}.
\newblock
\showISBNx{9798400705175}
\urldef\tempurl%
\url{https://doi.org/10.1145/3641235.3664438}
\showDOI{\tempurl}


\bibitem[Koedinger et~al\mbox{.}(2013)]%
        {koedinger_new_2013}
\bibfield{author}{\bibinfo{person}{Kenneth~R. Koedinger}, \bibinfo{person}{Emma Brunskill}, \bibinfo{person}{Ryan S. J.~d. Baker}, \bibinfo{person}{Elizabeth~A. McLaughlin}, {and} \bibinfo{person}{John Stamper}.} \bibinfo{year}{2013}\natexlab{}.
\newblock \showarticletitle{New {Potentials} for {Data}-{Driven} {Intelligent} {Tutoring} {System} {Development} and {Optimization}}.
\newblock \bibinfo{journal}{\emph{AI Magazine}} \bibinfo{volume}{34}, \bibinfo{number}{3} (\bibinfo{year}{2013}), \bibinfo{pages}{27--41}.
\newblock
\showISSN{2371-9621}
\urldef\tempurl%
\url{https://doi.org/10.1609/aimag.v34i3.2484}
\showDOI{\tempurl}
\newblock
\shownote{\_eprint: https://onlinelibrary.wiley.com/doi/pdf/10.1609/aimag.v34i3.2484}.


\bibitem[Kooli(2023)]%
        {kooli_chatbots_2023}
\bibfield{author}{\bibinfo{person}{Chokri Kooli}.} \bibinfo{year}{2023}\natexlab{}.
\newblock \showarticletitle{Chatbots in {Education} and {Research}: {A} {Critical} {Examination} of {Ethical} {Implications} and {Solutions}}.
\newblock \bibinfo{journal}{\emph{Sustainability}} \bibinfo{volume}{15}, \bibinfo{number}{7} (\bibinfo{date}{Jan.} \bibinfo{year}{2023}), \bibinfo{pages}{5614}.
\newblock
\showISSN{2071-1050}
\urldef\tempurl%
\url{https://doi.org/10.3390/su15075614}
\showDOI{\tempurl}
\newblock
\shownote{Number: 7 Publisher: Multidisciplinary Digital Publishing Institute}.


\bibitem[Kubullek et~al\mbox{.}(2024)]%
        {kubullek_understanding_2024}
\bibfield{author}{\bibinfo{person}{Ann-Kathrin Kubullek}, \bibinfo{person}{Nadire Kumaç}, {and} \bibinfo{person}{Aysegül Dogangün}.} \bibinfo{year}{2024}\natexlab{}.
\newblock \showarticletitle{Understanding the {Adoption} of {ChatGPT} in {Higher} {Education}: {A} {Comparative} {Study} with {Insights} from {STEM} and {Business} {Students}}. In \bibinfo{booktitle}{\emph{Proceedings of {Mensch} und {Computer} 2024}} \emph{(\bibinfo{series}{{MuC} '24})}. \bibinfo{publisher}{Association for Computing Machinery}, \bibinfo{address}{New York, NY, USA}, \bibinfo{pages}{684--689}.
\newblock
\showISBNx{9798400709982}
\urldef\tempurl%
\url{https://doi.org/10.1145/3670653.3677507}
\showDOI{\tempurl}


\bibitem[Kulik and Fletcher(2016)]%
        {kulik_effectiveness_2016}
\bibfield{author}{\bibinfo{person}{James~A. Kulik} {and} \bibinfo{person}{J.~D. Fletcher}.} \bibinfo{year}{2016}\natexlab{}.
\newblock \showarticletitle{Effectiveness of {Intelligent} {Tutoring} {Systems}: {A} {Meta}-{Analytic} {Review}}.
\newblock \bibinfo{journal}{\emph{Review of Educational Research}} \bibinfo{volume}{86}, \bibinfo{number}{1} (\bibinfo{date}{March} \bibinfo{year}{2016}), \bibinfo{pages}{42--78}.
\newblock
\showISSN{0034-6543}
\urldef\tempurl%
\url{https://doi.org/10.3102/0034654315581420}
\showDOI{\tempurl}
\newblock
\shownote{Publisher: American Educational Research Association}.


\bibitem[Kumar et~al\mbox{.}(2024a)]%
        {kumar_guiding_2024}
\bibfield{author}{\bibinfo{person}{Harsh Kumar}, \bibinfo{person}{Ilya Musabirov}, \bibinfo{person}{Mohi Reza}, \bibinfo{person}{Jiakai Shi}, \bibinfo{person}{Xinyuan Wang}, \bibinfo{person}{Joseph~Jay Williams}, \bibinfo{person}{Anastasia Kuzminykh}, {and} \bibinfo{person}{Michael Liut}.} \bibinfo{year}{2024}\natexlab{a}.
\newblock \showarticletitle{Guiding {Students} in {Using} {LLMs} in {Supported} {Learning} {Environments}: {Effects} on {Interaction} {Dynamics}, {Learner} {Performance}, {Confidence}, and {Trust}}.
\newblock \bibinfo{journal}{\emph{Proc. ACM Hum.-Comput. Interact.}} \bibinfo{volume}{8}, \bibinfo{number}{CSCW2} (\bibinfo{date}{Nov.} \bibinfo{year}{2024}), \bibinfo{pages}{499:1--499:30}.
\newblock
\urldef\tempurl%
\url{https://doi.org/10.1145/3687038}
\showDOI{\tempurl}


\bibitem[Kumar et~al\mbox{.}(2023)]%
        {kumar_math_2023}
\bibfield{author}{\bibinfo{person}{Harsh Kumar}, \bibinfo{person}{David~M. Rothschild}, \bibinfo{person}{Daniel~G. Goldstein}, {and} \bibinfo{person}{Jake~M. Hofman}.} \bibinfo{year}{2023}\natexlab{}.
\newblock \bibinfo{title}{Math {Education} with {Large} {Language} {Models}: {Peril} or {Promise}?}
\newblock
\newblock
\urldef\tempurl%
\url{https://doi.org/10.2139/ssrn.4641653}
\showDOI{\tempurl}


\bibitem[Kumar et~al\mbox{.}(2024b)]%
        {kumar_supporting_2024}
\bibfield{author}{\bibinfo{person}{Harsh Kumar}, \bibinfo{person}{Ruiwei Xiao}, \bibinfo{person}{Benjamin Lawson}, \bibinfo{person}{Ilya Musabirov}, \bibinfo{person}{Jiakai Shi}, \bibinfo{person}{Xinyuan Wang}, \bibinfo{person}{Huayin Luo}, \bibinfo{person}{Joseph~Jay Williams}, \bibinfo{person}{Anna Rafferty}, \bibinfo{person}{John Stamper}, {and} \bibinfo{person}{Michael Liut}.} \bibinfo{year}{2024}\natexlab{b}.
\newblock \bibinfo{title}{Supporting {Self}-{Reflection} at {Scale} with {Large} {Language} {Models}: {Insights} from {Randomized} {Field} {Experiments} in {Classrooms}}.
\newblock
\newblock
\urldef\tempurl%
\url{https://doi.org/10.48550/arXiv.2406.07571}
\showDOI{\tempurl}
\newblock
\shownote{arXiv:2406.07571}.


\bibitem[Lan and Tung(2023)]%
        {lan_analyzing_2023}
\bibfield{author}{\bibinfo{person}{Duong~Hoai Lan} {and} \bibinfo{person}{Tran~Minh Tung}.} \bibinfo{year}{2023}\natexlab{}.
\newblock \showarticletitle{Analyzing the {Impact} of {Chat}-{GPT} {Usage} by {University} {Students} in {Vietnam}}.
\newblock \bibinfo{journal}{\emph{Migration Letters}} \bibinfo{volume}{20}, \bibinfo{number}{S10} (\bibinfo{date}{Nov.} \bibinfo{year}{2023}), \bibinfo{pages}{259--268}.
\newblock
\showISSN{1741-8992}
\urldef\tempurl%
\url{https://doi.org/10.59670/ml.v20iS10.5134}
\showDOI{\tempurl}
\newblock
\shownote{Number: S10}.


\bibitem[Lee et~al\mbox{.}(2023)]%
        {lee_evidence_2023}
\bibfield{author}{\bibinfo{person}{Douglas~G. Lee}, \bibinfo{person}{Jean Daunizeau}, {and} \bibinfo{person}{Giovanni Pezzulo}.} \bibinfo{year}{2023}\natexlab{}.
\newblock \showarticletitle{Evidence or {Confidence}: {What} {Is} {Really} {Monitored} during a {Decision}?}
\newblock \bibinfo{journal}{\emph{Psychonomic Bulletin \& Review}} (\bibinfo{date}{March} \bibinfo{year}{2023}).
\newblock
\showISSN{1531-5320}
\urldef\tempurl%
\url{https://doi.org/10.3758/s13423-023-02255-9}
\showDOI{\tempurl}


\bibitem[Lee et~al\mbox{.}(2024)]%
        {lee_conversational_2024}
\bibfield{author}{\bibinfo{person}{Soohwan Lee}, \bibinfo{person}{Seoyeong Hwang}, {and} \bibinfo{person}{Kyungho Lee}.} \bibinfo{year}{2024}\natexlab{}.
\newblock \bibinfo{title}{Conversational {Agents} as {Catalysts} for {Critical} {Thinking}: {Challenging} {Design} {Fixation} in {Group} {Design}}.
\newblock
\newblock
\urldef\tempurl%
\url{http://arxiv.org/abs/2406.11125}
\showURL{%
\tempurl}
\newblock
\shownote{arXiv:2406.11125}.


\bibitem[Liffiton et~al\mbox{.}(2023)]%
        {liffiton_codehelp_2023}
\bibfield{author}{\bibinfo{person}{Mark Liffiton}, \bibinfo{person}{Brad~E Sheese}, \bibinfo{person}{Jaromir Savelka}, {and} \bibinfo{person}{Paul Denny}.} \bibinfo{year}{2023}\natexlab{}.
\newblock \showarticletitle{{CodeHelp}: {Using} {Large} {Language} {Models} with {Guardrails} for {Scalable} {Support} in {Programming} {Classes}}. In \bibinfo{booktitle}{\emph{Proceedings of the 23rd {Koli} {Calling} {International} {Conference} on {Computing} {Education} {Research}}}. \bibinfo{publisher}{ACM}, \bibinfo{address}{Koli Finland}, \bibinfo{pages}{1--11}.
\newblock
\showISBNx{9798400716539}
\urldef\tempurl%
\url{https://doi.org/10.1145/3631802.3631830}
\showDOI{\tempurl}


\bibitem[Lowyck and Pöysä(2001)]%
        {lowyck_design_2001}
\bibfield{author}{\bibinfo{person}{J Lowyck} {and} \bibinfo{person}{J Pöysä}.} \bibinfo{year}{2001}\natexlab{}.
\newblock \showarticletitle{Design of collaborative learning environments}.
\newblock \bibinfo{journal}{\emph{Computers in Human Behavior}} \bibinfo{volume}{17}, \bibinfo{number}{5} (\bibinfo{date}{Sept.} \bibinfo{year}{2001}), \bibinfo{pages}{507--516}.
\newblock
\showISSN{0747-5632}
\urldef\tempurl%
\url{https://doi.org/10.1016/S0747-5632(01)00017-6}
\showDOI{\tempurl}


\bibitem[Mahon et~al\mbox{.}(2024)]%
        {mahon_guidelines_2024}
\bibfield{author}{\bibinfo{person}{Joyce Mahon}, \bibinfo{person}{Brian Mac~Namee}, {and} \bibinfo{person}{Brett~A. Becker}.} \bibinfo{year}{2024}\natexlab{}.
\newblock \showarticletitle{Guidelines for the {Evolving} {Role} of {Generative} {AI} in {Introductory} {Programming} {Based} on {Emerging} {Practice}}. In \bibinfo{booktitle}{\emph{Proceedings of the 2024 on {Innovation} and {Technology} in {Computer} {Science} {Education} {V}. 1}}. \bibinfo{publisher}{ACM}, \bibinfo{address}{Milan Italy}, \bibinfo{pages}{10--16}.
\newblock
\showISBNx{9798400706004}
\urldef\tempurl%
\url{https://doi.org/10.1145/3649217.3653602}
\showDOI{\tempurl}


\bibitem[Margulieux et~al\mbox{.}(2024)]%
        {margulieux_self-regulation_2024}
\bibfield{author}{\bibinfo{person}{Lauren~E. Margulieux}, \bibinfo{person}{James Prather}, \bibinfo{person}{Brent~N. Reeves}, \bibinfo{person}{Brett~A. Becker}, \bibinfo{person}{Gozde Cetin~Uzun}, \bibinfo{person}{Dastyni Loksa}, \bibinfo{person}{Juho Leinonen}, {and} \bibinfo{person}{Paul Denny}.} \bibinfo{year}{2024}\natexlab{}.
\newblock \showarticletitle{Self-{Regulation}, {Self}-{Efficacy}, and {Fear} of {Failure} {Interactions} with {How} {Novices} {Use} {LLMs} to {Solve} {Programming} {Problems}}. In \bibinfo{booktitle}{\emph{Proceedings of the 2024 on {Innovation} and {Technology} in {Computer} {Science} {Education} {V}. 1}} \emph{(\bibinfo{series}{{ITiCSE} 2024})}. \bibinfo{publisher}{Association for Computing Machinery}, \bibinfo{address}{New York, NY, USA}, \bibinfo{pages}{276--282}.
\newblock
\showISBNx{9798400706004}
\urldef\tempurl%
\url{https://doi.org/10.1145/3649217.3653621}
\showDOI{\tempurl}


\bibitem[Morales-Chan et~al\mbox{.}(2024)]%
        {morales-chan_ai-driven_2024}
\bibfield{author}{\bibinfo{person}{Miguel Morales-Chan}, \bibinfo{person}{Hector~R. Amado-Salvatierra}, {and} \bibinfo{person}{Rocael Hernandez-Rizzardini}.} \bibinfo{year}{2024}\natexlab{}.
\newblock \showarticletitle{{AI}-{Driven} {Content} {Creation}: {Revolutionizing} {Educational} {Materials}}. In \bibinfo{booktitle}{\emph{Proceedings of the {Eleventh} {ACM} {Conference} on {Learning} @ {Scale}}} \emph{(\bibinfo{series}{L@{S} '24})}. \bibinfo{publisher}{Association for Computing Machinery}, \bibinfo{address}{New York, NY, USA}, \bibinfo{pages}{556--558}.
\newblock
\showISBNx{9798400706332}
\urldef\tempurl%
\url{https://doi.org/10.1145/3657604.3664640}
\showDOI{\tempurl}


\bibitem[Olsen and Diekema(2012)]%
        {olsen_i_2012}
\bibfield{author}{\bibinfo{person}{M.~Whitney Olsen} {and} \bibinfo{person}{Anne~R. Diekema}.} \bibinfo{year}{2012}\natexlab{}.
\newblock \showarticletitle{“{I} just {Wikipedia} it”: {Information} behavior of first-year writing students}.
\newblock \bibinfo{journal}{\emph{Proceedings of the American Society for Information Science and Technology}} \bibinfo{volume}{49}, \bibinfo{number}{1} (\bibinfo{year}{2012}), \bibinfo{pages}{1--11}.
\newblock
\showISSN{1550-8390}
\urldef\tempurl%
\url{https://doi.org/10.1002/meet.14504901176}
\showDOI{\tempurl}
\newblock
\shownote{\_eprint: https://onlinelibrary.wiley.com/doi/pdf/10.1002/meet.14504901176}.


\bibitem[Park and Ahn(2024)]%
        {park_promise_2024}
\bibfield{author}{\bibinfo{person}{Hyanghee Park} {and} \bibinfo{person}{Daehwan Ahn}.} \bibinfo{year}{2024}\natexlab{}.
\newblock \showarticletitle{The {Promise} and {Peril} of {ChatGPT} in {Higher} {Education}: {Opportunities}, {Challenges}, and {Design} {Implications}}. In \bibinfo{booktitle}{\emph{Proceedings of the {CHI} {Conference} on {Human} {Factors} in {Computing} {Systems}}} \emph{(\bibinfo{series}{{CHI} '24})}. \bibinfo{publisher}{Association for Computing Machinery}, \bibinfo{address}{New York, NY, USA}, \bibinfo{pages}{1--21}.
\newblock
\showISBNx{9798400703300}
\urldef\tempurl%
\url{https://doi.org/10.1145/3613904.3642785}
\showDOI{\tempurl}


\bibitem[Passi et~al\mbox{.}(2024)]%
        {passi_appropriate_2024}
\bibfield{author}{\bibinfo{person}{Samir Passi}, \bibinfo{person}{Shipi Dhanorkar}, {and} \bibinfo{person}{Mihaela Vorvoreanu}.} \bibinfo{year}{2024}\natexlab{}.
\newblock \bibinfo{booktitle}{\emph{Appropriate {Reliance} on {Generative} {AI}: {Research} {Synthesis}}}.
\newblock \bibinfo{type}{Microsoft {Technical} {Report}} MSR-TR-2024-7. \bibinfo{institution}{Microsoft}.
\newblock
\urldef\tempurl%
\url{https://www.microsoft.com/en-us/research/uploads/prodnew/2024/03/GenAI_AppropriateReliance_Published2024-3-21.pdf}
\showURL{%
\tempurl}


\bibitem[Passi and Vorvoreanu(2022)]%
        {passi_overreliance_2022}
\bibfield{author}{\bibinfo{person}{Samir Passi} {and} \bibinfo{person}{Mihaela Vorvoreanu}.} \bibinfo{year}{2022}\natexlab{}.
\newblock \bibinfo{booktitle}{\emph{Overreliance on {AI}: {Literature} {Review}}}.
\newblock \bibinfo{type}{Microsoft {Technical} {Report}} MSR-TR-2022-12. \bibinfo{institution}{Microsoft}.
\newblock
\urldef\tempurl%
\url{https://www.microsoft.com/en-us/research/publication/overreliance-on-ai-literature-review/}
\showURL{%
\tempurl}


\bibitem[Petrovska et~al\mbox{.}(2024)]%
        {petrovska_incorporating_2024}
\bibfield{author}{\bibinfo{person}{Olga Petrovska}, \bibinfo{person}{Lee Clift}, \bibinfo{person}{Faron Moller}, {and} \bibinfo{person}{Rebecca Pearsall}.} \bibinfo{year}{2024}\natexlab{}.
\newblock \showarticletitle{Incorporating {Generative} {AI} into {Software} {Development} {Education}}. In \bibinfo{booktitle}{\emph{Proceedings of the 8th {Conference} on {Computing} {Education} {Practice}}} \emph{(\bibinfo{series}{{CEP} '24})}. \bibinfo{publisher}{Association for Computing Machinery}, \bibinfo{address}{New York, NY, USA}, \bibinfo{pages}{37--40}.
\newblock
\showISBNx{9798400709326}
\urldef\tempurl%
\url{https://doi.org/10.1145/3633053.3633057}
\showDOI{\tempurl}


\bibitem[Prather et~al\mbox{.}(2023a)]%
        {prather_robots_2023}
\bibfield{author}{\bibinfo{person}{James Prather}, \bibinfo{person}{Paul Denny}, \bibinfo{person}{Juho Leinonen}, \bibinfo{person}{Brett~A. Becker}, \bibinfo{person}{Ibrahim Albluwi}, \bibinfo{person}{Michelle Craig}, \bibinfo{person}{Hieke Keuning}, \bibinfo{person}{Natalie Kiesler}, \bibinfo{person}{Tobias Kohn}, \bibinfo{person}{Andrew Luxton-Reilly}, \bibinfo{person}{Stephen MacNeil}, \bibinfo{person}{Andrew Petersen}, \bibinfo{person}{Raymond Pettit}, \bibinfo{person}{Brent~N. Reeves}, {and} \bibinfo{person}{Jaromir Savelka}.} \bibinfo{year}{2023}\natexlab{a}.
\newblock \showarticletitle{The {Robots} {Are} {Here}: {Navigating} the {Generative} {AI} {Revolution} in {Computing} {Education}}. In \bibinfo{booktitle}{\emph{Proceedings of the 2023 {Working} {Group} {Reports} on {Innovation} and {Technology} in {Computer} {Science} {Education}}} \emph{(\bibinfo{series}{{ITiCSE}-{WGR} '23})}. \bibinfo{publisher}{Association for Computing Machinery}, \bibinfo{address}{New York, NY, USA}, \bibinfo{pages}{108--159}.
\newblock
\showISBNx{9798400704055}
\urldef\tempurl%
\url{https://doi.org/10.1145/3623762.3633499}
\showDOI{\tempurl}


\bibitem[Prather et~al\mbox{.}(2023b)]%
        {prather_its_2023}
\bibfield{author}{\bibinfo{person}{James Prather}, \bibinfo{person}{Brent~N. Reeves}, \bibinfo{person}{Paul Denny}, \bibinfo{person}{Brett~A. Becker}, \bibinfo{person}{Juho Leinonen}, \bibinfo{person}{Andrew Luxton-Reilly}, \bibinfo{person}{Garrett Powell}, \bibinfo{person}{James Finnie-Ansley}, {and} \bibinfo{person}{Eddie~Antonio Santos}.} \bibinfo{year}{2023}\natexlab{b}.
\newblock \bibinfo{title}{"{It}'s {Weird} {That} it {Knows} {What} {I} {Want}": {Usability} and {Interactions} with {Copilot} for {Novice} {Programmers}}.
\newblock
\newblock
\urldef\tempurl%
\url{https://doi.org/10.48550/arXiv.2304.02491}
\showDOI{\tempurl}
\newblock
\shownote{arXiv:2304.02491 [cs]}.


\bibitem[Prather et~al\mbox{.}(2024)]%
        {prather_widening_2024}
\bibfield{author}{\bibinfo{person}{James Prather}, \bibinfo{person}{Brent~N Reeves}, \bibinfo{person}{Juho Leinonen}, \bibinfo{person}{Stephen MacNeil}, \bibinfo{person}{Arisoa~S Randrianasolo}, \bibinfo{person}{Brett~A. Becker}, \bibinfo{person}{Bailey Kimmel}, \bibinfo{person}{Jared Wright}, {and} \bibinfo{person}{Ben Briggs}.} \bibinfo{year}{2024}\natexlab{}.
\newblock \showarticletitle{The {Widening} {Gap}: {The} {Benefits} and {Harms} of {Generative} {AI} for {Novice} {Programmers}}. In \bibinfo{booktitle}{\emph{Proceedings of the 2024 {ACM} {Conference} on {International} {Computing} {Education} {Research} - {Volume} 1}} \emph{(\bibinfo{series}{{ICER} '24}, Vol.~\bibinfo{volume}{1})}. \bibinfo{publisher}{Association for Computing Machinery}, \bibinfo{address}{New York, NY, USA}, \bibinfo{pages}{469--486}.
\newblock
\showISBNx{9798400704758}
\urldef\tempurl%
\url{https://doi.org/10.1145/3632620.3671116}
\showDOI{\tempurl}


\bibitem[Pudasaini et~al\mbox{.}(2024)]%
        {pudasaini_survey_2024}
\bibfield{author}{\bibinfo{person}{Shushanta Pudasaini}, \bibinfo{person}{Luis Miralles-Pechuán}, \bibinfo{person}{David Lillis}, {and} \bibinfo{person}{Marisa~Llorens Salvador}.} \bibinfo{year}{2024}\natexlab{}.
\newblock \bibinfo{title}{Survey on {Plagiarism} {Detection} in {Large} {Language} {Models}: {The} {Impact} of {ChatGPT} and {Gemini} on {Academic} {Integrity}}.
\newblock
\newblock
\urldef\tempurl%
\url{http://arxiv.org/abs/2407.13105}
\showURL{%
\tempurl}
\newblock
\shownote{arXiv:2407.13105 [cs]}.


\bibitem[Rahman(2019)]%
        {rahman_21st_2019}
\bibfield{author}{\bibinfo{person}{Md~Mehadi Rahman}.} \bibinfo{year}{2019}\natexlab{}.
\newblock \bibinfo{title}{21st {Century} {Skill} '{Problem} {Solving}': {Defining} the {Concept}}.
\newblock
\newblock
\urldef\tempurl%
\url{https://papers.ssrn.com/abstract=3660729}
\showURL{%
\tempurl}


\bibitem[Rajabi et~al\mbox{.}(2023)]%
        {rajabi_exploring_2023}
\bibfield{author}{\bibinfo{person}{Parsa Rajabi}, \bibinfo{person}{Parnian Taghipour}, \bibinfo{person}{Diana Cukierman}, {and} \bibinfo{person}{Tenzin Doleck}.} \bibinfo{year}{2023}\natexlab{}.
\newblock \showarticletitle{Exploring {ChatGPT}’s impact on post-secondary education: {A} qualitative study}. In \bibinfo{booktitle}{\emph{Proceedings of the 25th {Western} {Canadian} {Conference} on {Computing} {Education}}} \emph{(\bibinfo{series}{{WCCCE} '23})}. \bibinfo{publisher}{Association for Computing Machinery}, \bibinfo{address}{New York, NY, USA}, \bibinfo{pages}{1--6}.
\newblock
\showISBNx{9798400707896}
\urldef\tempurl%
\url{https://doi.org/10.1145/3593342.3593360}
\showDOI{\tempurl}


\bibitem[Reich(2020)]%
        {reich_failure_2020}
\bibfield{author}{\bibinfo{person}{Justin Reich}.} \bibinfo{year}{2020}\natexlab{}.
\newblock \bibinfo{booktitle}{\emph{Failure to {Disrupt}: {Why} {Technology} {Alone} {Can}’t {Transform} {Education}}}.
\newblock \bibinfo{publisher}{Harvard University Press}.
\newblock
\showISBNx{978-0-674-08904-4}
\urldef\tempurl%
\url{https://doi.org/10.2307/j.ctv322v4cp}
\showDOI{\tempurl}


\bibitem[Reich and Daccord(2015)]%
        {reich_best_2015}
\bibfield{author}{\bibinfo{person}{Justin Reich} {and} \bibinfo{person}{Tom Daccord}.} \bibinfo{year}{2015}\natexlab{}.
\newblock \bibinfo{booktitle}{\emph{Best {Ideas} for {Teaching} with {Technology}: {A} {Practical} {Guide} for {Teachers}, by {Teachers}}}.
\newblock \bibinfo{publisher}{Routledge}, \bibinfo{address}{New York}.
\newblock
\showISBNx{978-1-315-70616-0}
\urldef\tempurl%
\url{https://doi.org/10.4324/9781315706160}
\showDOI{\tempurl}


\bibitem[Rihoux and Ragin(2009)]%
        {rihoux_configurational_2009}
\bibfield{author}{\bibinfo{person}{B Rihoux} {and} \bibinfo{person}{C Ragin}.} \bibinfo{year}{2009}\natexlab{}.
\newblock \bibinfo{booktitle}{\emph{Configurational {Comparative} {Methods}: {Qualitative} {Comparative} {Analysis} ({QCA}) and {Related} {Techniques}}}.
\newblock \bibinfo{publisher}{SAGE Publications, Inc.}
\newblock
\showISBNx{978-1-4522-2656-9}
\urldef\tempurl%
\url{https://doi.org/10.4135/9781452226569}
\showDOI{\tempurl}


\bibitem[Rogers(2003)]%
        {rogers_diffusion_2003}
\bibfield{author}{\bibinfo{person}{Everett~M. Rogers}.} \bibinfo{year}{2003}\natexlab{}.
\newblock \bibinfo{booktitle}{\emph{Diffusion of {Innovations}, 5th {Edition}}}.
\newblock \bibinfo{publisher}{Simon and Schuster}.
\newblock
\showISBNx{978-0-7432-5823-4}
\newblock
\shownote{Google-Books-ID: 9U1K5LjUOwEC}.


\bibitem[Sackstein et~al\mbox{.}(2023)]%
        {sackstein_theories_2023}
\bibfield{author}{\bibinfo{person}{Suzanne Sackstein}, \bibinfo{person}{Machdel Matthee}, {and} \bibinfo{person}{Lizette Weilbach}.} \bibinfo{year}{2023}\natexlab{}.
\newblock \showarticletitle{Theories and {Models} {Employed} to {Understand} the {Use} of {Technology} in {Education}: {A} {Hermeneutic} {Literature} {Review}}.
\newblock \bibinfo{journal}{\emph{Education and Information Technologies}} \bibinfo{volume}{28}, \bibinfo{number}{5} (\bibinfo{date}{May} \bibinfo{year}{2023}), \bibinfo{pages}{5041--5081}.
\newblock
\showISSN{1360-2357, 1573-7608}
\urldef\tempurl%
\url{https://doi.org/10.1007/s10639-022-11345-5}
\showDOI{\tempurl}


\bibitem[Sackstein(2021)]%
        {sackstein_understanding_2021}
\bibfield{author}{\bibinfo{person}{Suzanne~Lee Sackstein}.} \bibinfo{year}{2021}\natexlab{}.
\newblock \emph{\bibinfo{title}{Understanding {Teachers}' {Beliefs}, {Professional} {Dispositions}, {Orientation} towards {Technology} and {Technology} {Use} in {South} {African} {Secondary} {Schools}: {A} {Longitudinal} {Micro}-, {Meso}- and {Meta}-{Theory} {Perspective}}}.
\newblock Thesis. \bibinfo{school}{University of Pretoria}.
\newblock
\urldef\tempurl%
\url{https://repository.up.ac.za/handle/2263/84906}
\showURL{%
\tempurl}
\newblock
\shownote{Accepted: 2022-04-25T06:41:13Z}.


\bibitem[Sarkar(2024)]%
        {sarkar_ai_2024}
\bibfield{author}{\bibinfo{person}{Advait Sarkar}.} \bibinfo{year}{2024}\natexlab{}.
\newblock \showarticletitle{{AI} {Should} {Challenge}, {Not} {Obey}}.
\newblock \bibinfo{journal}{\emph{Commun. ACM}} (\bibinfo{year}{2024}).
\newblock


\bibitem[Sheard et~al\mbox{.}(2024)]%
        {sheard_instructor_2024}
\bibfield{author}{\bibinfo{person}{Judy Sheard}, \bibinfo{person}{Paul Denny}, \bibinfo{person}{Arto Hellas}, \bibinfo{person}{Juho Leinonen}, \bibinfo{person}{Lauri Malmi}, {and} \bibinfo{person}{{Simon}}.} \bibinfo{year}{2024}\natexlab{}.
\newblock \showarticletitle{Instructor {Perceptions} of {AI} {Code} {Generation} {Tools} - {A} {Multi}-{Institutional} {Interview} {Study}}. In \bibinfo{booktitle}{\emph{Proceedings of the 55th {ACM} {Technical} {Symposium} on {Computer} {Science} {Education} {V}. 1}}. \bibinfo{publisher}{ACM}, \bibinfo{address}{Portland OR USA}, \bibinfo{pages}{1223--1229}.
\newblock
\showISBNx{9798400704239}
\urldef\tempurl%
\url{https://doi.org/10.1145/3626252.3630880}
\showDOI{\tempurl}


\bibitem[Sheese et~al\mbox{.}(2024)]%
        {sheese_patterns_2024}
\bibfield{author}{\bibinfo{person}{Brad Sheese}, \bibinfo{person}{Mark Liffiton}, \bibinfo{person}{Jaromir Savelka}, {and} \bibinfo{person}{Paul Denny}.} \bibinfo{year}{2024}\natexlab{}.
\newblock \showarticletitle{Patterns of {Student} {Help}-{Seeking} {When} {Using} a {Large} {Language} {Model}-{Powered} {Programming} {Assistant}}. In \bibinfo{booktitle}{\emph{Proceedings of the 26th {Australasian} {Computing} {Education} {Conference}}}. \bibinfo{publisher}{ACM}, \bibinfo{address}{Sydney NSW Australia}, \bibinfo{pages}{49--57}.
\newblock
\showISBNx{9798400716195}
\urldef\tempurl%
\url{https://doi.org/10.1145/3636243.3636249}
\showDOI{\tempurl}


\bibitem[Shoufan(2023)]%
        {shoufan_exploring_2023}
\bibfield{author}{\bibinfo{person}{Abdulhadi Shoufan}.} \bibinfo{year}{2023}\natexlab{}.
\newblock \showarticletitle{Exploring {Students}’ {Perceptions} of {ChatGPT}: {Thematic} {Analysis} and {Follow}-{Up} {Survey}}.
\newblock \bibinfo{journal}{\emph{IEEE Access}}  \bibinfo{volume}{11} (\bibinfo{year}{2023}), \bibinfo{pages}{38805--38818}.
\newblock
\showISSN{2169-3536}
\urldef\tempurl%
\url{https://doi.org/10.1109/ACCESS.2023.3268224}
\showDOI{\tempurl}
\newblock
\shownote{Conference Name: IEEE Access}.


\bibitem[Singh et~al\mbox{.}(2023b)]%
        {singh_exploring_2023}
\bibfield{author}{\bibinfo{person}{Harpreet Singh}, \bibinfo{person}{Mohammad-Hassan Tayarani-Najaran}, {and} \bibinfo{person}{Muhammad Yaqoob}.} \bibinfo{year}{2023}\natexlab{b}.
\newblock \showarticletitle{Exploring {Computer} {Science} {Students}’ {Perception} of {ChatGPT} in {Higher} {Education}: {A} {Descriptive} and {Correlation} {Study}}.
\newblock \bibinfo{journal}{\emph{Education Sciences}} \bibinfo{volume}{13}, \bibinfo{number}{9} (\bibinfo{date}{Sept.} \bibinfo{year}{2023}), \bibinfo{pages}{924}.
\newblock
\showISSN{2227-7102}
\urldef\tempurl%
\url{https://doi.org/10.3390/educsci13090924}
\showDOI{\tempurl}
\newblock
\shownote{Number: 9 Publisher: Multidisciplinary Digital Publishing Institute}.


\bibitem[Singh et~al\mbox{.}(2023a)]%
        {singh_students_2023}
\bibfield{author}{\bibinfo{person}{Pragya Singh}, \bibinfo{person}{Nidhi Phutela}, \bibinfo{person}{Priya Grover}, \bibinfo{person}{Deepti Sinha}, {and} \bibinfo{person}{Sachin Sinha}.} \bibinfo{year}{2023}\natexlab{a}.
\newblock \showarticletitle{Student’s {Perception} of {Chat} {GPT}}. In \bibinfo{booktitle}{\emph{2023 {International} {Conference} on {Electrical}, {Communication} and {Computer} {Engineering} ({ICECCE})}}. \bibinfo{pages}{1--6}.
\newblock
\urldef\tempurl%
\url{https://doi.org/10.1109/ICECCE61019.2023.10442033}
\showDOI{\tempurl}


\bibitem[Skitka et~al\mbox{.}(2000)]%
        {skitka_automation_2000}
\bibfield{author}{\bibinfo{person}{Linda~J. Skitka}, \bibinfo{person}{Kathleen~L. Mosier}, \bibinfo{person}{Mark Burdick}, {and} \bibinfo{person}{Bonnie Rosenblatt}.} \bibinfo{year}{2000}\natexlab{}.
\newblock \showarticletitle{Automation {Bias} and {Errors}: {Are} {Crews} {Better} {Than} {Individuals}?}
\newblock \bibinfo{journal}{\emph{The International Journal of Aviation Psychology}} \bibinfo{volume}{10}, \bibinfo{number}{1} (\bibinfo{date}{Jan.} \bibinfo{year}{2000}), \bibinfo{pages}{85--97}.
\newblock
\showISSN{1050-8414}
\urldef\tempurl%
\url{https://doi.org/10.1207/S15327108IJAP1001_5}
\showDOI{\tempurl}
\newblock
\shownote{Publisher: Taylor \& Francis \_eprint: https://doi.org/10.1207/S15327108IJAP1001\_5}.


\bibitem[Smolansky et~al\mbox{.}(2023)]%
        {smolansky_educator_2023}
\bibfield{author}{\bibinfo{person}{Adele Smolansky}, \bibinfo{person}{Andrew Cram}, \bibinfo{person}{Corina Raduescu}, \bibinfo{person}{Sandris Zeivots}, \bibinfo{person}{Elaine Huber}, {and} \bibinfo{person}{Rene~F. Kizilcec}.} \bibinfo{year}{2023}\natexlab{}.
\newblock \showarticletitle{Educator and {Student} {Perspectives} on the {Impact} of {Generative} {AI} on {Assessments} in {Higher} {Education}}. In \bibinfo{booktitle}{\emph{Proceedings of the {Tenth} {ACM} {Conference} on {Learning} @ {Scale}}} \emph{(\bibinfo{series}{L@{S} '23})}. \bibinfo{publisher}{Association for Computing Machinery}, \bibinfo{address}{New York, NY, USA}, \bibinfo{pages}{378--382}.
\newblock
\showISBNx{9798400700255}
\urldef\tempurl%
\url{https://doi.org/10.1145/3573051.3596191}
\showDOI{\tempurl}


\bibitem[Stone(2024)]%
        {stone_exploring_2024}
\bibfield{author}{\bibinfo{person}{Irene Stone}.} \bibinfo{year}{2024}\natexlab{}.
\newblock \showarticletitle{Exploring {Human}-{Centered} {Approaches} in {Generative} {AI} and {Introductory} {Programming} {Research}: {A} {Scoping} {Review}}. In \bibinfo{booktitle}{\emph{Proceedings of the 2024 {Conference} on {United} {Kingdom} \& {Ireland} {Computing} {Education} {Research}}}. \bibinfo{publisher}{ACM}, \bibinfo{address}{Manchester United Kingdom}, \bibinfo{pages}{1--7}.
\newblock
\showISBNx{9798400711770}
\urldef\tempurl%
\url{https://doi.org/10.1145/3689535.3689553}
\showDOI{\tempurl}


\bibitem[Stones(2005)]%
        {stones_structuration_2005}
\bibfield{author}{\bibinfo{person}{Rob Stones}.} \bibinfo{year}{2005}\natexlab{}.
\newblock \bibinfo{booktitle}{\emph{Structuration {Theory}}}.
\newblock \bibinfo{publisher}{Palgrave-Macmillan}.
\newblock


\bibitem[Tan and Subramonyam(2024)]%
        {tan_more_2024}
\bibfield{author}{\bibinfo{person}{Mei Tan} {and} \bibinfo{person}{Hari Subramonyam}.} \bibinfo{year}{2024}\natexlab{}.
\newblock \showarticletitle{More than {Model} {Documentation}: {Uncovering} {Teachers}' {Bespoke} {Information} {Needs} for {Informed} {Classroom} {Integration} of {ChatGPT}}. In \bibinfo{booktitle}{\emph{Proceedings of the 2024 {CHI} {Conference} on {Human} {Factors} in {Computing} {Systems}}} \emph{(\bibinfo{series}{{CHI} '24})}. \bibinfo{publisher}{Association for Computing Machinery}, \bibinfo{address}{New York, NY, USA}, \bibinfo{pages}{1--19}.
\newblock
\showISBNx{9798400703300}
\urldef\tempurl%
\url{https://doi.org/10.1145/3613904.3642592}
\showDOI{\tempurl}


\bibitem[Tankelevitch et~al\mbox{.}(2024)]%
        {tankelevitch_metacognitive_2024}
\bibfield{author}{\bibinfo{person}{Lev Tankelevitch}, \bibinfo{person}{Viktor Kewenig}, \bibinfo{person}{Auste Simkute}, \bibinfo{person}{Ava~Elizabeth Scott}, \bibinfo{person}{Advait Sarkar}, \bibinfo{person}{Abigail Sellen}, {and} \bibinfo{person}{Sean Rintel}.} \bibinfo{year}{2024}\natexlab{}.
\newblock \showarticletitle{The {Metacognitive} {Demands} and {Opportunities} of {Generative} {AI}}. In \bibinfo{booktitle}{\emph{Proceedings of the {CHI} {Conference} on {Human} {Factors} in {Computing} {Systems}}} \emph{(\bibinfo{series}{{CHI} '24})}. \bibinfo{publisher}{Association for Computing Machinery}, \bibinfo{address}{New York, NY, USA}, \bibinfo{pages}{1--24}.
\newblock
\showISBNx{9798400703300}
\urldef\tempurl%
\url{https://doi.org/10.1145/3613904.3642902}
\showDOI{\tempurl}


\bibitem[Thomas(2006)]%
        {thomas_general_2006}
\bibfield{author}{\bibinfo{person}{David~R. Thomas}.} \bibinfo{year}{2006}\natexlab{}.
\newblock \showarticletitle{A general inductive approach for analyzing qualitative evaluation data}.
\newblock \bibinfo{journal}{\emph{American Journal of Evaluation}} \bibinfo{volume}{27}, \bibinfo{number}{2} (\bibinfo{year}{2006}), \bibinfo{pages}{237--246}.
\newblock
\showISSN{1557-0878}
\urldef\tempurl%
\url{https://doi.org/10.1177/1098214005283748}
\showDOI{\tempurl}
\newblock
\shownote{Place: US Publisher: Sage Publications}.


\bibitem[Tlili et~al\mbox{.}(2023)]%
        {tlili_what_2023}
\bibfield{author}{\bibinfo{person}{Ahmed Tlili}, \bibinfo{person}{Boulus Shehata}, \bibinfo{person}{Michael~Agyemang Adarkwah}, \bibinfo{person}{Aras Bozkurt}, \bibinfo{person}{Daniel~T. Hickey}, \bibinfo{person}{Ronghuai Huang}, {and} \bibinfo{person}{Brighter Agyemang}.} \bibinfo{year}{2023}\natexlab{}.
\newblock \showarticletitle{What if the devil is my guardian angel: {ChatGPT} as a case study of using chatbots in education}.
\newblock \bibinfo{journal}{\emph{Smart Learning Environments}} \bibinfo{volume}{10}, \bibinfo{number}{1} (\bibinfo{date}{Feb.} \bibinfo{year}{2023}), \bibinfo{pages}{15}.
\newblock
\showISSN{2196-7091}
\urldef\tempurl%
\url{https://doi.org/10.1186/s40561-023-00237-x}
\showDOI{\tempurl}


\bibitem[Usdan et~al\mbox{.}(2024)]%
        {usdan_generative_2024}
\bibfield{author}{\bibinfo{person}{Jordan Usdan}, \bibinfo{person}{Allison Connell~Pensky}, {and} \bibinfo{person}{Harley Chang}.} \bibinfo{year}{2024}\natexlab{}.
\newblock \bibinfo{title}{Generative {AI}'s {Impact} on {Graduate} {Student} {Writing} {Productivity} and {Quality}}.
\newblock
\newblock
\urldef\tempurl%
\url{https://papers.ssrn.com/abstract=4941022}
\showURL{%
\tempurl}


\bibitem[VanLehn(2011)]%
        {vanlehn_relative_2011}
\bibfield{author}{\bibinfo{person}{Kurt VanLehn}.} \bibinfo{year}{2011}\natexlab{}.
\newblock \showarticletitle{The relative effectiveness of human tutoring, intelligent tutoring systems, and other tutoring systems}.
\newblock \bibinfo{journal}{\emph{Educational Psychologist}} \bibinfo{volume}{46}, \bibinfo{number}{4} (\bibinfo{year}{2011}), \bibinfo{pages}{197--221}.
\newblock
\showISSN{1532-6985}
\urldef\tempurl%
\url{https://doi.org/10.1080/00461520.2011.611369}
\showDOI{\tempurl}
\newblock
\shownote{Place: United Kingdom Publisher: Taylor \& Francis}.


\bibitem[Wang et~al\mbox{.}(2024)]%
        {wang_tutor_2024}
\bibfield{author}{\bibinfo{person}{Rose~E. Wang}, \bibinfo{person}{Ana~T. Ribeiro}, \bibinfo{person}{Carly~D. Robinson}, \bibinfo{person}{Susanna Loeb}, {and} \bibinfo{person}{Dora Demszky}.} \bibinfo{year}{2024}\natexlab{}.
\newblock \bibinfo{title}{Tutor {CoPilot}: {A} {Human}-{AI} {Approach} for {Scaling} {Real}-{Time} {Expertise}}.
\newblock
\newblock
\urldef\tempurl%
\url{http://arxiv.org/abs/2410.03017}
\showURL{%
\tempurl}
\newblock
\shownote{arXiv:2410.03017}.


\bibitem[Winne and Nesbit(2010)]%
        {winne_psychology_2010}
\bibfield{author}{\bibinfo{person}{Philip~H. Winne} {and} \bibinfo{person}{John~C. Nesbit}.} \bibinfo{year}{2010}\natexlab{}.
\newblock \showarticletitle{The psychology of academic achievement}.
\newblock \bibinfo{journal}{\emph{Annual Review of Psychology}}  \bibinfo{volume}{61} (\bibinfo{year}{2010}), \bibinfo{pages}{653--678}.
\newblock
\showISSN{1545-2085}
\urldef\tempurl%
\url{https://doi.org/10.1146/annurev.psych.093008.100348}
\showDOI{\tempurl}
\newblock
\shownote{Place: US Publisher: Annual Reviews}.


\bibitem[Zastudil et~al\mbox{.}(2023)]%
        {zastudil_generative_2023}
\bibfield{author}{\bibinfo{person}{Cynthia Zastudil}, \bibinfo{person}{Magdalena Rogalska}, \bibinfo{person}{Christine Kapp}, \bibinfo{person}{Jennifer Vaughn}, {and} \bibinfo{person}{Stephen MacNeil}.} \bibinfo{year}{2023}\natexlab{}.
\newblock \showarticletitle{Generative {AI} in {Computing} {Education}: {Perspectives} of {Students} and {Instructors}}. In \bibinfo{booktitle}{\emph{2023 {IEEE} {Frontiers} in {Education} {Conference} ({FIE})}}. \bibinfo{pages}{1--9}.
\newblock
\urldef\tempurl%
\url{https://doi.org/10.1109/FIE58773.2023.10343467}
\showDOI{\tempurl}
\newblock
\shownote{ISSN: 2377-634X}.


\end{thebibliography}

\appendix \label{appendix}

\section{Pre-interview survey questions} \label{app:survey-questions}

Pre-Interview Survey 

(1) First and last name:

(2) Email address: 

(only be used for scheduling interviews and delivering gift vouchers)  

(3) Age (Educators) 

\begin{itemize}
    \item 18-29
    \item 30-44 
    \item 45-59 
    \item 60 and over
    \item Prefer not to answer
\end{itemize}  

(4) Age (Students) 

\begin{itemize}
\item 18-21 
\item 22-25 
\item 26-29 
\item 29-32 
\item 32-35 
\item 36 or over 
\item Prefer not to answer 
\end{itemize} 

(5) Gender 

\begin{itemize}
    \item Man 
    \item Woman
    \item Non-binary/gender-diverse 
    \item Prefer not to answer
    \item Self-described
\end{itemize}

(6) University 

\begin{itemize}
    \item <Anonymized>
    \item <Anonymized>
\end{itemize}

(7) Year of Study

(8) Degree / Course Area? (E.g. Biology, Philosophy, Psychology, etc.)

(9) Is English your first language? 

\begin{itemize}
    \item Yes
    \item No
\end{itemize}

(10)How comfortable are you with writing in English?

\begin{itemize}
    \item Not at all comfortable
    \item A little bit
    \item Somewhat
    \item Quite a bit
    \item Very comfortable
\end{itemize}

(11) How comfortable are you with using technology? 

\begin{itemize}
    \item Not at all comfortable
    \item A little bit
    \item Somewhat
    \item Quite a bit
    \item Very comfortable
\end{itemize}

(12) How early or late an adopter are you of new technologies?  

\begin{itemize}
    \item Very early adopter
    \item Somewhat early adopter
    \item Neither early nor late
    \item Somewhat late adopter
    \item Very late adopter
\end{itemize}

(13) How frequently do you use ChatGPT or similar AI tools?  

\begin{itemize}
    \item N/A – haven’t heard of this before
    \item Never
    \item Less than once per month
    \item About once per month
    \item A few times per month
    \item A few times per week
    \item At least once per day
\end{itemize}  

(14) For what purposes do you use ChatGPT or similar AI tools? 

\begin{itemize}
    \item N/A - I do not use ChatGPT or similar AI tools
    \item Playing around / exploring it for fun
    \item Summarising written content from daily communications (e.g., emails, messages)
    \item Generating written content for daily communications (e.g., emails, messages)
    \item Summarising written content from longer documents (e.g., articles, books)
    \item Generating written content for longer documents (e.g., essays, reports, presentations)
    \item Generating images for educational/work purposes/
    \item Generating video for educational/work purposes/ Brainstorming ideas
    \item Researching information
    \item Studying information (e.g., doing quizzes)
    \item Practising skills
    \item Self-reflection
    \item Planning
    \item Coding
    \item Other (please specify)
\end{itemize}

(15) To what extent do you think that Artificial Intelligence can significantly enhance the learning process for students? 

\begin{itemize}
    \item Not at all
    \item A little bit
    \item Somewhat
    \item Quite a bit
    \item Very much
\end{itemize}

(16) To what extent do you think that Artificial Intelligence can significantly enhance the teaching process for teachers? 

\begin{itemize}
    \item Not at all
    \item A little bit
    \item Somewhat
    \item Quite a bit
    \item Very much
\end{itemize}

(17) How concerned are you about the increasing use of AI in Education?  

\begin{itemize}
    \item Not at all concerned
    \item A little bit
    \item Somewhat
    \item Quite a bit
    \item Very concerned
\end{itemize}

(18) [optional] Do you have any other comments you’d like to share about the use of ChatGPT or similar AI tools in education (or anything else related)?  

\section{Semi-structured interview protocol}\label{app:interview-protocol}

\textbf{Interview – Student participants} 

\textit{Introductions and hit record} 

We would like to speak to you about your experience with and usage of ChatGPT or similar AI tools. Our goal is to understand what some emerging unspoken rules or etiquettes are of how people interact with these kind of systems. There are no right or wrong answers, and we are not evaluating you – we are just interested in gathering a variety of people’s experiences. Please ask if anything is unclear.\vspace{0.3\baselineskip} 

\textbf{General use} 

\textit{Key questions} 

\begin{itemize}
    \item How often do you use the new AI tools like ChatGPT, Dall-E, Stable Diffusion, Midjourney, etc. and what tasks do you primarily use them for? 
    \item Have you encountered challenges or limitations when using these new AI tools? 
\end{itemize}\vspace{0.3\baselineskip} 

\textbf{General trust} 

\textit{Key question}

\begin{itemize}
    \item How much do you trust the answers provided by these new AI tools? 
\end{itemize}

\textit{Prompts} 

\begin{itemize}
    \item Have you ever felt misled by the responses of these new AI tools? If so, can you provide an example? 
    \item Do you double-check the answers or solutions provided by these new AI tools? If so, how often? 
    \item How does using these new AI tools make you feel about your own abilities? 
\end{itemize}\vspace{0.3\baselineskip} 

\textbf{Norms around educational use} 

\textit{Key question}

\begin{itemize}
    \item How do students discuss or share these new AI tools? 
\end{itemize}

\textit{Prompts} 

\begin{itemize}
    \item Are there norms emerging around the use of these AI tools amongst students at your university (e.g. do students talk about doing assignments with these new AI tools?) 
    \item Are there things about these new AI tools that you disagree on with other students? 
    \item Are there any written guidelines you follow to use these new AI tools? 
    \item Are there any "unspoken rules" you follow when using these new AI tools?
    \item Do you think using these new AI tools is like cheating? 
\end{itemize}\vspace{0.3\baselineskip} 

\textbf{Comparison with traditional educational methods} 

\textit{Key question}

\begin{itemize}
    \item In what ways have these new AI tools changed or affected your learning processes? 
\end{itemize}

\textit{Prompts} 

\begin{itemize}
    \item How does using these new AI tools compare to traditional methods of seeking information, like textbooks, asking a lecturer, or even searching the Internet? 
    \item Do you feel that these new AI tools complement, replace, or disrupts traditional learning tools? 
    \item Are there some kinds of learning assignments, where it feels more important to do all the work yourself, rather than using results from these new AI tools? 
    \item How do you decide when it's appropriate to use these new AI tools versus seeking help from a person? 
    \item Are there certain topics or areas where you feel these new AI tools are less effective or helpful? 
    \item Do you feel there are topics or questions that shouldn't be asked of these new AI tools? 
\end{itemize}\vspace{0.3\baselineskip} 

\textbf{Future Prospects and Changes} 

\textit{Key question}

\begin{itemize}
    \item How do you see the role of these new AI tools evolving at universities over the next few years? 
\end{itemize}

\textit{Prompts}

\begin{itemize}
    \item Are there features or changes you'd like to see in these new AI tools to make them more useful for educational purposes?
    \item Has anything changed about your perception of these new AI tools after this interview?  
\end{itemize} \vspace{0.3\baselineskip} 

\textbf{Interview – University lecturer/tutor} 

\textit{Introductions and hit record} 

We would like speak to you about your experience with and usage of ChatGPT or similar AI tools. Our goal is to understand what some emerging unspoken rules or etiquettes are of how people interact with these kind of systems. There are no right or wrong answers, and we are not evaluating you – we are just interested in gathering a variety of people’s experiences. Please ask if anything is unclear. \vspace{0.3\baselineskip} 

\textbf{General use}  

\textit{Key questions} 

\begin{itemize}
    \item How often do you use the new AI tools like ChatGPT, Dall-E, Stable Diffusion, Midjourney, etc. and what tasks do you primarily use them for? 
    \item Have you encountered challenges or limitations when using these new AI tools? 
\end{itemize}\vspace{0.3\baselineskip} 

\textbf{General trust}

\textit{Key question}

\begin{itemize}
    \item How much do you trust the answers provided by these new AI tools? 
\end{itemize}

\textit{Prompts} 

\begin{itemize}
    \item Have you ever felt misled by the responses of these new AI tools? If so, can you provide an example? 
    \item Do you double-check the answers or solutions provided by these new AI tools? If so, how often? 
    \item How does using these new AI tools make you feel about your own abilities? 
\end{itemize}\vspace{0.3\baselineskip} 

\textbf{Norms around educational use} 

\textit{Key questions} 

\begin{itemize}
    \item How do teaching staff (lecturers/tutors) discuss or share these new AI tools? 
    \item How do teaching staff talk about students using these new AI tools? 
\end{itemize}

\textit{Prompts} 

\begin{itemize}
    \item Are there norms emerging around the use of these AI tools amongst teaching staff or students? 
    \item Are there things about these new AI tools that you disagree on with other teaching staff? 
    \item Are there any written guidelines for using these new AI tools at your university? 
    \item Are there any "unspoken rules" for using these new AI tools among teaching staff or students? 
    \item Do you think using these new AI tools is like cheating by teaching staff or students? 
\end{itemize}\vspace{0.3\baselineskip} 

\textbf{Comparison with traditional educational methods} 

\textit{Key question}

\begin{itemize}
    \item In what ways have these new AI tools changed or affected your teaching processes? 
\end{itemize}

\textit{Prompts} 

\begin{itemize}
    \item How does using these new AI tools compare to your traditional methods of teaching (e.g. preparing resources, actual teaching, logistics of teaching)? 
    \item Do you feel that these new AI tools complement, replace, or disrupts traditional teaching or learning tools? 
    \item Are there some kinds of teaching work where it feels more important to you to do all the work yourself, rather than using results from these new AI tools? 
    \item How do you decide when it's appropriate to use these new AI tools versus seeking help from a person? 
    \item Are there certain topics or areas where you feel these new AI tools are less effective or helpful? 
    \item Do you feel there are topics or questions that shouldn't be asked of these new AI tools? 
\end{itemize}\vspace{0.3\baselineskip} 

\textbf{Future Prospects and Changes} 

\textit{Key question}

\begin{itemize}
    \item How do you see the role of these new AI tools evolving at universities over the next few years? 
\end{itemize}

\textit{Prompts}

\begin{itemize}
    \item Are there features or changes you'd like to see in these new AI tools to make them more useful for educational purposes? 
    \item Has anything changed about your perception of these new AI tools after this interview? 
\end{itemize}

\section{Survey responses} \label{app:survey-responses}
\begin{figure}[H]%
  \centering
  \includegraphics[width=\linewidth]{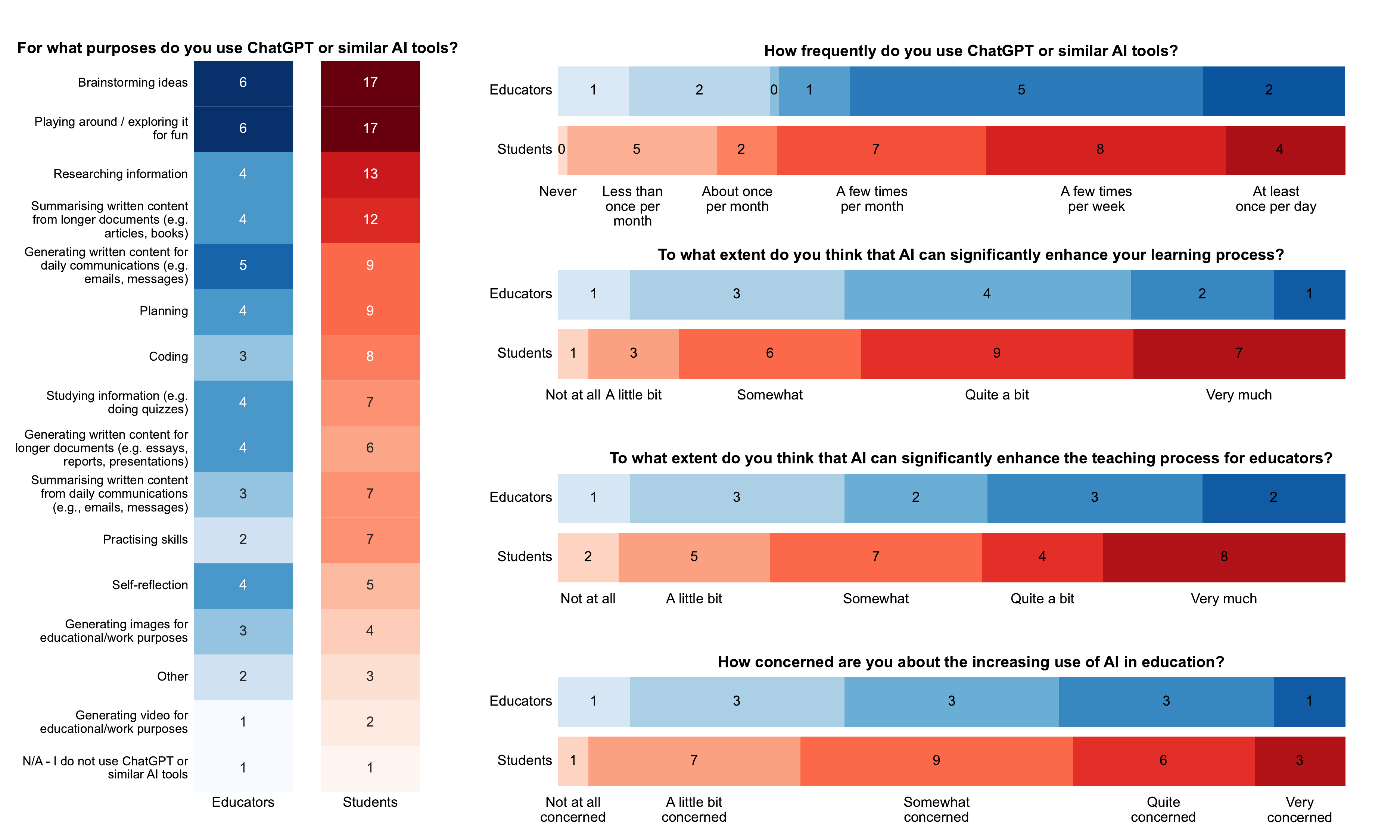}
  \caption{Survey findings. (A) Students' and educators' uses of GenAI. (B) Students' and educators' frequency of GenAI use. (C) Students' and educators' perceptions of whether GenAI can enhance the learning process. (D) Students' and educators' perceptions of whether GenAI can enhance the teaching process. (E) Students' and educators' level of concern about the increasing use of GenAI in education. Numbers indicate counts.}
  \Description{separated into educators and students}
  \label{FIG:survey-findings}
\end{figure}

Both educators and students are actively using AI tools like ChatGPT (\autoref{FIG:survey-findings}A). The most common purposes reported by both groups are brainstorming ideas and exploring for fun, but both groups also reported researching information, and summarising and generating written content (\autoref{FIG:survey-findings}B). Students are more optimistic about AI's potential to enhance learning, with 16 out of 26 students believing AI can enhance learning `quite a bit' or `very much', compared to 3 out of 11 educators (\autoref{FIG:survey-findings}C). Both groups show mixed opinions on AI's potential to enhance teaching, with slightly more positive views among students (\autoref{FIG:survey-findings}D). Concerns about AI in education are present in both groups (\autoref{FIG:survey-findings}E). %
Overall, the data suggest that while AI tools are viewed as useful and are being adopted by both groups, there remains some caution about their full integration into education.

\section{Open Research Questions}\label{app:futurequestions}

Our findings suggest a variety of challenges and opportunities for successfully integrating GenAI tools into education. %
The following table summarizes key open research questions identified from our research.

\renewcommand{\arraystretch}{1.4}
\begin{table*}[h!]
\small
  \caption{Open research questions related to GenAI tools in education}
  \label{tab:research-questions-genai}
  \begin{tabular}{p{0.2\linewidth}  p{0.7\linewidth}}
    \toprule
    \textbf{Area} & \textbf{Research questions} \\
    \midrule
    \textbf{Current conjuncture} & \\
    \hline
    \multicolumn{2}{l}{\textit{Unclear guidelines and fixation on plagiarism}} \\
    & How can universities develop clear, systematic guidelines for GenAI use that go beyond plagiarism, and how should these guidelines be effectively communicated to both educators and students? \\
    & What participatory design methods (e.g., involving educators and students in policy development) can increase the relevance and acceptance of GenAI guidelines? \\
    & How can creating a balanced and pragmatic approach to GenAI use help reduce "plagiarism anxiety" and promote a constructive dialogue about GenAI tools in education? \\
    \hline
    \multicolumn{2}{l}{\textit{Educators' lack of communication about GenAI}} \\
    & What are the barriers that prevent open communication about GenAI among educators, and how can these be overcome to foster consistent teaching practices? \\
    & How can universities empower educators to share their experiences and knowledge about integrating GenAI tools, especially when formal policies are not yet established? \\
    \midrule
    \textbf{Anticipated changes in \textit{external} structures} & \\
    \hline
    \multicolumn{2}{l}{\textit{GenAI literacy among students and educators}} \\
    & What types of multi-faceted GenAI literacy training programs (e.g., peer-assisted learning groups, error recognition tasks) can support effective GenAI tool use among students and educators? \\
    & How can formalizing collaborative GenAI learning environments enhance engagement and mitigate disparities in GenAI literacy? \\
    \hline
    \multicolumn{2}{l}{\textit{Academic integrity and innovating assessments}} \\
    & How can assessments be redesigned to incorporate GenAI tools while maintaining academic integrity and fostering skills such as critical thinking and creativity? \\
    & What safeguards (e.g., cognitive forcing techniques) are needed to prevent over-reliance on GenAI tools and protect essential skills development? \\
    \hline
    \multicolumn{2}{l}{\textit{Reimagining interpersonal relationships in education}} \\
    & How can the design of GenAI-enabled personalized learning include roles for educators and foster interpersonal interaction among students? \\
    & What strategies can ensure that GenAI complements, rather than replaces, the human aspects of education, such as mentorship and peer collaboration? \\
    \midrule
    \textbf{Anticipated changes in \textit{internal} structures} & \\
    \hline
    \multicolumn{2}{l}{\textit{Long-term impact on skill development}} \\
    & What are the long-term effects of GenAI use on students' skill development, and how can educational practices be adapted to balance these effects? \\
    & How can students' self-regulated learning strategies be supported to enhance critical thinking and prevent over-reliance on GenAI tools? \\
    \multicolumn{2}{l}{\textit{Integrating personalized learning with GenAI}} \\
    & What design principles should guide the development of GenAI tools that support personalized learning while maintaining the role of human educators and collaborative learning? \\
    & How can GenAI tools be designed to balance learner autonomy with external guidance to promote effective learning outcomes? \\
  \bottomrule
\end{tabular}
\end{table*}
\renewcommand{\arraystretch}{1}

\end{document}